\documentclass[11pt]{article}

\usepackage[a4paper,margin=1in]{geometry}
\usepackage[T1]{fontenc}
\usepackage[utf8]{inputenc}
\usepackage{lmodern}
\usepackage{amsmath,amssymb,amsfonts,amsthm,bm}
\usepackage{booktabs,longtable,array,multirow}
\usepackage{graphicx}
\usepackage[labelfont=bf,labelsep=space]{caption}
\usepackage[authoryear,round]{natbib}
\usepackage{placeins}
\usepackage{needspace}
\usepackage[hidelinks]{hyperref}

\newtheorem{assumption}{Assumption}
\newtheorem{proposition}{Proposition}

\newcommand{\E}{\mathbb{E}}
\newcommand{\Var}{\operatorname{Var}}
\newcommand{\Cov}{\operatorname{Cov}}
\newcommand{\indep}{\perp\!\!\!\perp}
\newcommand{\bX}{\bm{X}}
\newcommand{\bmu}{\bm{\mu}}
\newcommand{\bbeta}{\bm{\beta}}
\newcommand{\btheta}{\bm{\theta}}

\newcommand{\bPsi}{\bm{\Psi}}
\newcommand{\trans}{^{\mathsf T}}
\newcommand{\PairedRMSERange}{0.245--0.841}
\newcommand{\PairedLengthRange}{0.247--0.857}

\newcommand{\ReportCoverRange}{0.828--0.959}

\newcommand{\CalOLSNull}{0.054}
\newcommand{\CalOLSPower}{0.112}
\newcommand{\CalHuberFixNull}{0.057}
\newcommand{\CalHuberFixPower}{0.652}
\newcommand{\CalHuberSelNull}{0.052}
\newcommand{\CalHuberSelPower}{0.721}
\newcommand{\CalSemiNull}{0.052}
\newcommand{\CalSemiPower}{0.927}
\newcommand{\CalSemiSuccessRange}{0.960--0.967}
\newcommand{\CalSemiReportPower}{0.892}
\newcommand{\CalHuberSelContrast}{0.204}
\newcommand{\CalHuberSelContrastCI}{[0.172, 0.238]}

\title{Semiparametric Inference and Sensitivity Analysis for Causal Mediation in Linear Models with Unspecified Error Distributions}
\author{Mijeong Kim\\
\small Department of Statistics, Ewha Womans University, Seoul, Republic of Korea\\
\small Corresponding author: \href{mailto:m.kim@ewha.ac.kr}{m.kim@ewha.ac.kr}\\
\small ORCID: \href{https://orcid.org/0000-0002-3578-5413}{0000-0002-3578-5413}}
\date{}

\begin{document}
\maketitle

\begin{abstract}
Causal mediation analysis distinguishes pathways through which a treatment affects an outcome. In linear models, ordinary least squares may leave information in the error distribution unused.
We develop a semiparametric framework for joint inference in linear mediation models with treatment--mediator interaction and unspecified error distributions. A stacked estimating system integrates regression-error scores and propagates uncertainty from both regressions and the baseline-covariate distribution through natural-effect decompositions. Under the stated model and regularity conditions, we characterize the natural-effect influence functions and establish an OLS-relative asymptotic variance and local-power ordering.
We further develop density-score estimation and joint inference for residual-correlation sensitivity analysis of unmeasured mediator--outcome confounding. Reduced-form regressions and joint uncertainty propagation for their coefficients, residual variances and cross-covariances yield pointwise sensitivity curves and variance-explained contours under explicit working assumptions.
Simulations show reduced root mean squared error across several non-Gaussian settings and markedly higher power for a weak indirect effect under asymmetric-mixture errors. The power advantage over ordinary least squares and interaction-adapted Huber comparators persists after independent null calibration. Evaluations also examine interval coverage, numerical success, full sensitivity curves and variance misspecification.
An analysis of the randomized JOBS II behavioral intervention illustrates how natural-effect estimates, uncertainty and confounding sensitivity can be interpreted together. Complete estimating equations, a deterministic multi-start algorithm and reproducible numerical output accompany the analysis.
\end{abstract}

\noindent\textbf{Keywords:} causal mediation; natural effects; sensitivity analysis; semiparametric estimation; unspecified error distributions

\section{Introduction}

Natural direct and indirect effects provide counterfactual descriptions of how an exposure may operate through a mediator \citep{robins1992,pearl2001,imai2010a,vanderweele2015}.  In linear models these effects reduce to familiar functions of coefficients from a mediator regression and an outcome regression.  Ordinary least squares (OLS) offers a convenient way to estimate these coefficients without specifying an error distribution.  A further question is whether estimating the shape of that distribution can improve precision when the regression model is correctly specified.  Skewed and multimodal errors make this question relevant even within a simple linear mediation analysis.

With treatment--mediator interaction, the total effect admits two equally valid natural-effect decompositions \citep{valeri2013,vanderweele2015}.  The pure natural direct effect (PNDE) combines with the total natural indirect effect (TNIE), while the total natural direct effect (TNDE) combines with the pure natural indirect effect (PNIE).  Each component remains a single population quantity.  Distinguishing these decompositions is important both substantively and computationally, because all four components depend on coefficients from linked regressions.

Existing semiparametric mediation methods provide several routes to natural-effect inference.  \citet{tchetgen2012} derived efficiency bounds and multiply robust estimators, \citet{chan2016} developed nonparametric estimation through balancing weights, and \citet{farbmacher2022} combined efficient scores with sample splitting and machine learning for flexible covariate adjustment.  We address a complementary question: how estimated error-density scores can improve natural-effect inference within linked linear models.  The regression model in Section~\ref{sec:preliminaries} defines the corresponding efficiency benchmark and permits an explicit characterization of the information gain over OLS.

A separate literature addresses distributional robustness in regression-based mediation.  \citet{yuan2014} used median regression, while \citet{alfons2022} implemented robust regression and bootstrap procedures in \texttt{robmed}; \citet{alfons2025} review the different notions of robustness.  \citet{wang2025} developed a Huber-loss approach balancing robustness and efficiency.  Parametric structural equation models also accommodate skewness through specified non-Gaussian families \citep{asparouhov2016}.  Here the target is a mean-based natural effect, and the error density is estimated without selecting a parametric family.  The focus is on extracting error-shape information for precision, complementing methods designed for resistance to contamination or high-leverage observations.

Building on the regression scores of \citet{kim2023}, this article connects error-density information to joint mediation inference, precision and detection of weak effects.  Three contributions are central.  First, a joint estimating system for the two regressions and the baseline-covariate mean yields natural-effect influence functions and propagates uncertainty through both decompositions.  Second, an explicit asymptotic variance gap identifies when the feasible estimator improves on OLS and links that gain to local power.  Third, observation-level score equations, deterministic multi-start computation and joint covariance estimation provide a reproducible procedure.  Its evaluation covers precision, coverage and numerical success, including matched-data comparisons with OLS, the quasi-Bayesian and percentile-bootstrap procedures in the \textsf{R} package \texttt{mediation}, and natural effect models in the \textsf{R} package \texttt{medflex}.  Independently calibrated power comparisons also include interaction-adapted Huber estimators.  Section~\ref{sec:preliminaries} states the statistical and causal assumptions.

A further extension develops density-score estimation and joint inference for the residual-correlation sensitivity functional of \citet{imai2010a}. Related work establishes semiparametric sensitivity formulations \citep{tchetgen2012} and, more recently, debiased-machine-learning procedures with $R^2$ contours and robustness summaries \citep{liu2026}. Within this literature, our reduced-form construction accommodates unspecified marginal error densities and propagates the complete covariance of regression coefficients, residual variances, cross-covariances and baseline means through the sensitivity functional. It provides pointwise sensitivity curves, residual-variance contours and zero-boundary inference, together with a full-curve Monte Carlo assessment. The analysis combines efficient marginal regression scores under the working model with a joint influence-function representation for sensitivity inference.

Section~\ref{sec:methodology} develops the estimation, inference and OLS-relative precision and local-power results. Sections~\ref{sec:simulations} and \ref{sec:confounding} evaluate the procedure and its confounding-sensitivity extension, followed by the JOBS II application and discussion. The Supplementary Material follows the same sequence and supplies proofs, implementation details and complete numerical results.

\section{Semiparametric estimation and inference}
\label{sec:methodology}

\subsection{Model and natural-effect preliminaries}
\label{sec:preliminaries}

Let the observed data be independent copies of $O=(\bX,T,M,Y)$, where $T\in\{0,1\}$ is treatment, $M$ is a continuous mediator, $Y$ is a continuous outcome, and $\bX$ contains pretreatment covariates.  We posit
\begin{align}
M &= \alpha_2+\beta_2T+\bm{\xi}_2\trans\bX+\epsilon_M,
\label{eq:mediator-model}\\
Y &= \alpha_3+\beta_3T+\gamma M+\eta TM+\bm{\xi}_3\trans\bX+\epsilon_Y.
\label{eq:outcome-model}
\end{align}
For estimation we assume $\epsilon_M\indep(T,\bX)$, $\epsilon_Y\indep(T,M,\bX)$, zero error means, finite positive variances, and otherwise unknown continuously differentiable error densities.  Thus the errors are homoscedastic with freely varying non-Gaussian shapes.  These restrictions define the statistical model used for the efficiency calculation, while the causal interpretation follows from the identification assumptions below.

Write $M(t)$ for the potential mediator under treatment $t$ and $Y(t,m)$ for the potential outcome under treatment $t$ and mediator value $m$.  We use consistency, positivity, and sequential ignorability:
\begin{assumption}[Sequential ignorability]
For $t,t'\in\{0,1\}$ and relevant $(m,\bm{x})$,
\[
\{Y(t',m),M(t)\}\indep T\mid\bX,
\qquad
Y(t',m)\indep M(t)\mid T=t,\bX=\bm{x}.
\]
\end{assumption}
The second condition rules out unmeasured mediator--outcome confounding after conditioning on treatment and baseline covariates.  It remains substantive even when treatment is randomized.

Define
\begin{align*}
\delta(t)&=\E\{Y(t,M(1))-Y(t,M(0))\},\\
\zeta(t)&=\E\{Y(1,M(t))-Y(0,M(t))\},\\
\tau&=\E\{Y(1,M(1))-Y(0,M(0))\}.
\end{align*}
The conventional labels are pure natural indirect effect (PNIE), total natural indirect effect (TNIE), pure natural direct effect (PNDE), total natural direct effect (TNDE), and total effect (TE):
\[
\operatorname{PNIE}=\delta(0),\quad
\operatorname{TNIE}=\delta(1),\quad
\operatorname{PNDE}=\zeta(0),\quad
\operatorname{TNDE}=\zeta(1),\quad
\operatorname{TE}=\tau.
\]

Let $\bmu_X=\E(\bX)$, $\mu_0=\alpha_2+\bm{\xi}_2\trans\bmu_X$, and $\mu_1=\mu_0+\beta_2$. Under Assumption~1 and models \eqref{eq:mediator-model}--\eqref{eq:outcome-model}, standard mediation identification gives \citep{imai2010a,valeri2013,vanderweele2015}
\begin{align}
\operatorname{PNIE}&=\beta_2\gamma,
&\operatorname{TNIE}&=\beta_2(\gamma+\eta),\nonumber\\
\operatorname{PNDE}&=\beta_3+\eta\mu_0,
&\operatorname{TNDE}&=\beta_3+\eta\mu_1,
\label{eq:effect-map}\\
\operatorname{TE}&=\beta_3+\beta_2\gamma+\eta\mu_1.\nonumber
\end{align}
Consequently,
\begin{equation}
\operatorname{TE}=\operatorname{PNDE}+\operatorname{TNIE}
=\operatorname{TNDE}+\operatorname{PNIE}.
\label{eq:two-decompositions}
\end{equation}
When $\eta=0$, PNIE equals TNIE and PNDE equals TNDE, so the two decompositions coincide. Section~S1 of the Supplementary Material gives the identification calculation.

\subsection{Error-density scores and feasible estimation}
\label{sec:regression-scores}

The regression score derived by \citet[Theorem~1]{kim2023} supplies the building block for the joint mediation construction.  Consider a generic regression
\begin{equation}
R=m(\bm{Z};\bbeta)+\epsilon,\qquad
\E(\epsilon)=0,\quad \Var(\epsilon)=v,\quad \epsilon\indep\bm{Z},
\label{eq:generic-regression}
\end{equation}
where $f$ is the unknown error density and $\dot m_{\bbeta}(\bm{Z};\bbeta)=\partial m(\bm{Z};\bbeta)/\partial\bbeta$.  Put
\begin{equation}
q(\epsilon)=\epsilon^2-v-\frac{\mu_3}{v}\epsilon,\qquad
\mu_3=\E(\epsilon^3),\qquad Q_2=\E\{q(\epsilon)^2\}.
\label{eq:q-definition}
\end{equation}
We assume $f$ is positive and continuously differentiable on the relevant support, $\E(\epsilon^4)<\infty$, $0<Q_2<\infty$, the location Fisher information is finite, the regression design has full-rank information, and the population efficient-score equation has a locally unique root.  These conditions ensure nondegenerate score information and local identification of the regression parameters.
The efficient score of \citet{kim2023} is
\begin{align}
S_{\bbeta}^{\mathrm{eff}}
={}&-\frac{f'(\epsilon)}{f(\epsilon)}
 \left[\dot m_{\bbeta}(\bm{Z};\bbeta)
 -\E\{\dot m_{\bbeta}(\bm{Z};\bbeta)\}\right]\nonumber\\
&+\left\{\frac{\epsilon}{v}-\frac{\mu_3q(\epsilon)}{vQ_2}\right\}
\E\{\dot m_{\bbeta}(\bm{Z};\bbeta)\},
\label{eq:efficient-score-beta}\\
S_v^{\mathrm{eff}}
={}&\frac{q(\epsilon)}{Q_2}.
\label{eq:efficient-score-v}
\end{align}
We solve for $\lambda=\log v$, so unconstrained updates preserve $v=\exp(\lambda)>0$. By the chain rule, the corresponding efficient score is
\[
S_\lambda^{\mathrm{eff}}
=\frac{dv}{d\lambda}S_v^{\mathrm{eff}}
=vS_v^{\mathrm{eff}}.
\]
Because $v$ is positive and common within a regression, $\sum_i vS_{v,i}^{\mathrm{eff}}=0$ and $\sum_i S_{v,i}^{\mathrm{eff}}=0$ have the same roots, including after plug-in substitution. The estimate is recovered as $\widehat v=\exp(\widehat\lambda)$; separate valid-root checks assess convergence and covariance validity.

At a candidate parameter, the residuals are $e_i=R_i-m(\bm{Z}_i;\bbeta)$.  We replace expectations in \eqref{eq:q-definition}--\eqref{eq:efficient-score-v} by empirical moments and use the leave-one-out Gaussian-kernel estimators
\begin{align}
\widehat f_{-i}(e_i)
&=\frac{1}{(n-1)h}\sum_{j\ne i}\phi\left(\frac{e_i-e_j}{h}\right),\nonumber\\
\widehat f'_{-i}(e_i)
&=-\frac{1}{(n-1)h^3}\sum_{j\ne i}(e_i-e_j)
\phi\left(\frac{e_i-e_j}{h}\right),
\label{eq:kernel-score}
\end{align}
where $h=1.06\,s_e n^{-1/5}$.  At the same candidate value, we set
\[
\widehat\mu_3=n^{-1}\sum_i e_i^3,\qquad
\widehat\mu_4=n^{-1}\sum_i e_i^4,\qquad
\widehat Q_2=\widehat\mu_4-v^2-\widehat\mu_3^2/v,
\]
and substitute these quantities into \eqref{eq:efficient-score-beta}--\eqref{eq:efficient-score-v}.  At the model truth, $Q_2=\mu_4-v^2-\mu_3^2/v$; the displayed expression is its plug-in counterpart at a candidate value, rather than the literal sample mean of $q(e_i)^2$ away from the root.  This deterministic normal-reference rule follows the implementation in \citet{kim2023} and the standard density-estimation rule of \citet{silverman1986}; it also satisfies $h\rightarrow0$ and $nh^3\rightarrow\infty$, as required for pointwise consistency of a first-derivative kernel estimator.  The leave-one-out form avoids each residual contributing a kernel centered on itself.  For numerical stabilization, we use $\kappa_n=\min(10^{-10},n^{-2})$ as the density floor when forming $\widehat f'_{-i}/\widehat f_{-i}$ and reject candidates with $\widehat Q_2\leq\kappa_n$.  At every sample size used in this article $\kappa_n=10^{-10}$, so this asymptotic formulation reproduces the reported computation exactly while allowing the safeguard to vanish as $n\rightarrow\infty$.  For either regression, the estimate solves
\begin{equation}
\frac{1}{n}\sum_{i=1}^n
\begin{pmatrix}
\widehat S_{\bbeta,i}^{\mathrm{eff}}\\
\widehat S_{\lambda,i}^{\mathrm{eff}}
\end{pmatrix}=\bm{0}.
\label{eq:one-regression-equation}
\end{equation}
Equations \eqref{eq:efficient-score-beta}--\eqref{eq:one-regression-equation} specify the feasible estimator in the supplied code. Under the smoothness, tail, bandwidth and unique-root conditions of \citet[Theorem~2 and its proof]{kim2023}, together with the implementation conditions in Section~S3 of the Supplementary Material, the stabilized leave-one-out estimator has the same first-order expansion as their Gaussian-kernel score root. Section~S3 establishes this equivalence and sufficient conditions for consistent multi-start root selection, connecting the asymptotic results below to the implemented estimator.

\subsection{The stacked mediation system}

Let $\btheta_M=(\bbeta_M\trans,\lambda_M)\trans$ and $\btheta_Y=(\bbeta_Y\trans,\lambda_Y)\trans$ collect the mediator and outcome regression parameters, including log variances.  Applying the generic regression model in \eqref{eq:generic-regression} to the response--covariate-vector pairs $(R,\bm{Z})=(M,(1,T,\bX\trans)\trans)$ and $(R,\bm{Z})=(Y,(1,T,M,TM,\bX\trans)\trans)$, respectively, defines the population score vectors $S_{M,i}^{\mathrm{eff}}(\btheta_M)$ and $S_{Y,i}^{\mathrm{eff}}(\btheta_Y)$.  Each has the form $(S_{\bbeta}^{\mathrm{eff}\,\mathsf T},vS_v^{\mathrm{eff}})\trans$ given in \eqref{eq:efficient-score-beta}--\eqref{eq:efficient-score-v}.

The scores depend on unknown error densities and design moments. For linear regressions, $\E\{\dot m_{\bbeta}(\bm{Z};\bbeta)\}=\E(\bm{Z})$, estimated by sample means including $\overline{\bX}$. These plug-in quantities support the coefficient and log-variance equations; a separate moment equation represents estimation of the population mean $\bmu_X$.

To represent estimation of both the regression parameters and the population baseline mean, define the observation-level stacked estimating function by
\begin{equation*}
\bPsi_i(\bm{\vartheta})=
\begin{pmatrix}
S_{M,i}^{\mathrm{eff}}(\btheta_M)\\
S_{Y,i}^{\mathrm{eff}}(\btheta_Y)\\
\bX_i-\bmu_X
\end{pmatrix},\qquad
\bm{\vartheta}=(\btheta_M\trans,\btheta_Y\trans,\bmu_X\trans)\trans,
\end{equation*}
Here $\bPsi_i$ denotes the population-score version.  In computation, $\widehat{\bPsi}_i$ replaces each unknown density, moment, and design expectation by the sample quantities described above, and $\widehat{\bm{\vartheta}}$ solves $\sum_i\widehat{\bPsi}_i(\bm{\vartheta})=\bm{0}$.  This connects the general estimating system directly to the two regression fits and to population averaging over $\bX$.

The last block is the moment equation for the baseline mean:
\[
\frac{1}{n}\sum_{i=1}^n(\bX_i-\bmu_X)=\bm{0}
\quad\Longrightarrow\quad
\widehat{\bmu}_X=\overline{\bX}.
\]
This is the same empirical mean used in the regression scores. Its explicit inclusion in the stack carries the variability of $\overline{\bX}$ and its covariance with the regression estimators into effect inference, without changing the regression fits.

For example, $\operatorname{PNDE}=\beta_3+\eta(\alpha_2+\bm{\xi}_2\trans\bmu_X)$, whose first-order expansion includes $\eta\bm{\xi}_2\trans(\overline{\bX}-\bmu_X)$.  The same direct term enters TNDE and TE, but not PNIE or TNIE; it vanishes when $\eta=0$.  For effects standardized to the observed covariates, this population-mean term is omitted.

Define
\begin{equation*}
\widehat A=\frac{1}{n}\sum_{i=1}^n
\frac{\partial\widehat{\bPsi}_i(\widehat{\bm{\vartheta}})}
{\partial\bm{\vartheta}\trans},\qquad
\widehat B=\frac{1}{n}\sum_{i=1}^n
\widehat{\bPsi}_i(\widehat{\bm{\vartheta}})
\widehat{\bPsi}_i(\widehat{\bm{\vartheta}})\trans.
\end{equation*}
The regression blocks of $\widehat A$ use centered finite differences with coordinate step $d_n\max(1,|\widehat\vartheta_j|)$, where $d_n=\min(10^{-4},n^{-1/3})$; the baseline-mean block is $-I$. Thus $d_n=10^{-4}$ at the reported sample sizes and $d_n\rightarrow0$ asymptotically. Residuals, bandwidths, densities and moments are recomputed after each perturbation. Under Conditions (R1)--(R5) in Section~S3 of the Supplementary Material, covering score regularity, consistent feasible roots, derivative convergence and covariance consistency, the estimating-equation sandwich is \citep{vandervaart1998,tsiatis2006}
\begin{equation*}
\widehat\Sigma_{\vartheta}=n^{-1}\widehat A^{-1}
\widehat B\widehat A^{-\mathsf T}.
\end{equation*}
The empirical cross-products in $\widehat B$ retain finite-sample dependence between equations rather than forcing the off-diagonal blocks to zero.

Let $g(\bm{\vartheta})$ be the five-dimensional map in \eqref{eq:effect-map} and $G=\partial g/\partial\bm{\vartheta}\trans$.  The estimated effect covariance is $\widehat G\widehat\Sigma_{\vartheta}\widehat G\trans$.  Explicit gradients are given in Section~S2 of the Supplementary Material and are implemented directly in the code.

\begin{proposition}[Composition of the regression influence functions]
\label{prop:composition}
Suppose models \eqref{eq:mediator-model}--\eqref{eq:outcome-model} and Condition (R1) in Section~S3 of the Supplementary Material hold, and the joint density factorizes into the law of $(\bX,T)$ and the two independent regression-error densities.  For $k\in\{M,Y\}$, let $I_k=\E\{S_k^{\mathrm{eff}}S_k^{\mathrm{eff}\,\mathsf T}\}$ and $\varphi_k=I_k^{-1}S_k^{\mathrm{eff}}$, and partition $G=(G_M,G_Y,G_X)$.  Then the efficient influence function for $g(\bm{\vartheta})$ is
\begin{equation}
\varphi_g(O)=G_M\varphi_M(O)+G_Y\varphi_Y(O)
+G_X\{\bX-\bmu_X\}.
\label{eq:effect-eif}
\end{equation}
If, in addition, Conditions (R2)--(R4) hold, then the natural-effect plug-in estimator is asymptotically linear with influence function \eqref{eq:effect-eif} and attains the corresponding efficiency bound.  Condition (R5) additionally ensures consistency of the propagated covariance and justifies marginal Wald inference for scalar effects with positive asymptotic variance.
\end{proposition}

The proposition combines regression-score efficiency with the pathwise chain rule \citep{kim2023,bickel1993,tsiatis2006}. Section~S3 proves component orthogonality, the fitted-estimator expansion and covariance consistency. The $G_X(\bX-\bmu_X)$ term accounts for population averaging; the vanishing numerical safeguards preserve this first-order representation under the stated conditions.

The reported intervals target marginal 95\% coverage for each effect.  The two identities in \eqref{eq:two-decompositions} imply that the joint effect covariance has rank at most three; it is propagated without inverting a five-dimensional effect covariance.  A scalar Wald interval additionally requires positive first-order asymptotic variance.  The regular delta approximation therefore applies away from the joint-zero point $\beta_2=\gamma=0$, where the PNIE gradient vanishes; inference at that nonregular point calls for a dedicated higher-order or resampling analysis.

\subsection{Asymptotic precision and local power relative to OLS}
\label{sec:precision-power}

The feasible-root result connects the regression efficiency theory of \citet{kim2023} to an explicit natural-effect variance comparison for the supplied kernel-score algorithm.  For $k\in\{M,Y\}$, write the corresponding regression as $R_k=a_k+\bm b_k\trans\bm W_k+\epsilon_k$, where $\bm W_M=(T,\bX\trans)\trans$ and $\bm W_Y=(T,M,TM,\bX\trans)\trans$.  Let $\bm\omega_k=\E(\bm W_k)$, $C_k=\Var(\bm W_k)$, $s_k(e)=-f_k'(e)/f_k(e)$, $J_k=\E\{s_k(\epsilon_k)^2\}$, and
\[
H_k=\begin{pmatrix}-\bm\omega_k\trans\\ I\end{pmatrix}.
\]
The matrix $C_k$ is assumed nonsingular.  For natural effect $g_j$, let $d_{j,k}=\partial g_j/\partial\bbeta_k$ be the column gradient for $\bbeta_k=(a_k,\bm b_k\trans)\trans$, and let $V_{j,r}$ denote the limiting variance of $\sqrt n(\widehat g_{j,r}-g_j)$ for method $r$.

\Needspace{8\baselineskip}
\begin{proposition}[OLS-relative asymptotic variance]
\label{prop:ols-precision-power}
Under the regression model and Conditions (R1)--(R5) in Section~S3 of the Supplementary Material, for a natural effect with positive first-order variance,
\begin{equation}
V_{j,\mathrm{OLS}}-V_{j,\mathrm{semi}}
=\sum_{k\in\{M,Y\}}(v_k-J_k^{-1})
 (H_k\trans d_{j,k})\trans C_k^{-1}(H_k\trans d_{j,k})\geq0.
\label{eq:ols-efficiency-gap}
\end{equation}
The inequality is strict if, for at least one regression, $v_kJ_k>1$ and $H_k\trans d_{j,k}\ne0$.
\end{proposition}

\Needspace{13\baselineskip}
\begin{proof}
Integration by parts gives $\E\{\epsilon_k s_k(\epsilon_k)\}=1$; hence Cauchy--Schwarz gives $v_kJ_k\geq1$.  Joint inversion of the coefficient and variance score blocks in \eqref{eq:efficient-score-beta}--\eqref{eq:efficient-score-v} gives the regression-coefficient influence functions
\begin{align*}
\varphi_{\bbeta,k}^{\mathrm{semi}}
&=e_{0,k}\epsilon_k+H_kC_k^{-1}(\bm W_k-\bm\omega_k)
  \frac{s_k(\epsilon_k)}{J_k},\\
\varphi_{\bbeta,k}^{\mathrm{OLS}}
&=e_{0,k}\epsilon_k+H_kC_k^{-1}(\bm W_k-\bm\omega_k)\epsilon_k,
\end{align*}
where $e_{0,k}$ selects the intercept.  Independence of $\epsilon_k$ and $\bm W_k$ makes their coefficient covariance difference $(v_k-J_k^{-1})H_kC_k^{-1}H_k\trans$.  The mediator and outcome coefficient influence functions are mutually orthogonal and orthogonal to $\bX-\bmu_X$ by conditional mean zero.  The baseline-mean variance is therefore common to both methods, and applying the effect gradient yields \eqref{eq:ols-efficiency-gap}.  Section~S3 of the Supplementary Material gives the score-block calculation.
\end{proof}

For PNIE, the comparison simplifies to
\begin{equation*}
V_{\mathrm{PNIE},\mathrm{OLS}}-V_{\mathrm{PNIE},\mathrm{semi}}
=\gamma^2(v_M-J_M^{-1})[C_M^{-1}]_{TT}
 +\beta_2^2(v_Y-J_Y^{-1})[C_Y^{-1}]_{MM}.
\end{equation*}
Thus information in either error density can improve precision for an indirect effect through its corresponding regression coefficient.  Equality holds for Gaussian errors, for which $J_k=1/v_k$.  This ordering characterizes asymptotic precision relative to OLS for regular effects under the independent-error model; the joint-zero PNIE point requires the separate nonregular analysis noted above.  Finite-sample performance, comparisons with robust methods and properties of the confounding-sensitivity estimators are evaluated in their respective experiments.

The variance ordering also has a test-theoretic implication.  For two-sided Wald tests of $g_j=0$ at the same asymptotic level $\alpha_{\mathrm{test}}$, consider a regular local alternative $g_{j,n}=h/\sqrt n$ with $h\ne0$, along which the asymptotic linear expansions and variance consistency hold.  The limiting power of method $r$ is
\begin{equation*}
\pi_{j,r}(h)=1-\Phi\!\left(z_{1-\alpha_{\mathrm{test}}/2}
 -\frac{h}{\sqrt{V_{j,r}}}\right)
 +\Phi\!\left(-z_{1-\alpha_{\mathrm{test}}/2}
 -\frac{h}{\sqrt{V_{j,r}}}\right),
\end{equation*}
where $\Phi$ is the standard-normal distribution function and $z_p$ its $p$-quantile.  As a function of $|h|/\sqrt{V_{j,r}}$, this power is increasing; hence Proposition~\ref{prop:ols-precision-power} gives $\pi_{j,\mathrm{semi}}(h)\geq\pi_{j,\mathrm{OLS}}(h)$, strictly when \eqref{eq:ols-efficiency-gap} is strict.  This comparison links variance reduction to local power for regular alternatives at a common asymptotic test size and motivates the independently calibrated finite-sample evaluation below.

\subsection{Deterministic multi-start computation}

Let $(\widetilde\bbeta,\widetilde v)$ be the OLS pilot from the same regression and let
\begin{equation*}
d_j=\max\{0.05,\;c\max(1,|\widetilde\beta_j|)\},
\qquad c=0.10.
\end{equation*}
The lower bound prevents an exactly zero or small OLS coefficient from producing indistinguishable starts; scaling by $\max(1,|\widetilde\beta_j|)$ makes the perturbation relative for larger coefficients.  With $\bm{s}=(1,-1,1,-1,\ldots)\trans$, the seven starts are
\begin{align*}
&(\widetilde\bbeta,\log\widetilde v),
(\widetilde\bbeta+\bm d,\log\widetilde v),
(\widetilde\bbeta-\bm d,\log\widetilde v),\\
&(\widetilde\bbeta+\bm s\circ\bm d,\log\widetilde v),
(\widetilde\bbeta-\bm s\circ\bm d,\log\widetilde v),
(\widetilde\bbeta,\log(1.25\widetilde v)),
(\widetilde\bbeta,\log(0.80\widetilde v)).
\end{align*}
The all-positive, all-negative, and alternating perturbations explore distinct coefficient directions without random restarts; the final two starts separately probe the variance coordinate and are symmetric on the log-variance scale because $0.80=1/1.25$.  With the reported values of $c$, the $0.05$ floor is redundant because $c\max(1,|\widetilde\beta_j|)\geq0.05$ and becomes active only in implementations using a smaller $c$.  These rules are designed to make root exploration deterministic, transparent and reproducible.

Each start uses \texttt{rootSolve::multiroot()} with at most 200 iterations and relative and absolute tolerance $a_n=\min(10^{-9},n^{-1})$. Valid roots require finite parameters, maximum absolute mean score below $b_n=\min(10^{-6},n^{-1})$, Jacobian reciprocal condition number above $10^{-10}$ and finite covariance with positive diagonal. At the reported sample sizes, $a_n=10^{-9}$ and $b_n=10^{-6}$; asymptotically, $b_n=o(n^{-1/2})$ supplies the accuracy in Condition (R2). Duplicate roots are removed after rounding parameters to seven decimal places. Section~S4 gives the complete algorithm.

If more than one valid root remains, we choose the one minimizing
\begin{equation*}
D(\btheta)=
\frac{(\bbeta-\widetilde\bbeta)\trans(\bm Z\trans\bm Z/n)
(\bbeta-\widetilde\bbeta)}{\widetilde v}
+\{\lambda-\log(\widetilde v)\}^2.
\end{equation*}
The first term measures the change in fitted linear predictor on the observed design scale, while the second compares log variances.  Under the correctly specified linear mean and zero conditional error mean, OLS provides a consistent coefficient pilot for non-Gaussian errors and a principled anchor for root selection.  OLS supplies initialization and the selection reference; all reported proposed-method estimates solve the density-score equations.  We repeat all starts with $c\in\{0.05,0.10,0.20\}$ in the algorithm-sensitivity study.

\paragraph{Use of generative AI.}
OpenAI Codex was used to assist with manuscript drafting and revision, mathematical derivations and exposition, and the development and debugging of analysis code. The author takes full responsibility for the mathematical arguments, code, numerical results, interpretations and final manuscript.

\section{Simulation studies}
\label{sec:simulations}

\subsection{Data-generating mechanisms}

For the main experiment, $X\sim N(0,1)$ and $T\sim\operatorname{Bernoulli}(0.5)$ independently, with
\begin{align*}
M&=0.2+0.4T+0.3X+\epsilon_M,\\
Y&=0.5T-0.8M+1.0TM+0.4X+\epsilon_Y.
\end{align*}
The errors are independently drawn from the same standardized distribution.  The true effect vector $(\operatorname{PNIE},\operatorname{TNIE},\operatorname{PNDE},\operatorname{TNDE},\operatorname{TE})$ is $(-0.32,0.08,0.70,1.10,0.78)$.

Four zero-mean, unit-variance errors are considered.  Let $\phi$ and $\Phi$ denote the standard normal density and distribution function.  For each nonstandard raw variable $U$ below, we generate $\epsilon=(U-\mu_U)/\sigma_U$; equivalently, its final density is $f_\epsilon(e)=\sigma_U f_U(\mu_U+\sigma_U e)$.
\begin{enumerate}
\item Gaussian: $f_\epsilon(e)=\phi(e)$.
\item Skew-normal: $f_U(u)=2\phi(u)\Phi(6u)$.  With $\delta=6/\sqrt{37}$, the exact constants are $\mu_U=\delta\sqrt{2/\pi}=0.7870$ and $\sigma_U=\sqrt{1-2\delta^2/\pi}=0.6169$.
\item Asymmetric mixture:
\[
f_U(u)=\frac{0.82}{0.35}\phi\left(\frac{u+0.7}{0.35}\right)
+\frac{0.18}{0.75}\phi\left(\frac{u-3.2}{0.75}\right),
\]
with $\mu_U=0.002$ and $\sigma_U=1.5642$.
\item Symmetric bimodal mixture:
\[
f_U(u)=\frac{0.5}{0.45}\phi\left(\frac{u+1.7}{0.45}\right)
+\frac{0.5}{0.45}\phi\left(\frac{u-1.7}{0.45}\right),
\]
with $\mu_U=0$ and $\sigma_U=\sqrt{0.45^2+1.7^2}=1.7586$.
\end{enumerate}
The densities are shown in Figure~S1 of the Supplementary Material.

The main experiment uses $n\in\{200,500\}$ and 1,000 Monte Carlo replications in every cell.  The smaller size probes finite-sample behavior of the estimated density derivative, while the larger size assesses stabilization at a still-moderate sample size.  We report bias, root mean squared error (RMSE), 95\% Wald coverage, average interval length, numerical success rate, and the Monte Carlo standard error (MCSE) of coverage.  Failures remain in the denominator of the reported success rate but are excluded from performance summaries, whose effective replication counts are reported in the complete tables.

A separate comparator experiment uses $n=300$, intermediate between the two main-study sizes, and 1,000 Monte Carlo replications per error distribution.  The four distributions yield 4,000 data sets and 20,000 attempted method fits.  The experiment compares OLS delta intervals and the proposed method with (i) the quasi-Bayesian approximation of \citet{imai2010a,imai2010b} using 1,000 draws, (ii) their nonparametric percentile bootstrap using 499 resamples, both through the \textsf{R} package \texttt{mediation} \citep{tingley2014}, and (iii) an imputation-based natural effect model with sandwich standard errors through the \textsf{R} package \texttt{medflex} \citep{steen2017}.  All procedures target the same five effects and use compatible linear mean structures.  For \texttt{medflex}, outcome imputation uses $Y\sim T*M+T*X$, and the natural effect model uses $Y\sim T_0*T_1+T_0*X$ with centered $X$.  The $T_0X$ term is required by the interaction-induced covariate dependence of the natural direct effects.  The additional $TX$ imputation term satisfies the package's working-model compatibility check; its true coefficient is zero in the simulation.  Thus this comparator has a slightly richer outcome working model, as detailed in Online Resource~1.

The two \texttt{mediation} entries are inferential options within its \emph{model-based} approach.  The package's single-experiment design-based routine is formulated for a discrete mediator, and its parallel-design routine is formulated for mediator manipulation \citep{tingley2014}.  The model-based routines therefore provide the like-for-like comparison for the continuous mediator and target used here.

The OLS delta analysis uses an empirical estimating-equation sandwich for the two regressions and the baseline mean, providing likelihood-free covariance estimation based on first two moments.  The proposed estimator additionally uses the shape of a non-Gaussian error density.  All procedures target the same population effect definitions, with method-specific uncertainty calculations: quasi-Bayesian simulation conditions on observed covariates, the \texttt{medflex} decomposition uses package-native standard errors evaluated at the centered covariate reference, the stacked calculation adds baseline-mean variability, and the nonparametric bootstrap captures that variability through observation-level resampling.

\subsection{Main and comparator results}

\begin{table}[t]
\centering
\caption{Main Monte Carlo results based on 1,000 attempts per cell and summarized over valid fits; counts and success rates appear in Table~S1. G, Gaussian; SN, skew-normal; AM, asymmetric mixture; BM, symmetric bimodal. Boldface marks lower RMSE and shorter length within each design and effect; these descriptive highlights must be interpreted jointly with coverage and numerical success. Bias is signed}
\label{tab:main_simulation}
\footnotesize
\setlength{\tabcolsep}{3.5pt}
\begin{tabular}{rllrrrrrrrr}
\toprule
& & & \multicolumn{4}{c}{PNIE} & \multicolumn{4}{c}{TE} \\
\cmidrule(lr){4-7}\cmidrule(lr){8-11}
$n$ & Error & Method & Bias & RMSE & Cov. & Length & Bias & RMSE & Cov. & Length \\
\midrule
200 & G & OLS delta & 0.003 & \textbf{0.121} & 0.947 & \textbf{0.467} & -0.001 & \textbf{0.162} & 0.949 & \textbf{0.643} \\
200 & G & Proposed & 0.004 & 0.130 & 0.932 & 0.495 & -0.001 & 0.175 & 0.938 & 0.682 \\
\midrule
200 & SN & OLS delta & 0.005 & 0.122 & 0.945 & 0.469 & 0.000 & 0.171 & 0.937 & 0.645 \\
200 & SN & Proposed & 0.005 & \textbf{0.100} & 0.951 & \textbf{0.388} & 0.000 & \textbf{0.144} & 0.936 & \textbf{0.553} \\
\midrule
200 & AM & OLS delta & 0.001 & 0.126 & 0.942 & 0.475 & 0.008 & 0.168 & 0.946 & 0.646 \\
200 & AM & Proposed & -0.001 & \textbf{0.055} & 0.940 & \textbf{0.120} & 0.002 & \textbf{0.088} & 0.937 & \textbf{0.327} \\
\midrule
200 & BM & OLS delta & 0.004 & 0.119 & 0.940 & 0.470 & 0.002 & 0.157 & 0.957 & 0.646 \\
200 & BM & Proposed & 0.000 & \textbf{0.034} & 0.929 & \textbf{0.121} & 0.002 & \textbf{0.110} & 0.950 & \textbf{0.327} \\
\midrule
500 & G & OLS delta & 0.001 & \textbf{0.074} & 0.956 & \textbf{0.297} & -0.006 & \textbf{0.107} & 0.946 & \textbf{0.409} \\
500 & G & Proposed & 0.001 & 0.078 & 0.958 & 0.311 & -0.005 & 0.111 & 0.950 & 0.424 \\
\midrule
500 & SN & OLS delta & 0.000 & 0.077 & 0.955 & 0.298 & -0.001 & 0.109 & 0.948 & 0.409 \\
500 & SN & Proposed & 0.004 & \textbf{0.059} & 0.942 & \textbf{0.233} & -0.001 & \textbf{0.087} & 0.944 & \textbf{0.337} \\
\midrule
500 & AM & OLS delta & -0.004 & 0.075 & 0.946 & 0.299 & 0.007 & 0.106 & 0.947 & 0.409 \\
500 & AM & Proposed & 0.000 & \textbf{0.019} & 0.949 & \textbf{0.074} & 0.002 & \textbf{0.054} & 0.948 & \textbf{0.205} \\
\midrule
500 & BM & OLS delta & -0.003 & 0.076 & 0.945 & 0.298 & -0.003 & 0.101 & 0.955 & 0.409 \\
500 & BM & Proposed & 0.000 & \textbf{0.019} & 0.959 & \textbf{0.076} & -0.001 & \textbf{0.053} & 0.941 & \textbf{0.206} \\
\bottomrule
\end{tabular}
\end{table}

Table~\ref{tab:main_simulation} reports PNIE and TE; Table~S1 of the Supplementary Material gives all five effects and diagnostics. Across the 30 non-Gaussian sample-size-by-effect cells, proposed-to-OLS RMSE ratios ranged from 0.245 to 0.841 and interval-length ratios from 0.247 to 0.857. Proposed coverage ranged from 0.918 to 0.959. Numerical success was 1.000 under the skew-normal error, 0.887 and 0.921 for the asymmetric and bimodal mixtures at $n=200$, and 0.999 and 1.000 for those mixtures at $n=500$.

The Gaussian setting provides the reference in which OLS attains its efficiency benchmark.  There, the proposed-to-OLS RMSE ratios were 1.035--1.078 and interval-length ratios were 1.037--1.063, showing slightly larger RMSE and interval lengths for the feasible density-score estimator at this benchmark.

On common successful data sets, the 30 non-Gaussian main-study cells gave paired RMSE ratios of \PairedRMSERange\ and interval-length ratios of \PairedLengthRange, confirming the same precision pattern (Table~S3). The fraction of \emph{all attempted data sets} returning a proposed interval that covered the truth ranged from \ReportCoverRange. This report-and-cover measure combines numerical success with conditional coverage; Section~S5 gives its definition and the paired-ratio Monte Carlo standard errors.

\begin{figure}[t]
\centering
\includegraphics[width=0.96\textwidth]{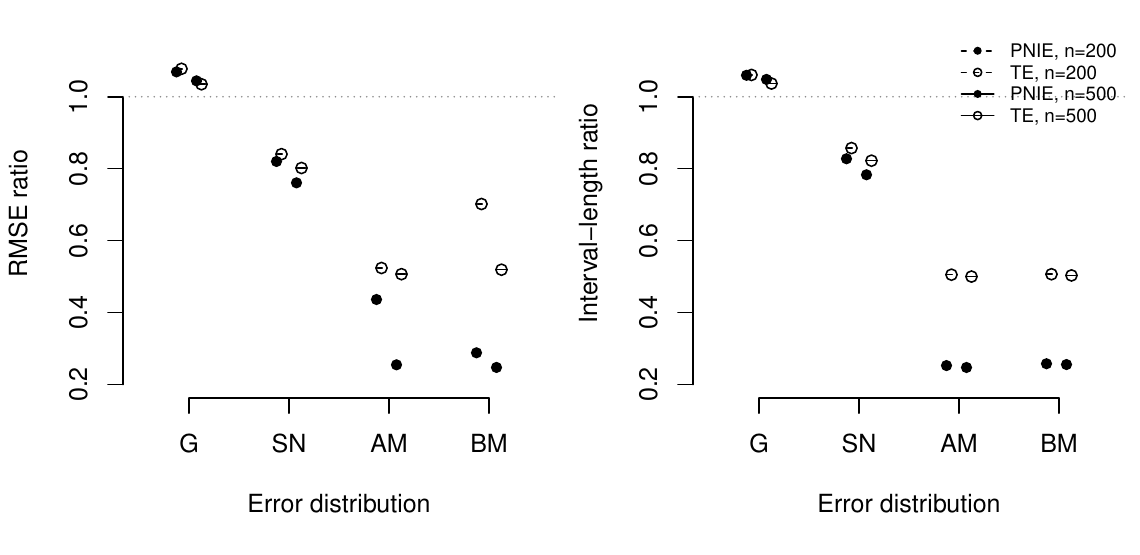}
\caption{Ratios of proposed-method RMSE and average 95\% interval length to their OLS counterparts, using method-specific summaries over numerically valid fits.  G, SN, AM, and BM denote Gaussian, skew-normal, asymmetric-mixture, and symmetric-bimodal errors; values below one favor the proposed method.  Point shape identifies PNIE or TE, and line type identifies sample size}
\label{fig:efficiency-ratios}
\end{figure}

Each point in Figure~\ref{fig:efficiency-ratios} is the proposed-to-OLS ratio for one error distribution, sample size and effect; a value of 0.5 means half the RMSE or average interval length of OLS. The figure uses method-specific valid fits, complemented by the common-data comparison above. Interval length is interpreted jointly with coverage and numerical success.

\begin{table}[t]
\centering
\caption{Comparator study at $n=300$ based on 1,000 attempts per cell, summarized over valid fits. The Imai procedures use 1,000 quasi-Bayesian draws or 499 bootstrap resamples per data set. Error abbreviations are as in Table~\ref{tab:main_simulation}. Boldface marks the smallest displayed RMSE and shortest length within each design and effect; these descriptive highlights are interpreted jointly with coverage and numerical success}
\label{tab:comparators}
\small
\begin{tabular}{llrrrrrr}
\toprule
& & \multicolumn{3}{c}{PNIE} & \multicolumn{3}{c}{TE} \\
\cmidrule(lr){3-5}\cmidrule(lr){6-8}
Error & Method & RMSE & Cov. & Length & RMSE & Cov. & Length \\
\midrule
G & OLS delta & \textbf{0.100} & 0.952 & \textbf{0.383} & \textbf{0.135} & 0.955 & 0.527 \\
G & Imai quasi-Bayes & \textbf{0.100} & 0.950 & 0.387 & \textbf{0.135} & 0.962 & 0.573 \\
G & Imai bootstrap & \textbf{0.100} & 0.946 & 0.384 & 0.147 & 0.966 & 0.565 \\
G & \texttt{medflex} NEM & \textbf{0.100} & 0.950 & 0.385 & \textbf{0.135} & 0.954 & \textbf{0.522} \\
G & Proposed & 0.107 & 0.942 & 0.406 & 0.139 & 0.948 & 0.552 \\
\midrule
SN & OLS delta & 0.099 & 0.950 & 0.385 & 0.133 & 0.952 & 0.527 \\
SN & Imai quasi-Bayes & 0.100 & 0.949 & 0.388 & 0.133 & 0.970 & 0.572 \\
SN & Imai bootstrap & 0.099 & 0.939 & 0.386 & 0.146 & 0.968 & 0.566 \\
SN & \texttt{medflex} NEM & 0.099 & 0.950 & 0.386 & 0.133 & 0.950 & 0.522 \\
SN & Proposed & \textbf{0.079} & 0.955 & \textbf{0.310} & \textbf{0.112} & 0.947 & \textbf{0.443} \\
\midrule
AM & OLS delta & 0.101 & 0.944 & 0.387 & 0.141 & 0.935 & 0.527 \\
AM & Imai quasi-Bayes & 0.102 & 0.944 & 0.389 & 0.141 & 0.954 & 0.573 \\
AM & Imai bootstrap & 0.101 & 0.945 & 0.390 & 0.153 & 0.953 & 0.567 \\
AM & \texttt{medflex} NEM & 0.102 & 0.948 & 0.388 & 0.140 & 0.935 & 0.522 \\
AM & Proposed & \textbf{0.026} & 0.938 & \textbf{0.096} & \textbf{0.069} & 0.944 & \textbf{0.264} \\
\midrule
BM & OLS delta & 0.096 & 0.956 & 0.385 & 0.133 & 0.951 & 0.527 \\
BM & Imai quasi-Bayes & 0.096 & 0.956 & 0.388 & 0.134 & 0.960 & 0.574 \\
BM & Imai bootstrap & 0.096 & 0.957 & 0.386 & 0.148 & 0.961 & 0.566 \\
BM & \texttt{medflex} NEM & 0.096 & 0.955 & 0.386 & 0.134 & 0.948 & 0.522 \\
BM & Proposed & \textbf{0.029} & 0.949 & \textbf{0.098} & \textbf{0.079} & 0.941 & \textbf{0.266} \\
\bottomrule
\end{tabular}
\end{table}

Table~\ref{tab:comparators} extends the evaluation to the package-based comparators.  Among the 15 non-Gaussian distribution-by-effect cells, the proposed method had the smallest RMSE in all 15, with coverage from 0.924 to 0.959.  Its numerical success was 1.000, 0.962, and 0.978 under the skew-normal, asymmetric-mixture, and bimodal errors, respectively; all four comparators succeeded in every attempted fit.  Under Gaussian errors, RMSE was comparable to or slightly larger than that of the comparators.  The two \texttt{mediation} procedures quantify uncertainty differently, while \texttt{medflex} estimates a natural effect model; the proposal additionally uses an estimated regression-error density within the model of Section~\ref{sec:preliminaries}.

\FloatBarrier

\subsection{Null size, power, and numerical sensitivity}

For null-size and power assessment, we use $n=300$, standardized asymmetric-mixture errors, $\gamma=-0.26$, $\eta=0.8$ and $\beta_2\in\{0,0.1,0.2,0.3,0.4\}$. The true PNIE is $-0.26\beta_2$, with the null at $\beta_2=0$. Each setting has 1,000 attempted replications of the two-sided test of $H_0:\operatorname{PNIE}=0$. At the nominal 5\% level, rejection occurs when the 95\% Wald interval excludes zero, equivalently when $|\widehat{\operatorname{PNIE}}/\widehat{\operatorname{SE}}|>z_{0.975}\approx1.96$. This standard-normal reference concerns the test statistic, not a Gaussian assumption for the regression errors.

Let $\mathcal V$ be the set of numerically valid fits and $[L_r,U_r]$ the interval in replication $r$.  The empirical rejection probability and its binomial Monte Carlo standard error are
\[
\widehat p_{\mathrm{rej}}
=\frac{1}{|\mathcal V|}\sum_{r\in\mathcal V}
\mathbf{1}\{L_r>0\ \text{or}\ U_r<0\},
\qquad
\operatorname{MCSE}(\widehat p_{\mathrm{rej}})
=\sqrt{\frac{\widehat p_{\mathrm{rej}}(1-\widehat p_{\mathrm{rej}})}{|\mathcal V|}}.
\]
At $\beta_2=0$, rejection estimates the conditional type~I error rate: closeness to the nominal 0.05, rather than a larger value, is desirable. At $\beta_2\ne0$, it estimates power among valid fits, so higher rejection is preferable at comparable type~I error rates. Coverage measures containment of the true effect; rejection equals one minus coverage only at the null, using the same valid fits.

The original nominal experiment gave null rejection proportions among valid fits of 0.016 for OLS and 0.056 for the proposal (Table~S4), indicating conservative OLS rejection and closer-to-nominal proposed rejection. For power comparisons at similar null sizes, we calibrate method-specific cutoffs to 5\% using 5,000 independent null data sets and freeze them for evaluation on 1,000 new data sets per effect value. The OLS cutoff decreases from 1.96 to 1.669, easing rejection, while the proposed cutoff is nearly unchanged at 1.967 (Table~S5). The held-out null checks attained size; Section~S6 details the order-statistic rule and two-stage Monte Carlo uncertainty.

The comparison includes Huber-FIX and Huber-SEL, motivated by \citet{wang2025}, using the same regressions and PNIE slope-product target. Huber-FIX uses 1.345 times a LAD-based residual scale; Huber-SEL minimizes an estimated variance factor over the authors' tuning grid. Under independent errors, the intercept absorbs the Huber location shift, preserving the slope target even for asymmetric errors. Section~S6 derives this result and the interaction-adapted covariance; Tables~S5--S7 give cutoffs, held-out rejection and matched-data contrasts.

\begin{figure}[t]
\centering
\includegraphics[width=0.96\textwidth]{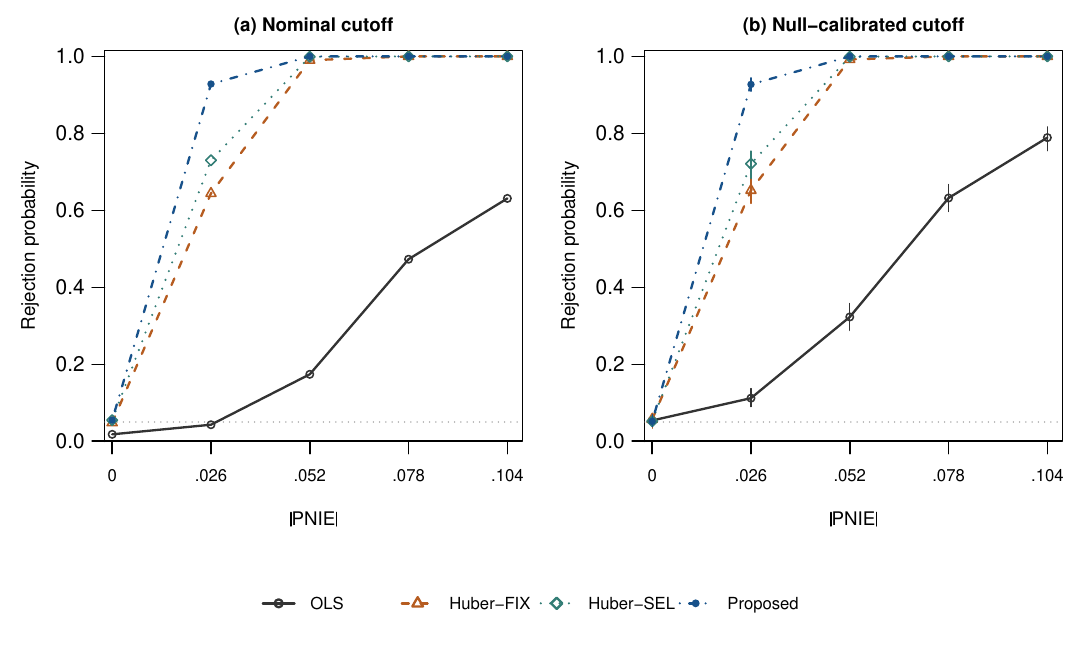}
\caption{Held-out PNIE rejection probabilities under the asymmetric-mixture error at $n=300$. Panel (a) uses the common normal cutoff; panel (b) uses method-specific cutoffs from 5,000 independent null-calibration data sets. Each point uses 1,000 new evaluation data sets, shared by the four methods. The ordinate is type~I error at $|\operatorname{PNIE}|=0$ and power elsewhere, conditional on a valid fit. Vertical segments in (b) are pointwise 95\% Monte Carlo uncertainty intervals from resampling both simulation stages; the horizontal line marks 0.05}
\label{fig:size-power}
\end{figure}

Figure~\ref{fig:size-power} compares nominal and calibrated rejection on the same new evaluation samples. These are distinct from the original samples in Table~S4, so the nominal rejection proportions need not coincide exactly. After calibration, held-out null rejection is \CalOLSNull\ for OLS, \CalHuberFixNull\ for Huber-FIX, \CalHuberSelNull\ for Huber-SEL and \CalSemiNull\ for the proposal. At $\beta_2=0.1$, corresponding to PNIE $=-0.026$, calibrated conditional power is \CalOLSPower, \CalHuberFixPower, \CalHuberSelPower\ and \CalSemiPower, respectively. Thus the weak-effect power gain persists after addressing OLS conservatism and extends to these Huber comparators in the same design. On data sets where all four procedures succeed, the proposed-minus-Huber-SEL power difference is \CalHuberSelContrast\ with a two-stage Monte Carlo 95\% interval of \CalHuberSelContrastCI.

Proposed numerical success ranges from \CalSemiSuccessRange. At PNIE $=-0.026$, the all-attempt fraction returning a valid fit and rejecting is \CalSemiReportPower, exceeding both Huber rates. For larger effects, near-unit Huber conditional power and complete numerical success yield higher all-attempt detection, locating the additional advantage at the weak effect. The calibration diagnoses this DGP-specific comparison; applications use the working-model inference above. Because $\gamma=-0.26$ at the null, the experiment concerns the regular one-component null, with the joint-zero case requiring separate nonregular analysis.

Using 300 common data sets for each mixture, we also fit starting-value scales $c=0.05$, $0.10$, and $0.20$.  Success was 0.957, 0.900, and 0.860 for the asymmetric mixture and 0.950, 0.940, and 0.910 for the bimodal mixture as $c$ increased.  Mean PNIE stayed between $-0.321$ and $-0.318$, and valid-root counts averaged near one.  Table~S8 shows stable effect estimates among valid fits alongside lower success at wider perturbations; unsuccessful fits were recorded as failures rather than replaced by OLS estimates.

\subsection{Sensitivity to variance misspecification}

We separately examine departures from the independent-error restriction without changing the conditional means or the causal identification conditions. At $n=300$, let $\epsilon_M=s(T,X)U_M$ and $\epsilon_Y=s(T,X)U_Y$, where the standardized innovations are mutually independent and independent of $(T,X)$. We use Gaussian and asymmetric-mixture innovations from the main study. The two variance mechanisms are
\[
s_X^2(X)=\exp(\kappa X-\kappa^2/2),\qquad
s_T^2(T)=\frac{\exp\{\kappa(2T-1)\}}{\cosh(\kappa)},
\]
with $\kappa\in\{0,0.15,0.30,0.60\}$. Both satisfy $\E(s^2)=1$, preserving marginal error variance. Sharing the $\kappa=0$ reference gives 14 configurations, each with 500 attempts: 7,000 data sets and 14,000 method fits. OLS uses its stacked sandwich; the proposed estimator and numerical settings are unchanged. Section~S7 provides the structural identification argument, variance contrasts and Monte Carlo uncertainty.

The results distinguish the two variance mechanisms. Under asymmetric-mixture innovations with covariate-dependent variance, proposed PNIE coverage was 0.959, 0.963 and 0.948 as $\kappa$ increased, with numerical success of 0.920 at $\kappa=0.60$. Treatment-dependent variance produced PNIE biases of 0.026, 0.065 and 0.180 and conditional coverages of 0.802, 0.327 and 0.000; corresponding OLS coverages were 0.952, 0.954 and 0.952. At the strongest treatment-dependent setting, proposed numerical success was 0.798 and total-effect coverage was 0.393. Coverage departures are evident at a treatment-to-control variance ratio of about 1.35 ($\kappa=0.15$) in this skewed setting. The paired-fit comparisons show the same pattern on common retained data sets. These results provide a practical reason to assess treatment-group variance differences alongside error shape and identify covariate-dependent error distributions as a priority for extending the homoscedastic framework.

Table~S9 and Figure~S2 document these calibration changes under preserved causal identification but misspecified error variances.

\FloatBarrier

\section{Density-score confounding-sensitivity analysis}
\label{sec:confounding}

We extend the density-score strategy of Section~\ref{sec:methodology} to residual-correlation sensitivity analysis. Allowing structural outcome errors to depend on $M$ prevents direct reuse of the outcome-regression score under its original independence assumption. We therefore use reduced-form regressions with unspecified marginal error densities and propagate their joint uncertainty through the sensitivity functional.

\subsection{Sensitivity target and reduced-form model}

We retain treatment exchangeability, consistency, positivity and a linear structural outcome model, while allowing mediator--outcome confounding. The pooled mediator and treatment-specific reduced-outcome means are
\[
\E(M\mid T,\bX)=\alpha_2+\beta_2T+\bm\xi_2\trans\bX,
\qquad q_t(\bX)=\E(Y\mid T=t,\bX)=a_t+\bm d_t\trans\bX.
\]
Let $U=M-\E(M\mid T,\bX)$ be independent of $(T,\bX)$ and let $W_t=Y-q_t(\bX)$ within $T=t$. The joint law of $(U,W_t)$ is assumed not to depend on $\bX$ within that stratum, while dependence between $U$ and $W_t$ is allowed. These restrictions permit marginal density-score regression without assuming independence between regression blocks. Write $v_M=\Var(U)$, $v_t=\Var(W_t\mid T=t)$ and $\omega_t=\Cov(U,W_t\mid T=t)$, with $v_Mv_t>\omega_t^2$. The mediator's error law is common across treatment groups; outcome reduced-form error laws may differ. Together with the linear means, these assumptions define the reduced-form working model.

For a structural mediator slope $b_t$ in treatment stratum $t$, the structural outcome error is $E_t=W_t-b_tU$. Specify $\rho_t=\operatorname{Corr}(U,E_t\mid T=t)$ as a sensitivity parameter. Solving the covariance identities gives
\begin{equation}
b_t(\rho_t)=\frac{\omega_t}{v_M}
-\frac{\rho_t}{\sqrt{1-\rho_t^2}}
\sqrt{\frac{v_t}{v_M}-\frac{\omega_t^2}{v_M^2}},
\qquad -1<\rho_t<1.
\label{eq:sensitivity-slope}
\end{equation}
Equation~\eqref{eq:sensitivity-slope} is the stratum-specific identity of \citet[Section~5 and Appendix~D]{imai2010a}. Because it uses only second-moment identities, it requires no Gaussian error assumption and applies to the stated semiparametric model with unspecified error densities. Its causal interpretation requires an additive outcome error invariant to mediator intervention and a constant mediator slope within each stratum; $b_1-b_0$ represents treatment--mediator interaction. Section~S8 of the Supplementary Material gives the structural derivation.

Let $\tau=(a_1-a_0)+(\bm d_1-\bm d_0)\trans\bmu_X$. The sensitivity functionals are
\begin{align}
\delta_t(\rho_t)&=\beta_2 b_t(\rho_t),\nonumber\\
\zeta_0(\rho_1)&=\tau-\delta_1(\rho_1),\qquad
\zeta_1(\rho_0)=\tau-\delta_0(\rho_0).
\label{eq:sensitivity-effects}
\end{align}
The illustrations follow the analyst-specified path $\rho_0=\rho_1=\rho$ through the two-correlation space; these correlations are not identified from observed data alone. At the population level, $\rho=0$ recovers the usual linear mean-effect decomposition, but zero correlation need not imply full sequential ignorability with non-Gaussian errors.

\subsection{Semiparametric estimation and joint inference}

To estimate \eqref{eq:sensitivity-effects}, we apply the density scores of Section~\ref{sec:regression-scores} to $M$ on $(1,T,\bX)$ and to $Y$ on $(1,\bX)$ within each treatment group, with an unspecified marginal error density for each regression. The OLS reduced-form benchmark uses the same regressions and an empirical sandwich, with residual second-moment equations for the variances. Both implementations append $1(T=t)(UW_t-\omega_t)$ for $t=0,1$ and $\bX-\bmu_X$ to their stacked equations. Cross-product moments and cross-block empirical score covariance account for dependence between the reduced-form errors. Full equations, bread derivatives and effect gradients are given in Section~S8 of the Supplementary Material.

Write the resulting nuisance vector as $\theta$ and the map in \eqref{eq:sensitivity-effects} as $g(\theta;\rho)$. The sensitivity estimator is $g(\widehat\theta;\rho)$. For fixed $\rho$, if the fitted marginal regressions and appended moments satisfy the joint asymptotic-linearity conditions stated in Section~S8, then the ordinary delta method gives
\[
\varphi_{g,\rho}=G_\rho\varphi_\theta,
\qquad
\widehat\Var\{g(\widehat\theta;\rho)\}
=\widehat G_\rho\widehat\Var(\widehat\theta)\widehat G_\rho\trans.
\]
We call the implementations Ordinary Least Squares Reduced-Form (OLS-RF) and Density-Score Reduced-Form (DS-RF). Efficient marginal scores under the working model supply the DS-RF components, with mediator pooling exploiting the common error shape to estimate $\beta_2$.

For $\beta_2\ne0$, solving the sensitivity identity for the point-estimate zero boundary gives
\[
\rho_t^\dagger=\omega_t/\sqrt{v_Mv_t}.
\]
Its uncertainty uses the full nuisance covariance and a Fisher-transform delta interval. We distinguish the point-estimate zero $\widehat\rho_t^\dagger$ from the correlation where a pointwise effect interval first includes zero, starting with the interval at $\rho=0$. The former has a common first-order residual-moment representation for both methods; density-score gains in curve estimation arise primarily through regression coefficients and their joint covariance.

\subsection{Numerical evaluation}

The extension is evaluated at $n=300$, with 500 attempts per configuration, using Gaussian and asymmetric-mixture innovations and $\rho_{\rm true}\in\{-0.3,0,0.3\}$. The original structural means are retained. Independent standardized innovations generate $\epsilon_M=U_M$ and $\epsilon_Y=\rho_{\rm true}U_M+\sqrt{1-\rho_{\rm true}^2}U_Y$, preserving natural effects while introducing confounding at nonzero correlation. Both methods are evaluated at $\rho=\rho_{\rm true}$ and the unadjusted reference $\rho=0$. The six configurations yield 3,000 data sets and 6,000 method attempts. Table~S10 gives representative effects at the true correlation; Section~S8 details verification and complete-output reporting.

At the correct assumed correlation, DS-RF succeeds in 498, 500 and 499 of 500 asymmetric-mixture attempts at $\rho_{\rm true}=-0.3,0,0.3$, respectively; all Gaussian and OLS-RF attempts succeed. On paired successful mixture samples, the PNIE RMSE ratios (DS-RF/OLS-RF) are 1.011, 0.706 and 0.559, and the interval-length ratios are 1.014, 0.698 and 0.524. DS-RF PNIE coverage is 0.958, 0.952 and 0.946. These results identify how the PNIE precision gain varies with the generating correlation. TNIE RMSE ratios are 0.794, 0.712 and 0.746, with DS-RF coverage between 0.928 and 0.944; TNDE RMSE is higher at the two lower correlations. Total-effect precision and zero-boundary interval lengths are nearly unchanged, and Gaussian results are consistent with the OLS-efficient reference. These patterns connect effect-specific precision gains to curve estimation and distinguish them from estimation of the moment-based zero boundary.

Evaluation at the unadjusted reference $\rho=0$ quantifies the consequences of leaving confounding unaccounted for: for mixture errors at true correlations $-0.3$ and $0.3$, DS-RF PNIE biases are $-0.119$ and $0.120$, and coverage is 0.512 and 0.158. This comparison highlights the complementary roles of nuisance-estimation precision and specification of the sensitivity value in recovering the causal effect. Monte Carlo uncertainty for the 500 attempts is quantified through complete effect-specific standard errors; paired summaries, failures and all-attempted report-and-cover fractions are also supplied in the project repository.

\subsection{Performance along sensitivity curves}

We evaluate the sensitivity curves at $\rho=-0.8,-0.75,\ldots,0.8$, using the same 500 data sets per configuration and identical estimator settings. Curve evaluations remain correlated within each replication; the Monte Carlo sample size is 500, not the number of grid evaluations.

The evaluation uses two complementary targets. If the generating correlation is $r$, the population PNIE/TNIE sensitivity curves in this design are
\begin{equation}
\delta_t^{(r)}(\rho)
=0.4\left\{-0.8+t+r-\frac{\rho}{\sqrt{1-\rho^2}}\sqrt{1-r^2}\right\}.
\label{eq:simulation-sensitivity-truth}
\end{equation}
Curve bias, RMSE and pointwise coverage target \eqref{eq:simulation-sensitivity-truth}; inclusion of the fixed generating effect is recorded separately. The targets coincide at $\rho=r$, distinguishing curve calibration from causal recovery when a different correlation is assumed.

\begin{figure}[htbp]
\centering
\includegraphics[width=0.98\textwidth]{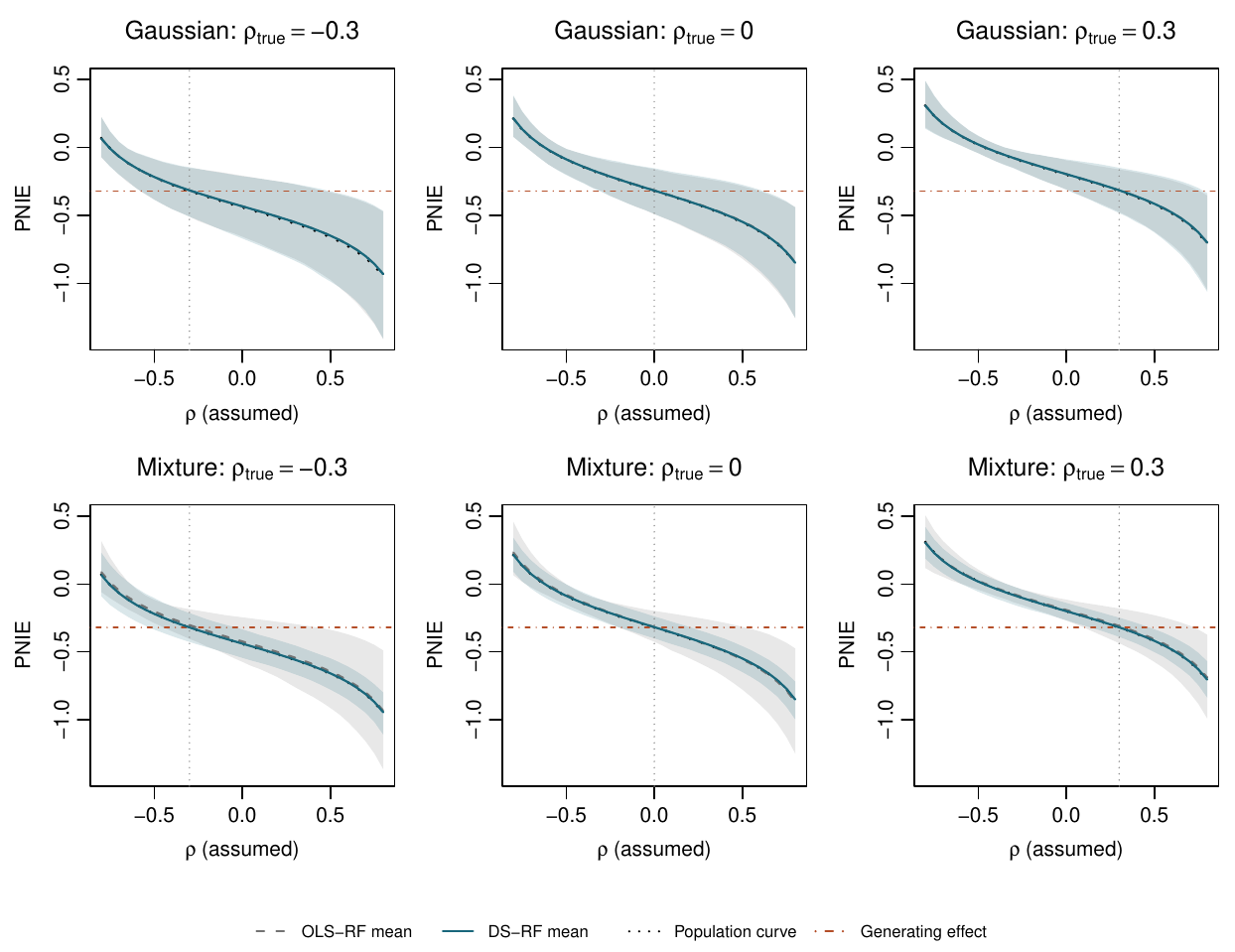}
\caption{Simulated PNIE sensitivity curves at $n=300$, with 500 attempted replications for each generating correlation and innovation density. Lines show Monte Carlo mean estimates and the population sensitivity curve. Gray and blue shading show central 90\% empirical sampling ranges of the OLS-RF and DS-RF estimates across successful replications, respectively. The horizontal reference is the generating PNIE, $-0.32$, and the vertical reference is the generating correlation. All six configurations are shown.}
\label{fig:simulation-sensitivity-curves}
\end{figure}

Figure~\ref{fig:simulation-sensitivity-curves} displays PNIE curve estimates and sampling ranges. Figures~S3--S4 and Table~S11 add pointwise coverage, paired RMSE and zero-boundary inference; the project repository retains all effects, correlations, Monte Carlo errors and all-attempted summaries.

Across all three generating correlations and the 33-point grid, paired PNIE RMSE ratios range from 0.985 to 1.054 under Gaussian innovations and from 0.341 to 1.363 under asymmetric-mixture innovations. The asymmetric-mixture results identify regions with substantial precision gains alongside regions with comparable or higher RMSE. Mixture TNIE ratios similarly range from 0.367 to 1.085. This pattern is explained by the nuisance components that dominate the effect gradient: near a zero of $b_t(\rho)$, the derivative of $\delta_t$ with respect to $\beta_2$ is small, whereas regions with a larger derivative can benefit more from precise mediator treatment-effect estimation.

For mixture innovations, DS-RF pointwise curve coverage ranges from 0.924 to 0.966 for PNIE and from 0.904 to 0.964 for TNIE, quantifying the calibration accompanying the precision pattern across the grid. Gaussian coverage likewise varies with the assumed correlation. Zero-boundary RMSE and interval lengths are nearly identical between methods (Table~S11), as predicted by their shared first-order residual-moment representation. Together, the results characterize three complementary features of sensitivity analysis: precision along the assumed-correlation curve, pointwise interval calibration and the point-estimate zero boundary. The separately retained generating-effect inclusion frequencies quantify how recovery of the causal effect depends on the specified confounding value.

\FloatBarrier

\section{JOBS II behavioral-intervention illustration}
\label{sec:application}

The JOBS II data distributed with version 4.5.1 of the \textsf{R} package \texttt{mediation} contain 899 complete observations from a randomized intervention designed to strengthen job-search skills and protect mental health among unemployed workers \citep{vinokur1997,imai2010b,tingley2014}.  Treatment is randomized assignment to the intervention, the mediator is job-search self-efficacy, and the outcome is post-intervention depressive symptoms.  We use the package sample as a methodological illustration of natural-effect estimation and sensitivity analysis.

Both mean models adjust for pretreatment depressive symptoms, economic hardship, sex and age, and the outcome model includes treatment--mediator interaction.  Every package observation is complete for these variables and is retained.  Effects are standardized to the empirical baseline-covariate distribution.  All five procedures use the same 899 observations.  The two analyses implemented in the \textsf{R} package \texttt{mediation} use 5,000 quasi-Bayesian draws and 1,999 percentile-bootstrap resamples, respectively; the \textsf{R} package \texttt{medflex} supplies the natural effect model.

\begin{table}[t]
\centering
\caption{Proposed density-score estimates of JOBS II natural effects and nominal 95\% limits ($n=899$). Negative values indicate lower post-intervention depressive symptoms under assignment to the intervention}
\label{tab:application_effects}
\small
\begin{tabular}{lrrr}
\toprule
Effect & Estimate & Lower & Upper \\
\midrule
PNIE & -0.0109 & -0.0205 & -0.0013 \\
TNIE & -0.0072 & -0.0144 & -0.0001 \\
PNDE & -0.0272 & -0.0892 & 0.0349 \\
TNDE & -0.0235 & -0.0835 & 0.0364 \\
TE & -0.0344 & -0.0962 & 0.0274 \\
\bottomrule
\end{tabular}
\end{table}

\begin{figure}[t]
\centering
\includegraphics[width=0.80\textwidth]{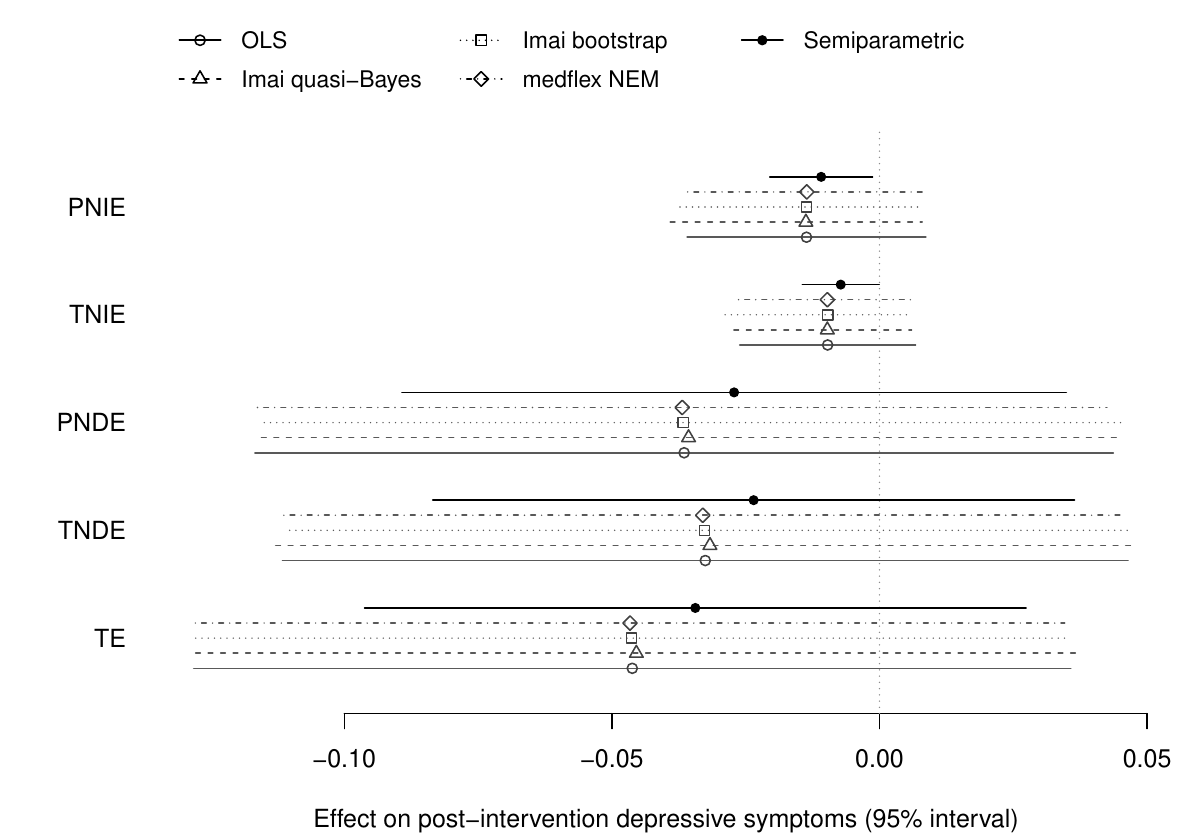}
\caption{JOBS II natural-effect estimates for post-intervention depressive symptoms.  Points denote estimates, horizontal bars nominal 95\% intervals, and the vertical line zero.  Negative values indicate lower depressive symptoms under assignment to the intervention}
\label{fig:application-effects}
\end{figure}

Table~\ref{tab:application_effects} reports the proposed estimates, while Figure~\ref{fig:application-effects} compares all five methods; their complete numerical results appear in Table~S12 of the Supplementary Material.  The four conventional analyses give similar small negative estimates.  OLS gives PNIE $-0.0136$ with nominal 95\% interval $[-0.0360,0.0087]$, whereas the proposal gives $-0.0109$ with $[-0.0205,-0.0013]$.  The corresponding TNIE estimates are $-0.0097$ and $-0.0072$.  The proposed-to-OLS standard-error ratios are 0.430 for PNIE, 0.437 for TNIE, and 0.754--0.774 for the direct and total effects.  Thus the largest precision difference occurs for the indirect effects, whose proposed nominal intervals exclude zero under the working model.

Section~S9 supplies the complete estimates and diagnostics. OLS and the proposal target the same effect definitions through different estimating equations; under misspecification, their limiting targets may also differ. The empirical comparison is therefore interpreted with diagnostics, while comparative bias and efficiency are assessed against known targets in the simulations.

The default fit selected the OLS-centered start in both equations, with one distinct valid root and mean-score norms below $3\times10^{-12}$.  Scales $c=0.05,0.10,0.20$ gave identical displayed results (Table~S14).  Removing pretreatment depression from the adjustment set changed the proposed PNIE only from $-0.0109$ to $-0.0122$; both nominal intervals excluded zero (Table~S15).  This stability check supports the numerical pattern while retaining the baseline-depression-adjusted specification as primary.

Model checks guide the working-model interpretation.  OLS residual skewness is $-0.781$ for the mediator and $1.083$ for the outcome, providing substantial non-Gaussian shape information.  Full-design residual-variance diagnostics give $p=0.0228$ and $p=7.67\times10^{-5}$, respectively \citep{koenker1981}.  Full diagnostics are in Table~S16 and Figure~S5.  We therefore interpret the reported intervals jointly with both the non-Gaussian residual shapes and evidence of variance heterogeneity.  The subsequent sensitivity analysis addresses unmeasured mediator--outcome confounding as a separate dimension of the illustration.

\FloatBarrier

\subsection{Confounding-sensitivity analysis for JOBS II}

Using the same participants and baseline variables, we fit the reduced-form extension over $\rho\in[-0.8,0.8]$. The three-regression stack and its complete nuisance covariance provide a self-contained calculation of the sensitivity functionals and their uncertainty. Its finite-sample $\rho=0$ results complement the primary conditional-outcome estimates in Table~\ref{tab:application_effects} as a distinct reduced-form analysis.

\begin{table}[htbp]\centering\small\setlength{\tabcolsep}{4pt}
\caption{JOBS II reduced-form sensitivity analysis. Nominal intervals are pointwise working-model intervals. The zero-effect correlation is conditional on a nonzero mediator treatment effect.}
\label{tab:confounding-thresholds}
\begin{tabular}{llrrr}\toprule
Effect & Method & Effect at $\rho=0$ (95\% CI) & $\widehat\rho^\dagger$ (95\% CI) & $(\widehat\rho^\dagger)^2$\\\midrule
PNIE & OLS-RF & -0.012 [-0.032, 0.008] & -0.231 [-0.334, -0.122] & 0.053\\
TNIE & OLS-RF & -0.010 [-0.028, 0.007] & -0.213 [-0.292, -0.130] & 0.045\\
PNIE & DS-RF & -0.015 [-0.028, -0.002] & -0.244 [-0.345, -0.137] & 0.059\\
TNIE & DS-RF & -0.013 [-0.025, -0.002] & -0.222 [-0.300, -0.142] & 0.049\\
\bottomrule\end{tabular}\end{table}

\begin{figure}[htbp]
\centering
\includegraphics[width=0.92\textwidth]{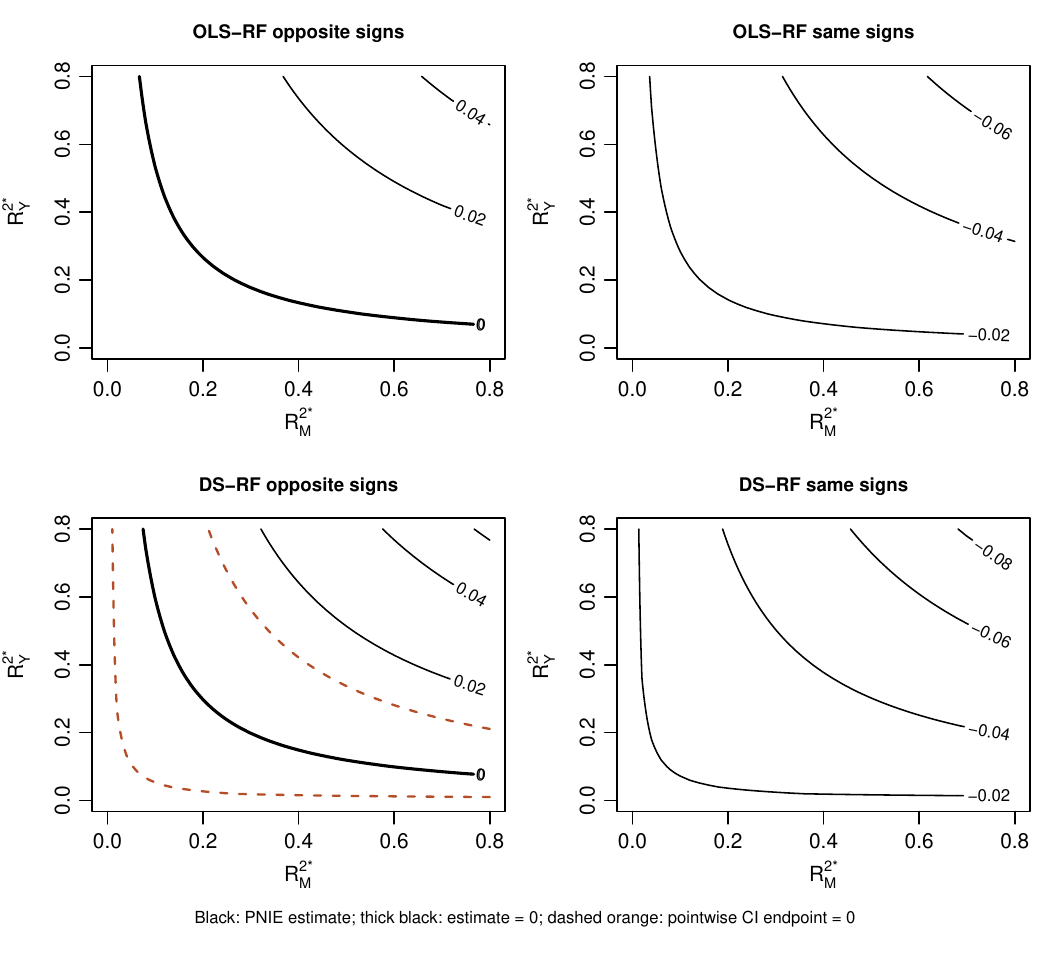}
\caption{JOBS II PNIE sensitivity contours from OLS-RF and DS-RF. Each black curve gives combinations of hypothetical residual-variance fractions yielding the labeled point estimate. The thick zero contour concerns the point estimate when crossed in the displayed grid; dashed orange curves mark a pointwise nominal interval endpoint equal to zero. Both confounding signs are shown. The contours are interpreted under the reduced-form working model and alongside the residual-variance diagnostics.}
\label{fig:confounding-contours}
\end{figure}

We use the additional variance-decomposition interpretation of \citet{imai2010a}. Let $R_M^{2*}$ and $R_Y^{2*}$ denote the residual variance fractions explained by a hypothetical pretreatment confounder. The calibration is $\rho=s\sqrt{R_M^{2*}R_Y^{2*}}$, where $s\in\{-1,1\}$ specifies opposite or matching confounder-effect signs. Figure~\ref{fig:confounding-contours} shows both signs. The axes quantify residual-variance fractions and are distinct from the total-variance quantities $\widetilde R_M^2,\widetilde R_Y^2$. This working calibration translates specified latent variance contributions into sensitivity values; the one-dimensional PNIE/TNIE curves appear in Figure~S6.

At $\rho=0$, OLS-RF gives PNIE $-0.0116$ with nominal interval $[-0.0316,0.0084]$, whereas DS-RF gives $-0.0149$ with $[-0.0284,-0.0015]$. The corresponding DS-RF TNIE is $-0.0132$ with $[-0.0246,-0.0018]$. The estimated correlations at which the point estimates cross zero are $-0.231$ for OLS-RF and $-0.244$ for DS-RF for PNIE, corresponding to residual-$R^2$ products of 0.053 and 0.059; TNIE gives $-0.213$ and $-0.222$. Thus opposite-sign confounder associations explaining roughly 5--6\% of the two residual variances in product provide a useful calibration for the estimated zero boundary. The close boundary estimates agree with their shared first-order residual-moment representation, while the shorter DS-RF intervals reflect the regression-coefficient contribution to curve estimation. Reduced-form variance diagnostics give reference $p$-values of 0.024 for the pooled mediator model and 0.009 and 0.006 for the two DS-RF outcome strata. The sensitivity curves are therefore interpreted as pointwise working-model summaries alongside the variance diagnostics and causal assumptions.

\FloatBarrier

\section{Discussion}
\label{sec:discussion}

The framework connects error-density information to natural-effect inference and confounding-sensitivity analysis. For regular natural effects under the independent-error mediation model and stated regularity conditions, the joint influence-function representation links regression-level information gains to lower asymptotic variance and higher local power relative to OLS. Simulations support these gains in several non-Gaussian settings, with Gaussian errors providing the OLS-efficient reference.

The independently calibrated experiment connects precision to detection of a weak indirect effect: the advantage persists after accounting for OLS conservatism and against interaction-adapted Huber estimators in the asymmetric-mixture design. Conditional power and all-attempt detection distinguish statistical gains from the probability of returning a usable result.

For the sensitivity functional of \citet{imai2010a}, density-score nuisance estimation and joint covariance propagation improve curve precision in regions identified by the effect gradient. The shared residual-moment representation explains why these gains coexist with nearly unchanged point-estimate zero boundaries. Efficient marginal scores and the assembled estimator's joint influence function support this distinction between curve estimation and confounding sensitivity.

Variance misspecification and unmeasured confounding are separate diagnostic questions. In the asymmetric-mixture experiment, treatment-dependent scaling has the larger impact on calibration. The confounding extension instead relaxes mediator--outcome ignorability while retaining its reduced-form error restrictions. JOBS II illustrates why both dimensions should accompany the effect estimates and nominal intervals.

The deterministic multi-start algorithm and retained fit diagnostics make numerical reliability part of the reproducible evaluation. Covariate-dependent error distributions, incomplete outcomes and comparisons with flexible mediation estimators are directions for further development. The present framework provides a theoretical and computational basis for using error-shape information in mediation inference under explicit assumptions.

\section*{Statements and Declarations}

\paragraph{Competing interests.}
The author declares no financial or non-financial competing interests.

\paragraph{Funding.}
No external funding was received for this study.

\paragraph{Data availability.}
The JOBS II illustration data are distributed with version 4.5.1 of the \textsf{R} package \texttt{mediation}. The exact data object used for this analysis is included in the project repository identified below.

\paragraph{Ethics approval and consent to participate.}
This methodological study uses publicly distributed, de-identified JOBS II illustration data.  It involved no new participant recruitment or participant identification; the original intervention is described by \citet{vinokur1997}.

\paragraph{Code availability.}
The code and reproducibility materials are publicly available at \url{https://github.com/mijeong-kim/semi_causal_med}. The repository contains the analysis scripts and seeds, the JOBS II data object, retained numerical results, tables, figures, package versions and session information. Its \texttt{README.md} provides the software requirements and commands for reproducing the analyses and rebuilding the tables and figures from the retained results.

\paragraph{Supplementary information.}
The Supplementary Material contains proofs, implementation details, complete numerical results and diagnostics. Code and reproducibility materials are provided through the GitHub repository listed above.

\paragraph{Author contributions.}
Mijeong Kim developed the statistical methodology, conducted the analyses, and wrote the manuscript.

\clearpage
\newgeometry{margin=0.8in}
\pdfbookmark[0]{Supplementary Material}{supplementary-material}
\section*{Supplementary Material}
\captionsetup[figure]{name=Figure}
\let\suppsection\section
\renewcommand{\section}{\FloatBarrier\suppsection}
\setcounter{section}{0}
\renewcommand{\thesection}{S\arabic{section}}
\def\theHsection{supp.\arabic{section}}
\setcounter{equation}{0}
\renewcommand{\theequation}{S\arabic{equation}}
\def\theHequation{supp.\arabic{equation}}
\setcounter{table}{0}
\renewcommand{\thetable}{S\arabic{table}}
\def\theHtable{supp.\arabic{table}}
\setcounter{figure}{0}
\renewcommand{\thefigure}{S\arabic{figure}}
\def\theHfigure{supp.\arabic{figure}}
\setcounter{proposition}{0}
\renewcommand{\theproposition}{S\arabic{proposition}}
\def\theHproposition{supp.\arabic{proposition}}
\setcounter{theorem}{0}
\renewcommand{\thetheorem}{S\arabic{theorem}}
\def\theHtheorem{supp.\arabic{theorem}}

This supplement follows the main article's order, supplying theoretical derivations, implementation details, complete simulation results and JOBS II diagnostics. The main article gives the principal interpretation; the tables here retain effect-specific estimates, calibration and numerical-reliability measures. Section~S10 describes reproduction of the analyses.

\section{Identification calculation}

Under sequential ignorability, consistency, and positivity, the mediation formula identifies the mean nested potential outcomes.  The linear models imply
\[
\E\{M(t)\mid\bX\}=\alpha_2+\beta_2t+\bm{\xi}_2\trans\bX
\]
and
\[
\E\{Y(t,m)\mid\bX\}=\alpha_3+\beta_3t+\gamma m+\eta tm
+\bm{\xi}_3\trans\bX.
\]
Therefore, conditional on $\bX$,
\begin{align*}
\E[&Y\{t,M(1)\}-Y\{t,M(0)\}\mid\bX]
 =(\gamma+\eta t)\beta_2,\\
\E[&Y\{1,M(t)\}-Y\{0,M(t)\}\mid\bX]
 =\beta_3+\eta\{\alpha_2+\beta_2t+\bm{\xi}_2\trans\bX\}.
\end{align*}
Averaging over $\bX$ yields
\begin{align*}
\operatorname{PNIE}&=\beta_2\gamma,
&\operatorname{TNIE}&=\beta_2(\gamma+\eta),\\
\operatorname{PNDE}&=\beta_3+\eta\mu_0,
&\operatorname{TNDE}&=\beta_3+\eta\mu_1,
\end{align*}
where $\mu_0=\alpha_2+\bm{\xi}_2\trans\bmu_X$ and $\mu_1=\mu_0+\beta_2$.  Direct addition gives
\[
\operatorname{PNDE}+\operatorname{TNIE}
=\operatorname{TNDE}+\operatorname{PNIE}
=\beta_3+\beta_2\gamma+\eta\mu_1=\operatorname{TE}.
\]

\section{Gradient of the natural-effect map}

Write the parameter blocks as
\[
\theta_M=(\alpha_2,\beta_2,\bm{\xi}_2\trans,v_M)\trans,
\qquad
\theta_Y=(\alpha_3,\beta_3,\gamma,\eta,\bm{\xi}_3\trans,v_Y)\trans,
\]
where the implementation uses log variances without changing any effect derivative.  Gradients not listed below are zero.  With differentials used to display the full map,
\begin{align}
d\operatorname{PNIE}
&=\gamma\,d\beta_2+\beta_2\,d\gamma,\nonumber\\
d\operatorname{TNIE}
&=(\gamma+\eta)\,d\beta_2+\beta_2\,d\gamma+\beta_2\,d\eta,\nonumber\\
d\operatorname{PNDE}
&=\eta\,d\alpha_2+\eta\bmu_X\trans d\bm{\xi}_2
+\eta\bm{\xi}_2\trans d\bmu_X+d\beta_3+\mu_0\,d\eta,\nonumber\\
d\operatorname{TNDE}
&=\eta\,d\alpha_2+\eta\,d\beta_2
+\eta\bmu_X\trans d\bm{\xi}_2+\eta\bm{\xi}_2\trans d\bmu_X
+d\beta_3+\mu_1\,d\eta,\nonumber\\
d\operatorname{TE}
&=\eta\,d\alpha_2+(\gamma+\eta)\,d\beta_2
+\eta\bmu_X\trans d\bm{\xi}_2+\eta\bm{\xi}_2\trans d\bmu_X
+d\beta_3+\beta_2\,d\gamma+\mu_1\,d\eta.
\label{eq:full-gradient}
\end{align}
Equation~\eqref{eq:full-gradient} is the analytic matrix $G$ used to propagate the stacked covariance.  In particular, the terms involving $d\bmu_X$ are retained rather than treating the sample covariate mean as fixed.

\section{Regularity conditions and influence-function composition}
\label{sec:regularity}

Let $W=(\bX,T)$.  Under the statistical assumptions in the main article, the observed-data density has the factorization
\begin{equation}
p(o)=p_W(w)f_M\{m-m_M(w;\theta_M)\}
f_Y\{y-m_Y(w,m;\theta_Y)\}.
\label{eq:factorization}
\end{equation}
The nuisance components are the unrestricted law of $W$ and the two error densities subject to their moment restrictions.  Let $S_M^{\mathrm{eff}}$ and $S_Y^{\mathrm{eff}}$ be the efficient regression scores stated in the main text.

\paragraph{Regularity conditions.}
For $k\in\{M,Y\}$, let $\theta_{k0}$ denote the true regression parameter, including the log variance, and let $\widehat S_{k,i}$ denote its feasible score with the density, moments and design expectations replaced by their estimates.  Under the regression model and factorization \eqref{eq:factorization}, we use the following conditions.  Condition (R1) concerns the population score representation; (R2)--(R4) support asymptotic linearity of the fitted estimators; and (R5) supports covariance estimation and Wald inference.

\begin{enumerate}
\renewcommand{\labelenumi}{(R\arabic{enumi})}
\renewcommand{\theenumi}{R\arabic{enumi}}
\item\label{cond:score-model}\textbf{Sampling, regression scores and moments.}
The observations are independent and identically distributed, and $\E\|\bX\|^2<\infty$.  For each regression, the error density is positive and continuously differentiable on its support, the error has mean zero, positive finite variance and a finite fourth moment, and $0<Q_{2,k}<\infty$.  The location Fisher information is finite, the design has full-rank information, and the population score equation has a locally unique root.  The boundary and integrability requirements for the mean-zero score identities and the regression-score calculation hold.  In particular, $\E(S_k^{\mathrm{eff}})=0$ and $I_k=\E\{S_k^{\mathrm{eff}}S_k^{\mathrm{eff}\,\mathsf T}\}$ is finite and positive definite.  These are the population regression-score requirements stated in Section~2.2 of the main article, together with the moment condition for population averaging over $\bX$.

\item\label{cond:consistent-root}\textbf{Consistent selected roots.}
With probability tending to one, each regression has a selected root $\widehat\theta_k$ satisfying $\widehat\theta_k\xrightarrow{p}\theta_{k0}$, where $\theta_{k0}$ is an interior parameter point.  The sample mean score at the selected root is $o_p(n^{-1/2})$.  An exact root satisfies the latter requirement automatically; for a numerical root it is a condition on the asymptotic solver accuracy.

\item\label{cond:score-replacement}\textbf{Negligible score-replacement error.}
At the true parameter,
\begin{equation}
n^{-1/2}\sum_i\{\widehat S_{k,i}(\theta_{k0})
              -S_{k,i}^{\mathrm{eff}}(\theta_{k0})\}=o_p(1).
\label{eq:score-replacement}
\end{equation}
This condition covers the combined replacement of the error density, its derivative, residual moments and design expectations.  It controls their effect on the aggregate estimating equation, beyond pointwise consistency of a density or density-derivative estimator.  The Gaussian-kernel and leave-one-out construction used in the code is connected to this condition in Proposition~\ref{prop:feasible-root-bridge} below.

\item\label{cond:score-derivative}\textbf{Local differentiability and derivative convergence.}
The sample mean score is continuously differentiable on a neighborhood containing the line segment between $\theta_{k0}$ and $\widehat\theta_k$, and
\begin{equation}
\frac{\partial}{\partial\theta_k^\mathsf T}
  \left\{n^{-1}\sum_i\widehat S_{k,i}(\theta_k)\right\}
  =-I_k+o_p(1)
\label{eq:local-bread}
\end{equation}
uniformly on that segment.

\item\label{cond:covariance}\textbf{Covariance consistency and nondegenerate scalar inference.}
At the fitted parameters, the empirical stacked score second moments satisfy $\widehat B\xrightarrow{p}B=\E(\Psi_i\Psi_i\trans)$, and the implemented bread, including its numerical regression derivatives, satisfies $\widehat A\xrightarrow{p}A=\operatorname{diag}(-I_M,-I_Y,-I_{d_X})$, where $d_X$ is the dimension of $\bX$.  The effect map $g$ is continuously differentiable.  For each scalar effect reported with a regular Wald interval, its limiting variance $G_jA^{-1}BA^{-\mathsf T}G_j\trans$ is strictly positive, where $G_j$ is the corresponding row of the effect gradient at the truth.  The full effect covariance need not be nonsingular.
\end{enumerate}

\paragraph{Feasible kernel scores and deterministic root selection.}
To connect the implemented root to Conditions~(\ref{cond:consistent-root})--(\ref{cond:score-derivative}), let $\overline S_{k,n}^{\mathrm{full}}(\theta)$ be the feasible full-kernel equation of \citet[Theorem~2]{kim2023} and $\overline S_{k,n}^{\mathrm{loo}}(\theta)$ its stabilized leave-one-out counterpart. The following proposition gives conditions for first-order equivalence and consistent root selection.

\begin{proposition}[Feasible-root bridge]
\label{prop:feasible-root-bridge}
For each $k\in\{M,Y\}$, suppose the kernel smoothness, boundary, tail, moment and unique-root conditions of \citet[Theorem~2 and its proof]{kim2023} hold uniformly in a neighborhood of $\theta_{k0}$, $s_{e,k}\xrightarrow{p}\sqrt{v_k}$, and $h_{k,n}=1.06s_{e,k}n^{-1/5}$.  Suppose also that there are expanding central sets $\mathcal E_{k,n}$ and constants $c_{k,n}>0$ such that, uniformly in that neighborhood, the true and estimated densities on $\mathcal E_{k,n}$ are at least $c_{k,n}$, the estimated first density derivative is $O_p(1)$, $\kappa_n=\min(10^{-10},n^{-2})=o(c_{k,n})$, and
\begin{equation}
\frac{\sqrt n}{nh_{k,n}c_{k,n}^2}\longrightarrow0.
\label{eq:central-density-rate}
\end{equation}
Assume that the contributions from $\mathcal E_{k,n}^{c}$ to the sample-mean score and its derivative are $o_p(n^{-1/2})$ and $o_p(1)$, respectively.  Finally, suppose the solver initialized at OLS returns the locally unique valid root with probability tending to one and any other valid root is asymptotically separated from $\theta_{k0}$ in the OLS-design distance.  Then the selected stabilized leave-one-out root satisfies Conditions~(\ref{cond:consistent-root})--(\ref{cond:score-derivative}) and
\begin{equation}
\sqrt n(\widehat\theta_k-\theta_{k0})
=I_k^{-1}n^{-1/2}\sum_{i=1}^n
S_{k,i}^{\mathrm{eff}}(\theta_{k0})+o_p(1).
\label{eq:feasible-root-expansion}
\end{equation}
Consequently, the natural-effect expansion and the OLS-relative variance and local-power comparisons in the main article apply to the feasible estimator returned by the supplied algorithm.
\end{proposition}

\begin{proof}
Write $K$ for the standard Gaussian density and let $\widehat f_i$ and $\widehat f'_i$ be the full-sample kernel estimators evaluated at residual $e_i$.  Because $K'(0)=0$, direct algebra gives
\begin{equation*}
\widehat f_{-i}
=\frac{n}{n-1}\left\{\widehat f_i-\frac{K(0)}{nh_{k,n}}\right\},
\qquad
\widehat f'_{-i}=\frac{n}{n-1}\widehat f'_i,
\end{equation*}
and hence
\[
\frac{\widehat f'_{-i}}{\widehat f_{-i}}
=\frac{\widehat f'_i}
 {\widehat f_i-K(0)/(nh_{k,n})}.
\]
On $\mathcal E_{k,n}$, deleting the self-density term therefore changes an
observation-level density score by
\[
O_p\!\left\{(nh_{k,n}c_{k,n}^{2})^{-1}\right\}.
\]
Equation~\eqref{eq:central-density-rate} makes its root-$n$ aggregate
contribution $o_p(1)$.  The relation $\kappa_n=o(c_{k,n})$ makes the numerical
floor asymptotically inactive on these central sets, and the stated tail
condition controls their complements.  The same calculation uniformly in a
local neighborhood, together with the derivative condition from the
kernel-root argument of \citet[Theorem~2]{kim2023}, gives
\begin{align*}
\sqrt n\{\overline S_{k,n}^{\mathrm{loo}}(\theta_{k0})
 -\overline S_{k,n}^{\mathrm{full}}(\theta_{k0})\}&=o_p(1),\\
\sup_{\theta}\left\|
\frac{\partial\overline S_{k,n}^{\mathrm{loo}}(\theta)}
     {\partial\theta\trans}
-\frac{\partial\overline S_{k,n}^{\mathrm{full}}(\theta)}
     {\partial\theta\trans}
\right\|&=o_p(1),
\end{align*}
where the supremum is over the shrinking neighborhood traversed by the root expansion.  Thus the leave-one-out root has the same first-order expansion as the full-kernel root in \citet[Theorem~2]{kim2023}.

OLS is consistent under Condition~(\ref{cond:score-model}), so the distance between the OLS pilot and the local consistent root is $o_p(1)$.  By the assumed separation, every valid nonlocal root has distance bounded away from zero with probability tending to one.  The minimum-distance rule therefore selects the local root.  A mean-value expansion of its score equation yields \eqref{eq:feasible-root-expansion}, which is precisely Conditions~(\ref{cond:consistent-root})--(\ref{cond:score-derivative}).  Stacking the two expansions with the sample-mean expansion and applying the effect gradient gives the final claim.
\end{proof}

\paragraph{Population influence-function composition.}
Under Condition~(\ref{cond:score-model}) and \eqref{eq:factorization}, every component of the mediator score satisfies
\[
\E\{S_M^{\mathrm{eff}}\mid W\}=0,
\]
because both the error-density score and the location--scale correction have zero expectation under $f_M$.  Similarly,
\[
\E\{S_Y^{\mathrm{eff}}\mid W,M\}=0.
\]
It follows by iterated expectation that
\begin{align*}
\E\{S_M^{\mathrm{eff}}S_Y^{\mathrm{eff}\,\mathsf T}\}&=0,\\
\E[S_M^{\mathrm{eff}}\{\bX-\bmu_X\}^{\mathsf T}]&=0,\\
\E[S_Y^{\mathrm{eff}}\{\bX-\bmu_X\}^{\mathsf T}]&=0.
\end{align*}
Thus the two regression scores and the centered baseline covariate are mutually orthogonal in the tangent space generated by \eqref{eq:factorization}.  The efficient influence functions for $\theta_M$, $\theta_Y$, and $\bmu_X$ are respectively
\[
\varphi_M=I_M^{-1}S_M^{\mathrm{eff}},\qquad
\varphi_Y=I_Y^{-1}S_Y^{\mathrm{eff}},\qquad
\varphi_X=\bX-\bmu_X.
\]

The natural-effect vector is a continuously differentiable finite-dimensional functional,
\[
g=g(\theta_M,\theta_Y,\bmu_X).
\]
Applying the pathwise chain rule in any regular parametric submodel gives
\[
\varphi_g=G_M\varphi_M+G_Y\varphi_Y+G_X\varphi_X.
\]
This function lies in the tangent space of the model and is orthogonal to its nuisance tangent space while representing the pathwise derivative of $g$; hence it is the canonical gradient.  Its variance is the population efficiency bound.  The result specializes established regression efficiency and the pathwise chain rule to the linked mediation effect map \citep{kim2023,bickel1993,tsiatis2006}.

\paragraph{OLS-relative variance and local power.}
Here we give the score-block calculation underlying Proposition~2 in the main article.  In one of the regressions, write $R=a+\bm b\trans\bm W+\epsilon$, with $\omega=\E(\bm W)$, $C=\Var(\bm W)$, $s(\epsilon)=-f'(\epsilon)/f(\epsilon)$, and $J=\E\{s(\epsilon)^2\}$.  Set $q=\epsilon^2-v-(\mu_3/v)\epsilon$, $Q_2=\E(q^2)$, and $t=\epsilon/v-\mu_3q/(vQ_2)$.  The efficient score for the intercept is $t$, the slope score is $s(\epsilon)(\bm W-\omega)+t\omega$, and the log-variance score is $S_\lambda=vq/Q_2$.  Centering the slope parameter by the fixed design mean $\omega$ separates its score $s(\epsilon)(\bm W-\omega)$ from the intercept--variance block.  By independence, this slope block has information $JC$ and is orthogonal to the other two scores.

Since $\E(\epsilon q)=0$, direct calculation gives
\[
\E(t^2)=\frac1v+\frac{\mu_3^2}{v^2Q_2},\qquad
\E(tS_\lambda)=-\frac{\mu_3}{Q_2},\qquad
\E(S_\lambda^2)=\frac{v^2}{Q_2}.
\]
Residualizing $t$ against $S_\lambda$ therefore yields
\[
t-\frac{\E(tS_\lambda)}{\E(S_\lambda^2)}S_\lambda
=t+\frac{\mu_3}{v^2}S_\lambda
=\frac{\epsilon}{v},
\]
whose information is $1/v$.  Joint information inversion thus gives the efficient slope influence function
\[
\varphi_{\bm b}^{\mathrm{semi}}
=C^{-1}(\bm W-\omega)\frac{s(\epsilon)}{J},
\]
and the original intercept influence function
\[
\varphi_a^{\mathrm{semi}}
=\epsilon-\omega\trans\varphi_{\bm b}^{\mathrm{semi}}.
\]
The OLS normal equations give the same two expressions with $s(\epsilon)/J$ replaced by $\epsilon$.  Equivalently, with $e_0=(1,0,\ldots,0)\trans$ and $H=\begin{pmatrix}-\omega\trans\\ I\end{pmatrix}$,
\begin{align*}
\Var(\varphi_{\bbeta}^{\mathrm{semi}})
&=v e_0e_0\trans+J^{-1}HC^{-1}H\trans,\\
\Var(\varphi_{\bbeta}^{\mathrm{OLS}})
&=v e_0e_0\trans+v HC^{-1}H\trans.
\end{align*}
The score identity $\E\{\epsilon s(\epsilon)\}=-\int e f'(e)\,de=1$ holds under the boundary condition in (R1).  Hence $1\leq\{vJ\}^{1/2}$ and $v-J^{-1}\geq0$; equality requires $s(\epsilon)=\epsilon/v$ almost surely, the Gaussian density under the stated smoothness and support conditions.

For either estimator, the influence function of natural effect $g_j$ is the sum of the mediator-coefficient, outcome-coefficient, and baseline-mean contributions.  The conditional mean-zero properties established above make these three blocks orthogonal for both OLS and the efficient regression estimator.  The baseline-mean contribution is identical for the two methods, so subtracting their effect variances yields the OLS-relative variance formula in Proposition~2 of the main article.  At a regular local alternative $g_{j,n}=h/\sqrt n$, assume the asymptotic linear expansions and variance consistency hold along the contiguous sequence.  Studentization gives a limiting $N(h/\sqrt{V_{j,r}},1)$ statistic.  With $z=z_{1-\alpha_{\mathrm{test}}/2}$ and $\lambda=|h|/\sqrt{V_{j,r}}$, its two-sided rejection probability is $1-\Phi(z-\lambda)+\Phi(-z-\lambda)$, whose derivative in $\lambda>0$ equals $\phi(z-\lambda)-\phi(z+\lambda)>0$.  This proves the local-power ordering at a common asymptotic size.  The argument does not cover the joint-zero PNIE point, where the first-order gradient vanishes.

\paragraph{Asymptotic inference for the fitted estimators.}
Under Conditions~(\ref{cond:consistent-root})--(\ref{cond:score-derivative}), score expansion using \eqref{eq:score-replacement}--\eqref{eq:local-bread} yields \eqref{eq:feasible-root-expansion}. Combining the two regression expansions with
$\sqrt n(\overline{\bX}-\bmu_X)=n^{-1/2}\sum_i(\bX_i-\bmu_X)$
and applying the multivariate delta method proves the estimator part of Proposition~1 in the main article. Condition~(\ref{cond:covariance}) gives sandwich consistency and marginal Wald inference for each scalar effect with positive asymptotic variance, while permitting a singular joint effect covariance.

\paragraph{Scope of the conditions and numerical implementation.}
Proposition~\ref{prop:feasible-root-bridge} requires the OLS-started solver to reach the local valid root with probability tending to one, with other valid roots separated in the selection distance. Local uniqueness, nonsingular score derivatives and a consistent OLS start provide the local setting; finite-sample diagnostics record convergence, conditioning and root multiplicity. The acceptance threshold $b_n=\min(10^{-6},n^{-1})$ gives mean-score error $o_p(n^{-1/2})$, and the finite-difference factor $d_n=\min(10^{-4},n^{-1/3})$ vanishes. Together with the derivative and moment conditions, these safeguards support Condition~(\ref{cond:covariance}).

\section{Complete numerical algorithm}

For each mediator or outcome regression, the supplied function \texttt{fit\_semiparametric\_lm()} performs the following deterministic steps.
\begin{enumerate}
\item Fit OLS and compute $(\widetilde\bbeta,\widetilde v)$ using $n^{-1}\sum_i\widetilde e_i^2$.
\item Form coefficient increments $d_j=\max\{0.05,c\max(1,|\widetilde\beta_j|)\}$ and the seven starts described in the multi-start subsection of the main article.
\item At every solver evaluation, recompute residuals, the leave-one-out density and derivative, the third and fourth residual moments, $q(e_i)$, and all observation-level efficient scores.
\item Run \texttt{rootSolve::multiroot()} with relative and absolute tolerances $a_n=\min(10^{-9},n^{-1})$ and at most 200 iterations.
\item Retain a root if it is finite, its maximum absolute mean score is at most $b_n=\min(10^{-6},n^{-1})$, its centered finite-difference Jacobian (coordinate step $d_n\max\{1,|\theta_j|\}$, $d_n=\min(10^{-4},n^{-1/3})$) has reciprocal condition number at least $10^{-10}$, and its sandwich covariance is finite with positive diagonal.
\item Remove duplicate roots after rounding parameter vectors to seven decimal places.
\item Select the valid root with minimum OLS-design distance defined in the main article.  When the candidate set is empty, record a numerical failure so every reported proposed-method estimate remains a solution of its density-score equations.
\item Stack the selected mediator and outcome scores with $\bX_i-\overline{\bX}$, form the empirical sandwich, and apply the analytic gradient in \eqref{eq:full-gradient}.
\end{enumerate}

The log-variance parameterization guarantees positive variance without post-solution truncation.  Score evaluation additionally uses $\kappa_n=\min(10^{-10},n^{-2})$ for the density floor and positive-$Q_2$ check stated in the main article.  At every sample size in the reported analyses, $(a_n,b_n,d_n,\kappa_n)=(10^{-9},10^{-6},10^{-4},10^{-10})$, so the asymptotic sequences reproduce the retained numerical results exactly.  Root acceptance requires score convergence, Jacobian invertibility and a defined covariance, with no empirically chosen coefficient or variance bounds.  These checks establish numerical validity of the returned solution; Proposition~\ref{prop:feasible-root-bridge} provides the corresponding first-order statistical link.

\section{Error distributions and complete simulation results}
\label{sec:simulation-results}

\subsection{Design and comparator specifications}

All figures in the article and this supplement were generated with base \textsf{R} graphics.

\begin{figure}[t]
\centering
\includegraphics[width=0.88\textwidth]{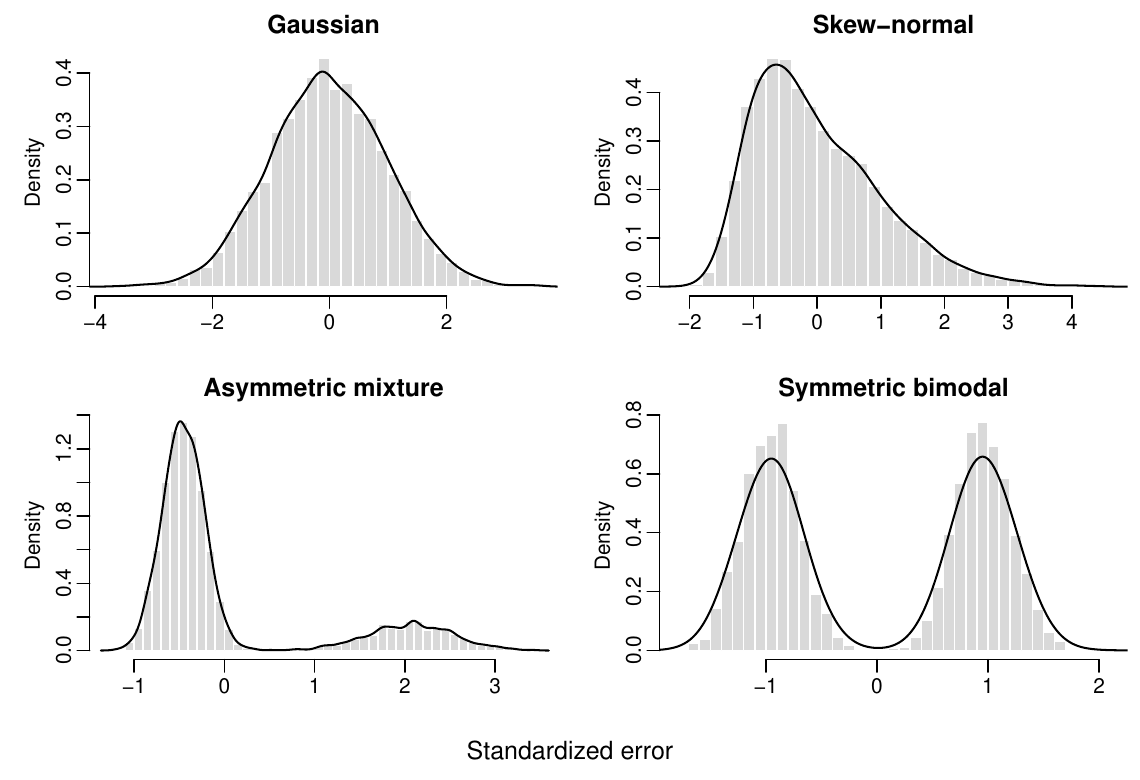}
\caption{Monte Carlo views of the four standardized error distributions.  Each panel uses 5,000 draws; the histogram is normalized to integrate to one and the line is a kernel density estimate}
\label{fig:error-distributions}
\end{figure}

The original simulation random-number seeds are set in \texttt{R/run\_simulations.R}.  The main experiment uses seed 20260827, the comparator experiment 20260828, the size--power experiment 20260829, and the algorithm-sensitivity experiment 20260830.  Each task receives a stored task-specific seed before parallel execution, so the results do not depend on the number of forked workers. The independently calibrated power experiment has separate seeds and retained replication-level output, as detailed in Section~\ref{sec:calibrated-power}.

The comparator experiment uses $n=300$, 1,000 replications per distribution, 1,000 quasi-Bayesian draws and 499 bootstrap resamples, as specified in the main article. The file \texttt{comparator\_seeds.csv} records the data-set seeds; \texttt{R/run\_comparator.R} supports independent execution and resumable checkpoints.

For clarity about the comparator specification, the linked linear model implies
\[
\E\{Y(t,M(t'))\mid\bX=\bm x\}
=\alpha_3+\beta_3t+(\gamma+\eta t)
 (\alpha_2+\beta_2t'+\bm\xi_2\trans\bm x)
+\bm\xi_3\trans\bm x.
\]
Thus the compatible natural effect model includes $t\bm x$ as well as $tt'$.  In both the comparator simulation and the application, \texttt{medflex} uses outcome imputation with $T*M$ and every $TX_j$ term, and a natural effect model with $T_0*T_1$ and every $T_0X_j$ interaction.  The extra $TX_j$ terms ensure that the imputation specification passes the package's working-model compatibility check; their true coefficients are zero in the simulation.  This gives the comparator a slightly more flexible outcome working model while preserving the same causal target.  Continuous and binary baseline covariates are centered at their empirical means.  The package's decomposition is evaluated at zero on that centered scale, which gives the plug-in population average under this linear structure.  Its native standard errors condition on the covariate reference, whereas the stacked estimator adds the baseline-mean influence term.

\subsection{Precision and interval calibration}

Tables~\ref{tab:main_simulation_full} and \ref{tab:comparator_full} extend the main article's representative results to all effects and distributions, including bias, coverage, interval length and numerical success. Section~\ref{sec:paired-audit} gives the common-data comparison.

\clearpage
\begingroup
\footnotesize
\setlength{\tabcolsep}{2.1pt}
\begin{longtable}{rlllrrrrrrr}
\caption{Complete main Monte Carlo results; Reps. is the number of numerically successful fits used for performance summaries, whereas Success uses all attempts; errors: G, Gaussian; SN, skew-normal; AM, asymmetric mixture; BM, symmetric bimodal}\label{tab:main_simulation_full}\\
\toprule
$n$ & Error & Method & Effect & Reps. & Bias & RMSE & Cov. & MCSE & Length & Success \\
\midrule
\endfirsthead
\toprule
$n$ & Error & Method & Effect & Reps. & Bias & RMSE & Cov. & MCSE & Length & Success \\
\midrule
\endhead
200 & G & OLS & PNIE & 1000 & 0.003 & 0.121 & 0.947 & 0.007 & 0.467 & 1.000 \\
200 & G & OLS & TNIE & 1000 & 0.001 & 0.053 & 0.925 & 0.008 & 0.197 & 1.000 \\
200 & G & OLS & PNDE & 1000 & -0.002 & 0.178 & 0.950 & 0.007 & 0.698 & 1.000 \\
200 & G & OLS & TNDE & 1000 & -0.003 & 0.176 & 0.946 & 0.007 & 0.699 & 1.000 \\
200 & G & OLS & TE & 1000 & -0.001 & 0.162 & 0.949 & 0.007 & 0.643 & 1.000 \\
200 & G & Proposed & PNIE & 1000 & 0.004 & 0.130 & 0.932 & 0.008 & 0.495 & 1.000 \\
200 & G & Proposed & TNIE & 1000 & 0.001 & 0.056 & 0.912 & 0.009 & 0.210 & 1.000 \\
200 & G & Proposed & PNDE & 1000 & -0.002 & 0.191 & 0.934 & 0.008 & 0.742 & 1.000 \\
200 & G & Proposed & TNDE & 1000 & -0.005 & 0.189 & 0.941 & 0.007 & 0.740 & 1.000 \\
200 & G & Proposed & TE & 1000 & -0.001 & 0.175 & 0.938 & 0.008 & 0.682 & 1.000 \\
200 & SN & OLS & PNIE & 1000 & 0.005 & 0.122 & 0.945 & 0.007 & 0.469 & 1.000 \\
200 & SN & OLS & TNIE & 1000 & 0.001 & 0.051 & 0.911 & 0.009 & 0.194 & 1.000 \\
200 & SN & OLS & PNDE & 1000 & -0.001 & 0.185 & 0.936 & 0.008 & 0.701 & 1.000 \\
200 & SN & OLS & TNDE & 1000 & -0.005 & 0.181 & 0.951 & 0.007 & 0.701 & 1.000 \\
200 & SN & OLS & TE & 1000 & 0.000 & 0.171 & 0.937 & 0.008 & 0.645 & 1.000 \\
200 & SN & Proposed & PNIE & 1000 & 0.005 & 0.100 & 0.951 & 0.007 & 0.388 & 1.000 \\
200 & SN & Proposed & TNIE & 1000 & 0.002 & 0.043 & 0.918 & 0.009 & 0.163 & 1.000 \\
200 & SN & Proposed & PNDE & 1000 & -0.001 & 0.155 & 0.941 & 0.007 & 0.594 & 1.000 \\
200 & SN & Proposed & TNDE & 1000 & -0.004 & 0.152 & 0.949 & 0.007 & 0.597 & 1.000 \\
200 & SN & Proposed & TE & 1000 & 0.000 & 0.144 & 0.936 & 0.008 & 0.553 & 1.000 \\
200 & AM & OLS & PNIE & 1000 & 0.001 & 0.126 & 0.942 & 0.007 & 0.475 & 1.000 \\
200 & AM & OLS & TNIE & 1000 & 0.000 & 0.050 & 0.895 & 0.010 & 0.190 & 1.000 \\
200 & AM & OLS & PNDE & 1000 & 0.009 & 0.183 & 0.948 & 0.007 & 0.702 & 1.000 \\
200 & AM & OLS & TNDE & 1000 & 0.008 & 0.186 & 0.933 & 0.008 & 0.699 & 1.000 \\
200 & AM & OLS & TE & 1000 & 0.008 & 0.168 & 0.946 & 0.007 & 0.646 & 1.000 \\
200 & AM & Proposed & PNIE & 887 & -0.001 & 0.055 & 0.940 & 0.008 & 0.120 & 0.887 \\
200 & AM & Proposed & TNIE & 887 & 0.000 & 0.021 & 0.939 & 0.008 & 0.050 & 0.887 \\
200 & AM & Proposed & PNDE & 887 & 0.002 & 0.093 & 0.935 & 0.008 & 0.332 & 0.887 \\
200 & AM & Proposed & TNDE & 887 & 0.004 & 0.093 & 0.933 & 0.008 & 0.333 & 0.887 \\
200 & AM & Proposed & TE & 887 & 0.002 & 0.088 & 0.937 & 0.008 & 0.327 & 0.887 \\
200 & BM & OLS & PNIE & 1000 & 0.004 & 0.119 & 0.940 & 0.008 & 0.470 & 1.000 \\
200 & BM & OLS & TNIE & 1000 & -0.002 & 0.050 & 0.902 & 0.009 & 0.193 & 1.000 \\
200 & BM & OLS & PNDE & 1000 & 0.005 & 0.169 & 0.953 & 0.007 & 0.700 & 1.000 \\
200 & BM & OLS & TNDE & 1000 & -0.002 & 0.177 & 0.944 & 0.007 & 0.698 & 1.000 \\
200 & BM & OLS & TE & 1000 & 0.002 & 0.157 & 0.957 & 0.006 & 0.646 & 1.000 \\
200 & BM & Proposed & PNIE & 921 & 0.000 & 0.034 & 0.929 & 0.008 & 0.121 & 0.921 \\
200 & BM & Proposed & TNIE & 921 & 0.000 & 0.031 & 0.928 & 0.008 & 0.050 & 0.921 \\
200 & BM & Proposed & PNDE & 921 & 0.002 & 0.120 & 0.948 & 0.007 & 0.335 & 0.921 \\
200 & BM & Proposed & TNDE & 921 & 0.002 & 0.114 & 0.934 & 0.008 & 0.334 & 0.921 \\
200 & BM & Proposed & TE & 921 & 0.002 & 0.110 & 0.950 & 0.007 & 0.327 & 0.921 \\
\pagebreak
500 & G & OLS & PNIE & 1000 & 0.001 & 0.074 & 0.956 & 0.006 & 0.297 & 1.000 \\
500 & G & OLS & TNIE & 1000 & 0.000 & 0.031 & 0.941 & 0.007 & 0.122 & 1.000 \\
500 & G & OLS & PNDE & 1000 & -0.006 & 0.114 & 0.958 & 0.006 & 0.444 & 1.000 \\
500 & G & OLS & TNDE & 1000 & -0.007 & 0.117 & 0.931 & 0.008 & 0.443 & 1.000 \\
500 & G & OLS & TE & 1000 & -0.006 & 0.107 & 0.946 & 0.007 & 0.409 & 1.000 \\
500 & G & Proposed & PNIE & 1000 & 0.001 & 0.078 & 0.958 & 0.006 & 0.311 & 1.000 \\
500 & G & Proposed & TNIE & 1000 & 0.001 & 0.033 & 0.930 & 0.008 & 0.127 & 1.000 \\
500 & G & Proposed & PNDE & 1000 & -0.006 & 0.119 & 0.953 & 0.007 & 0.462 & 1.000 \\
500 & G & Proposed & TNDE & 1000 & -0.006 & 0.122 & 0.934 & 0.008 & 0.461 & 1.000 \\
500 & G & Proposed & TE & 1000 & -0.005 & 0.111 & 0.950 & 0.007 & 0.424 & 1.000 \\
500 & SN & OLS & PNIE & 1000 & 0.000 & 0.077 & 0.955 & 0.007 & 0.298 & 1.000 \\
500 & SN & OLS & TNIE & 1000 & 0.001 & 0.031 & 0.930 & 0.008 & 0.122 & 1.000 \\
500 & SN & OLS & PNDE & 1000 & -0.003 & 0.116 & 0.941 & 0.007 & 0.444 & 1.000 \\
500 & SN & OLS & TNDE & 1000 & -0.002 & 0.118 & 0.942 & 0.007 & 0.444 & 1.000 \\
500 & SN & OLS & TE & 1000 & -0.001 & 0.109 & 0.948 & 0.007 & 0.409 & 1.000 \\
500 & SN & Proposed & PNIE & 1000 & 0.004 & 0.059 & 0.942 & 0.007 & 0.233 & 1.000 \\
500 & SN & Proposed & TNIE & 1000 & 0.000 & 0.024 & 0.938 & 0.008 & 0.095 & 1.000 \\
500 & SN & Proposed & PNDE & 1000 & -0.001 & 0.093 & 0.943 & 0.007 & 0.361 & 1.000 \\
500 & SN & Proposed & TNDE & 1000 & -0.005 & 0.094 & 0.944 & 0.007 & 0.361 & 1.000 \\
500 & SN & Proposed & TE & 1000 & -0.001 & 0.087 & 0.944 & 0.007 & 0.337 & 1.000 \\
500 & AM & OLS & PNIE & 1000 & -0.004 & 0.075 & 0.946 & 0.007 & 0.299 & 1.000 \\
500 & AM & OLS & TNIE & 1000 & 0.000 & 0.031 & 0.928 & 0.008 & 0.120 & 1.000 \\
500 & AM & OLS & PNDE & 1000 & 0.007 & 0.116 & 0.940 & 0.008 & 0.444 & 1.000 \\
500 & AM & OLS & TNDE & 1000 & 0.012 & 0.111 & 0.956 & 0.006 & 0.443 & 1.000 \\
500 & AM & OLS & TE & 1000 & 0.007 & 0.106 & 0.947 & 0.007 & 0.409 & 1.000 \\
500 & AM & Proposed & PNIE & 999 & 0.000 & 0.019 & 0.949 & 0.007 & 0.074 & 0.999 \\
500 & AM & Proposed & TNIE & 999 & 0.000 & 0.008 & 0.943 & 0.007 & 0.030 & 0.999 \\
500 & AM & Proposed & PNDE & 999 & 0.002 & 0.055 & 0.948 & 0.007 & 0.209 & 0.999 \\
500 & AM & Proposed & TNDE & 999 & 0.003 & 0.055 & 0.948 & 0.007 & 0.209 & 0.999 \\
500 & AM & Proposed & TE & 999 & 0.002 & 0.054 & 0.948 & 0.007 & 0.205 & 0.999 \\
500 & BM & OLS & PNIE & 1000 & -0.003 & 0.076 & 0.945 & 0.007 & 0.298 & 1.000 \\
500 & BM & OLS & TNIE & 1000 & 0.000 & 0.032 & 0.934 & 0.008 & 0.122 & 1.000 \\
500 & BM & OLS & PNDE & 1000 & -0.003 & 0.110 & 0.953 & 0.007 & 0.443 & 1.000 \\
500 & BM & OLS & TNDE & 1000 & 0.000 & 0.109 & 0.963 & 0.006 & 0.444 & 1.000 \\
500 & BM & OLS & TE & 1000 & -0.003 & 0.101 & 0.955 & 0.007 & 0.409 & 1.000 \\
500 & BM & Proposed & PNIE & 1000 & 0.000 & 0.019 & 0.959 & 0.006 & 0.076 & 1.000 \\
500 & BM & Proposed & TNIE & 1000 & 0.000 & 0.008 & 0.939 & 0.008 & 0.031 & 1.000 \\
500 & BM & Proposed & PNDE & 1000 & -0.001 & 0.054 & 0.937 & 0.008 & 0.210 & 1.000 \\
500 & BM & Proposed & TNDE & 1000 & -0.001 & 0.053 & 0.955 & 0.007 & 0.211 & 1.000 \\
500 & BM & Proposed & TE & 1000 & -0.001 & 0.053 & 0.941 & 0.007 & 0.206 & 1.000 \\
\bottomrule
\end{longtable}
\endgroup

\clearpage
\begingroup
\footnotesize
\setlength{\tabcolsep}{2.1pt}
\begin{longtable}{rlllrrrrrrr}
\caption{Complete comparator Monte Carlo results based on 1,000 attempts per cell; Reps. is the number of numerically successful fits used for performance summaries, whereas Success uses all attempts; QB denotes quasi-Bayes and Boot denotes bootstrap; other error abbreviations are defined in Table S1}\label{tab:comparator_full}\\
\toprule
$n$ & Error & Method & Effect & Reps. & Bias & RMSE & Cov. & MCSE & Length & Success \\
\midrule
\endfirsthead
\toprule
$n$ & Error & Method & Effect & Reps. & Bias & RMSE & Cov. & MCSE & Length & Success \\
\midrule
\endhead
300 & G & OLS & PNIE & 1000 & 0.001 & 0.100 & 0.952 & 0.007 & 0.383 & 1.000 \\
300 & G & OLS & TNIE & 1000 & -0.001 & 0.040 & 0.927 & 0.008 & 0.158 & 1.000 \\
300 & G & OLS & PNDE & 1000 & -0.001 & 0.146 & 0.956 & 0.006 & 0.571 & 1.000 \\
300 & G & OLS & TNDE & 1000 & -0.003 & 0.145 & 0.944 & 0.007 & 0.573 & 1.000 \\
300 & G & OLS & TE & 1000 & -0.002 & 0.135 & 0.955 & 0.007 & 0.527 & 1.000 \\
300 & G & Imai-QB & PNIE & 1000 & 0.001 & 0.100 & 0.950 & 0.007 & 0.387 & 1.000 \\
300 & G & Imai-QB & TNIE & 1000 & -0.001 & 0.040 & 0.944 & 0.007 & 0.162 & 1.000 \\
300 & G & Imai-QB & PNDE & 1000 & -0.001 & 0.146 & 0.964 & 0.006 & 0.613 & 1.000 \\
300 & G & Imai-QB & TNDE & 1000 & -0.003 & 0.145 & 0.958 & 0.006 & 0.615 & 1.000 \\
300 & G & Imai-QB & TE & 1000 & -0.002 & 0.135 & 0.962 & 0.006 & 0.573 & 1.000 \\
300 & G & Imai-Boot & PNIE & 1000 & 0.001 & 0.100 & 0.946 & 0.007 & 0.384 & 1.000 \\
300 & G & Imai-Boot & TNIE & 1000 & -0.001 & 0.040 & 0.940 & 0.008 & 0.161 & 1.000 \\
300 & G & Imai-Boot & PNDE & 1000 & -0.003 & 0.157 & 0.963 & 0.006 & 0.608 & 1.000 \\
300 & G & Imai-Boot & TNDE & 1000 & -0.005 & 0.158 & 0.958 & 0.006 & 0.609 & 1.000 \\
300 & G & Imai-Boot & TE & 1000 & -0.004 & 0.147 & 0.966 & 0.006 & 0.565 & 1.000 \\
300 & G & NEM & PNIE & 1000 & 0.001 & 0.100 & 0.950 & 0.007 & 0.385 & 1.000 \\
300 & G & NEM & TNIE & 1000 & -0.001 & 0.040 & 0.922 & 0.008 & 0.160 & 1.000 \\
300 & G & NEM & PNDE & 1000 & -0.001 & 0.146 & 0.951 & 0.007 & 0.567 & 1.000 \\
300 & G & NEM & TNDE & 1000 & -0.003 & 0.146 & 0.941 & 0.007 & 0.569 & 1.000 \\
300 & G & NEM & TE & 1000 & -0.002 & 0.135 & 0.954 & 0.007 & 0.522 & 1.000 \\
300 & G & Proposed & PNIE & 1000 & 0.000 & 0.107 & 0.942 & 0.007 & 0.406 & 1.000 \\
300 & G & Proposed & TNIE & 1000 & -0.001 & 0.042 & 0.922 & 0.008 & 0.166 & 1.000 \\
300 & G & Proposed & PNDE & 1000 & -0.001 & 0.151 & 0.945 & 0.007 & 0.599 & 1.000 \\
300 & G & Proposed & TNDE & 1000 & -0.002 & 0.154 & 0.942 & 0.007 & 0.602 & 1.000 \\
300 & G & Proposed & TE & 1000 & -0.002 & 0.139 & 0.948 & 0.007 & 0.552 & 1.000 \\
300 & SN & OLS & PNIE & 1000 & 0.004 & 0.099 & 0.950 & 0.007 & 0.385 & 1.000 \\
300 & SN & OLS & TNIE & 1000 & -0.001 & 0.042 & 0.904 & 0.009 & 0.156 & 1.000 \\
300 & SN & OLS & PNDE & 1000 & -0.007 & 0.143 & 0.954 & 0.007 & 0.572 & 1.000 \\
300 & SN & OLS & TNDE & 1000 & -0.012 & 0.141 & 0.951 & 0.007 & 0.572 & 1.000 \\
300 & SN & OLS & TE & 1000 & -0.008 & 0.133 & 0.952 & 0.007 & 0.527 & 1.000 \\
300 & SN & Imai-QB & PNIE & 1000 & 0.004 & 0.100 & 0.949 & 0.007 & 0.388 & 1.000 \\
300 & SN & Imai-QB & TNIE & 1000 & -0.001 & 0.042 & 0.933 & 0.008 & 0.162 & 1.000 \\
300 & SN & Imai-QB & PNDE & 1000 & -0.007 & 0.143 & 0.966 & 0.006 & 0.613 & 1.000 \\
300 & SN & Imai-QB & TNDE & 1000 & -0.012 & 0.142 & 0.971 & 0.005 & 0.614 & 1.000 \\
300 & SN & Imai-QB & TE & 1000 & -0.008 & 0.133 & 0.970 & 0.005 & 0.572 & 1.000 \\
300 & SN & Imai-Boot & PNIE & 1000 & 0.004 & 0.099 & 0.939 & 0.008 & 0.386 & 1.000 \\
300 & SN & Imai-Boot & TNIE & 1000 & -0.001 & 0.042 & 0.922 & 0.008 & 0.159 & 1.000 \\
300 & SN & Imai-Boot & PNDE & 1000 & -0.006 & 0.156 & 0.968 & 0.006 & 0.609 & 1.000 \\
300 & SN & Imai-Boot & TNDE & 1000 & -0.011 & 0.153 & 0.965 & 0.006 & 0.609 & 1.000 \\
300 & SN & Imai-Boot & TE & 1000 & -0.007 & 0.146 & 0.968 & 0.006 & 0.566 & 1.000 \\
300 & SN & NEM & PNIE & 1000 & 0.004 & 0.099 & 0.950 & 0.007 & 0.386 & 1.000 \\
300 & SN & NEM & TNIE & 1000 & -0.001 & 0.043 & 0.901 & 0.009 & 0.158 & 1.000 \\
300 & SN & NEM & PNDE & 1000 & -0.007 & 0.143 & 0.948 & 0.007 & 0.568 & 1.000 \\
300 & SN & NEM & TNDE & 1000 & -0.012 & 0.141 & 0.954 & 0.007 & 0.568 & 1.000 \\
300 & SN & NEM & TE & 1000 & -0.008 & 0.133 & 0.950 & 0.007 & 0.522 & 1.000 \\
300 & SN & Proposed & PNIE & 1000 & 0.002 & 0.079 & 0.955 & 0.007 & 0.310 & 1.000 \\
300 & SN & Proposed & TNIE & 1000 & 0.001 & 0.033 & 0.924 & 0.008 & 0.127 & 1.000 \\
300 & SN & Proposed & PNDE & 1000 & -0.006 & 0.121 & 0.949 & 0.007 & 0.476 & 1.000 \\
300 & SN & Proposed & TNDE & 1000 & -0.007 & 0.117 & 0.959 & 0.006 & 0.477 & 1.000 \\
300 & SN & Proposed & TE & 1000 & -0.006 & 0.112 & 0.947 & 0.007 & 0.443 & 1.000 \\
\pagebreak
300 & AM & OLS & PNIE & 1000 & 0.000 & 0.101 & 0.944 & 0.007 & 0.387 & 1.000 \\
300 & AM & OLS & TNIE & 1000 & 0.000 & 0.041 & 0.886 & 0.010 & 0.153 & 1.000 \\
300 & AM & OLS & PNDE & 1000 & -0.001 & 0.152 & 0.939 & 0.008 & 0.573 & 1.000 \\
300 & AM & OLS & TNDE & 1000 & -0.001 & 0.148 & 0.945 & 0.007 & 0.572 & 1.000 \\
300 & AM & OLS & TE & 1000 & -0.001 & 0.141 & 0.935 & 0.008 & 0.527 & 1.000 \\
300 & AM & Imai-QB & PNIE & 1000 & 0.000 & 0.102 & 0.944 & 0.007 & 0.389 & 1.000 \\
300 & AM & Imai-QB & TNIE & 1000 & 0.000 & 0.041 & 0.947 & 0.007 & 0.162 & 1.000 \\
300 & AM & Imai-QB & PNDE & 1000 & -0.001 & 0.152 & 0.958 & 0.006 & 0.615 & 1.000 \\
300 & AM & Imai-QB & TNDE & 1000 & -0.002 & 0.148 & 0.960 & 0.006 & 0.616 & 1.000 \\
300 & AM & Imai-QB & TE & 1000 & -0.001 & 0.141 & 0.954 & 0.007 & 0.573 & 1.000 \\
300 & AM & Imai-Boot & PNIE & 1000 & 0.000 & 0.101 & 0.945 & 0.007 & 0.390 & 1.000 \\
300 & AM & Imai-Boot & TNIE & 1000 & 0.000 & 0.041 & 0.920 & 0.009 & 0.154 & 1.000 \\
300 & AM & Imai-Boot & PNDE & 1000 & 0.000 & 0.164 & 0.951 & 0.007 & 0.610 & 1.000 \\
300 & AM & Imai-Boot & TNDE & 1000 & 0.000 & 0.159 & 0.964 & 0.006 & 0.611 & 1.000 \\
300 & AM & Imai-Boot & TE & 1000 & 0.000 & 0.153 & 0.953 & 0.007 & 0.567 & 1.000 \\
300 & AM & NEM & PNIE & 1000 & 0.001 & 0.102 & 0.948 & 0.007 & 0.388 & 1.000 \\
300 & AM & NEM & TNIE & 1000 & 0.000 & 0.042 & 0.890 & 0.010 & 0.155 & 1.000 \\
300 & AM & NEM & PNDE & 1000 & -0.001 & 0.152 & 0.937 & 0.008 & 0.568 & 1.000 \\
300 & AM & NEM & TNDE & 1000 & -0.002 & 0.149 & 0.941 & 0.007 & 0.567 & 1.000 \\
300 & AM & NEM & TE & 1000 & -0.001 & 0.140 & 0.935 & 0.008 & 0.522 & 1.000 \\
300 & AM & Proposed & PNIE & 962 & 0.001 & 0.026 & 0.938 & 0.008 & 0.096 & 0.962 \\
300 & AM & Proposed & TNIE & 962 & 0.000 & 0.012 & 0.949 & 0.007 & 0.039 & 0.962 \\
300 & AM & Proposed & PNDE & 962 & 0.000 & 0.070 & 0.945 & 0.007 & 0.270 & 0.962 \\
300 & AM & Proposed & TNDE & 962 & -0.001 & 0.071 & 0.948 & 0.007 & 0.270 & 0.962 \\
300 & AM & Proposed & TE & 962 & 0.000 & 0.069 & 0.944 & 0.007 & 0.264 & 0.962 \\
300 & BM & OLS & PNIE & 1000 & 0.002 & 0.096 & 0.956 & 0.006 & 0.385 & 1.000 \\
300 & BM & OLS & TNIE & 1000 & -0.002 & 0.038 & 0.928 & 0.008 & 0.157 & 1.000 \\
300 & BM & OLS & PNDE & 1000 & 0.004 & 0.144 & 0.948 & 0.007 & 0.571 & 1.000 \\
300 & BM & OLS & TNDE & 1000 & 0.001 & 0.138 & 0.952 & 0.007 & 0.572 & 1.000 \\
300 & BM & OLS & TE & 1000 & 0.002 & 0.133 & 0.951 & 0.007 & 0.527 & 1.000 \\
300 & BM & Imai-QB & PNIE & 1000 & 0.002 & 0.096 & 0.956 & 0.006 & 0.388 & 1.000 \\
300 & BM & Imai-QB & TNIE & 1000 & -0.002 & 0.038 & 0.950 & 0.007 & 0.161 & 1.000 \\
300 & BM & Imai-QB & PNDE & 1000 & 0.004 & 0.144 & 0.958 & 0.006 & 0.615 & 1.000 \\
300 & BM & Imai-QB & TNDE & 1000 & 0.000 & 0.138 & 0.965 & 0.006 & 0.616 & 1.000 \\
300 & BM & Imai-QB & TE & 1000 & 0.002 & 0.134 & 0.960 & 0.006 & 0.574 & 1.000 \\
300 & BM & Imai-Boot & PNIE & 1000 & 0.002 & 0.096 & 0.957 & 0.006 & 0.386 & 1.000 \\
300 & BM & Imai-Boot & TNIE & 1000 & -0.002 & 0.038 & 0.949 & 0.007 & 0.160 & 1.000 \\
300 & BM & Imai-Boot & PNDE & 1000 & 0.003 & 0.158 & 0.961 & 0.006 & 0.608 & 1.000 \\
300 & BM & Imai-Boot & TNDE & 1000 & -0.001 & 0.153 & 0.967 & 0.006 & 0.608 & 1.000 \\
300 & BM & Imai-Boot & TE & 1000 & 0.001 & 0.148 & 0.961 & 0.006 & 0.566 & 1.000 \\
300 & BM & NEM & PNIE & 1000 & 0.002 & 0.096 & 0.955 & 0.007 & 0.386 & 1.000 \\
300 & BM & NEM & TNIE & 1000 & -0.002 & 0.038 & 0.929 & 0.008 & 0.159 & 1.000 \\
300 & BM & NEM & PNDE & 1000 & 0.004 & 0.144 & 0.948 & 0.007 & 0.567 & 1.000 \\
300 & BM & NEM & TNDE & 1000 & 0.000 & 0.138 & 0.951 & 0.007 & 0.568 & 1.000 \\
300 & BM & NEM & TE & 1000 & 0.002 & 0.134 & 0.948 & 0.007 & 0.522 & 1.000 \\
300 & BM & Proposed & PNIE & 978 & -0.001 & 0.029 & 0.949 & 0.007 & 0.098 & 0.978 \\
300 & BM & Proposed & TNIE & 978 & 0.000 & 0.020 & 0.943 & 0.007 & 0.040 & 0.978 \\
300 & BM & Proposed & PNDE & 978 & -0.001 & 0.080 & 0.943 & 0.007 & 0.272 & 0.978 \\
300 & BM & Proposed & TNDE & 978 & 0.001 & 0.080 & 0.943 & 0.007 & 0.272 & 0.978 \\
300 & BM & Proposed & TE & 978 & -0.001 & 0.079 & 0.941 & 0.008 & 0.266 & 0.978 \\
\bottomrule
\end{longtable}
\endgroup

\clearpage

\subsection{Paired-fit and reporting reliability}
\label{sec:paired-audit}

The script \texttt{R/paired\_fit\_audit.R} joins OLS and proposed records by study, sample size, distribution, replication and effect, then recomputes RMSE and interval-length ratios on common successful fits. All OLS fits succeeded, so the paired count equals the proposed-method successful-fit count.

For each method we calculate conditional coverage among the paired fits.  For the proposal we additionally report
\[
\frac{\#\{\text{attempts producing a valid interval covering the truth}\}}
     {\#\{\text{all attempted data sets}\}}.
\]
This report-and-cover fraction equals success rate times conditional coverage.  It measures the reliability of the complete numerical procedure and complements nominal coverage conditional on a returned interval.

The full CSV output contains all 60 study-by-design-by-effect rows.  For $R=\sqrt{\overline{u}/\overline{v}}$, where $u_i$ and $v_i$ are the paired squared estimation errors of the proposed and OLS methods, the reported approximate Monte Carlo standard error is
\[
\widehat{\operatorname{MCSE}}(R)
=\frac{R}{\sqrt N}\operatorname{sd}
  \left\{\frac12\left(\frac{u_i}{\overline{u}}
                     -\frac{v_i}{\overline{v}}\right)\right\}.
\]
This paired delta-method MCSE is conditional on retained fits; all-attempted summaries separately represent fit availability.

\begin{table}[htbp]\centering\footnotesize\setlength{\tabcolsep}{4pt}
\caption{Paired-fit audit using identical data sets for OLS and the proposal. Main and comparator designs each have 1,000 attempts per cell. $N$ is the number of paired valid fits. Ratios use proposed over OLS; $C_O$ and $C_P$ are conditional coverages. $R\&C$ is the fraction of all attempts yielding a proposed interval that covers the truth, not conventional conditional coverage. G, SN, AM and BM denote the four error distributions. All five effects and ratio Monte Carlo standard errors are in the accompanying CSV file.}
\label{tab:paired-audit}
\begin{tabular}{lrllrrrrrr}\toprule
Study & $n$ & Error & Effect & $N$ & RMSE ratio & Length ratio & $C_O$ & $C_P$ & $R\&C$ \\
\midrule
Main & 200 & G & PNIE & 1000 & 1.069 & 1.060 & 0.947 & 0.932 & 0.932 \\
Main & 200 & G & TE & 1000 & 1.078 & 1.061 & 0.949 & 0.938 & 0.938 \\
\midrule
Main & 200 & SN & PNIE & 1000 & 0.820 & 0.828 & 0.945 & 0.951 & 0.951 \\
Main & 200 & SN & TE & 1000 & 0.840 & 0.857 & 0.937 & 0.936 & 0.936 \\
\midrule
Main & 200 & AM & PNIE & 887 & 0.444 & 0.251 & 0.949 & 0.940 & 0.834 \\
Main & 200 & AM & TE & 887 & 0.550 & 0.504 & 0.966 & 0.937 & 0.831 \\
\midrule
Main & 200 & BM & PNIE & 921 & 0.289 & 0.258 & 0.942 & 0.929 & 0.856 \\
Main & 200 & BM & TE & 921 & 0.731 & 0.506 & 0.966 & 0.950 & 0.875 \\
\midrule
Main & 500 & G & PNIE & 1000 & 1.044 & 1.048 & 0.956 & 0.958 & 0.958 \\
Main & 500 & G & TE & 1000 & 1.035 & 1.037 & 0.946 & 0.950 & 0.950 \\
\midrule
Main & 500 & SN & PNIE & 1000 & 0.761 & 0.783 & 0.955 & 0.942 & 0.942 \\
Main & 500 & SN & TE & 1000 & 0.802 & 0.822 & 0.948 & 0.944 & 0.944 \\
\midrule
Main & 500 & AM & PNIE & 999 & 0.255 & 0.247 & 0.947 & 0.949 & 0.948 \\
Main & 500 & AM & TE & 999 & 0.507 & 0.500 & 0.947 & 0.948 & 0.947 \\
\midrule
Main & 500 & BM & PNIE & 1000 & 0.247 & 0.255 & 0.945 & 0.959 & 0.959 \\
Main & 500 & BM & TE & 1000 & 0.519 & 0.503 & 0.955 & 0.941 & 0.941 \\
\midrule
Comparator & 300 & G & PNIE & 1000 & 1.071 & 1.058 & 0.952 & 0.942 & 0.942 \\
Comparator & 300 & G & TE & 1000 & 1.032 & 1.048 & 0.955 & 0.948 & 0.948 \\
\midrule
Comparator & 300 & SN & PNIE & 1000 & 0.797 & 0.807 & 0.950 & 0.955 & 0.955 \\
Comparator & 300 & SN & TE & 1000 & 0.846 & 0.840 & 0.952 & 0.947 & 0.947 \\
\midrule
Comparator & 300 & AM & PNIE & 962 & 0.261 & 0.249 & 0.948 & 0.938 & 0.902 \\
Comparator & 300 & AM & TE & 962 & 0.506 & 0.501 & 0.943 & 0.944 & 0.908 \\
\midrule
Comparator & 300 & BM & PNIE & 978 & 0.303 & 0.256 & 0.955 & 0.949 & 0.928 \\
Comparator & 300 & BM & TE & 978 & 0.598 & 0.505 & 0.954 & 0.941 & 0.920 \\
\bottomrule\end{tabular}\end{table}

\FloatBarrier

\clearpage
\section{Null size, power and numerical sensitivity}
\label{sec:calibrated-power}

\subsection{Original nominal comparison}

Table~\ref{tab:size_power} gives the original nominal comparison at $n=300$ with 1,000 attempts per effect value. The independently calibrated comparison below separates detection gains from differences in attained null size.

\begin{table}[htbp]
\centering
\caption{Empirical size and power for PNIE under the standardized asymmetric-mixture error, $n=300$, based on 1,000 attempted replications. Rejection and MCSE are conditional on numerically successful fits; Success uses all attempts}
\label{tab:size_power}
\begin{tabular}{rrlrrrr}
\toprule
$\beta_2$ & True PNIE & Method & Reps. & Success & Reject & MCSE \\
\midrule
0.00 & 0.000 & OLS delta & 1000 & 1.000 & 0.016 & 0.004 \\
0.00 & 0.000 & Proposed & 1000 & 0.964 & 0.056 & 0.007 \\
0.10 & -0.026 & OLS delta & 1000 & 1.000 & 0.052 & 0.007 \\
0.10 & -0.026 & Proposed & 1000 & 0.968 & 0.946 & 0.007 \\
0.20 & -0.052 & OLS delta & 1000 & 1.000 & 0.195 & 0.013 \\
0.20 & -0.052 & Proposed & 1000 & 0.962 & 1.000 & 0.000 \\
0.30 & -0.078 & OLS delta & 1000 & 1.000 & 0.431 & 0.016 \\
0.30 & -0.078 & Proposed & 1000 & 0.958 & 1.000 & 0.000 \\
0.40 & -0.104 & OLS delta & 1000 & 1.000 & 0.653 & 0.015 \\
0.40 & -0.104 & Proposed & 1000 & 0.960 & 1.000 & 0.000 \\
\bottomrule
\end{tabular}
\end{table}

\FloatBarrier

\subsection{Independent null calibration and Huber comparison}

The asymmetric-mixture design uses $n=300$, $\gamma=-0.26$, $\eta=0.8$ and $\beta_2\in\{0,0.1,0.2,0.3,0.4\}$, giving PNIE $=-0.26\beta_2$. All other parameters and the OLS and proposed estimators are unchanged. The design was recorded in \path{POWER_CALIBRATION_PLAN_20260923.md}; all four methods analyze common data within each replication.

\subsubsection*{Huber target and implementation}

The comparators are motivated by \citet{wang2025}, whose data-driven Huber tuning uses a preliminary least-absolute-deviation (LAD) regression. We fit the same mediator and outcome formulas as the main analysis, $M\sim T+X$ and $Y\sim T*M+X$, using \texttt{quantreg::rq} with quantile $0.5$ and method \texttt{br} for the pilot. For each equation, let $r_i^{\rm LAD}$ be the pilot residual and $s=\operatorname{median}(|r_i^{\rm LAD}|)/0.6745$. Huber-FIX sets the raw residual threshold to $k=1.345s$. Huber-SEL minimizes
\[
\widehat\tau(k)=
\frac{n^{-1}\sum_i\min\{(r_i^{\rm LAD})^2,k^2\}}
     {\{n^{-1}\sum_i\mathbf1(|r_i^{\rm LAD}|\le k)\}^2}
\]
over $k=0.20,0.21,\ldots,\operatorname{round}(3s,2)$, following the grid in the authors' public implementation. A zero denominator is assigned infinite objective; an empty grid is a recorded failure. For the chosen threshold, iteratively reweighted least squares minimizes the convex Huber loss, starting from the LAD fit. It stops when the largest coefficient change is below $10^{-10}(1+\|\widehat b\|_\infty)$, with a limit of 500 iterations and an additional absolute mean-score tolerance of $10^{-7}$.

Both comparators target the same PNIE despite the asymmetric errors. To see this, write either regression as $R=\alpha+Z\trans b+\epsilon$, where $Z$ contains all nonconstant regressors, and put $\psi_k(u)=\max\{-k,\min(k,u)\}$. Independence of $\epsilon$ and $Z$ implies that the population Huber equations are solved by slope $b$ and intercept $\alpha+d_k$, where $\E\{\psi_k(\epsilon-d_k)\}=0$. Thus the intercept absorbs a constant location shift and the slopes retain their structural meanings. Consequently, the product of the mediator's treatment slope and the outcome's mediator slope targets $\beta_2\gamma$. The experiment uses this slope-only target, rather than interpreting Huber intercepts as conditional-mean intercepts for direct or total effects.

For the covariance calculation define $C=\Var(Z)$ and $B_k=\E\{\mathbf1(|\epsilon-d_k|<k)\}$. Under the common independent-error law and usual nonsingularity and smoothness conditions, the slope influence function is
\[
\operatorname{IF}_b
=C^{-1}(Z-\E Z)\frac{\psi_k(\epsilon-d_k)}{B_k}.
\]
The centered-design slope equations have zero population derivative with respect to the intercept and the threshold: the relevant derivatives factor into $\E(Z-\E Z)=0$ and an error-only expectation. This orthogonality explains why estimating a convergent pilot-based threshold does not contribute a first-order slope term under this model. We evaluate the empirical influence function at the \emph{final Huber residuals}, using the centered empirical design covariance and the fitted inlier fraction. After centering the observation-level influences, PNIE has estimated influence
\[
\widehat{\operatorname{IF}}_{\rm PNIE,i}
=\widehat\gamma_H\widehat{\operatorname{IF}}_{\beta_2,i}
+\widehat\beta_{2,H}\widehat{\operatorname{IF}}_{\gamma,i},
\qquad
\widehat{\operatorname{SE}}_H
=\frac{\{\sum_i\widehat{\operatorname{IF}}_{\rm PNIE,i}^{\,2}\}^{1/2}}{n}.
\]
This includes the empirical cross-regression covariance. It is an interaction-adapted comparison, rather than a verbatim reproduction of the simple-mediation Sobel implementation in \citet{wang2025}. Evaluating the score variance at the fitted Huber location also distinguishes this calculation from a pilot-residual variance approximation when errors are asymmetric.

\subsubsection*{Independent calibration and evaluation}

For method $j$, let $Z_{jr}=|\widehat{\operatorname{PNIE}}_{jr}/\widehat{\operatorname{SE}}_{jr}|$. We first generate 5,000 independent null data sets at $\beta_2=0$. Among the $m_j$ numerically valid null fits, order the statistics and select
\[
c_j=Z_{j,(\lceil0.95(m_j+1)\rceil)}.
\]
We then freeze these four critical values and evaluate $\mathbf1(Z_{jr}>c_j)$ on 1,000 \emph{new} data sets at each of the five effect values. The held-out null is disjoint from calibration and provides the reported type~I error check. The nominal comparison on these same evaluation data uses $z_{0.975}$ for all methods. Master seed 20260923 generates 10,000 distinct task seeds, retained with every fit status and numerical result. The null-calibration target is 5\%; its held-out estimate need not be exactly 0.05 because both stages have Monte Carlo variation. These DGP-specific critical values diagnose whether the original power difference can be attributed to unequal null size. They are not proposed as data-analysis critical values for an unknown error law or as uniform calibration over the composite mediation null. In particular, $\gamma$ is fixed away from zero here, so the joint-zero nonregular case is outside this experiment.

Conditional rejection uses each method's valid-fit denominator. Report-and-reject divides the number of usable rejections by all 1,000 attempts. We additionally compare methods on the common subset where all four fits succeed. These matched comparisons retain the method-specific critical values obtained from calibration. For simulation uncertainty, 1,000 bootstrap replications (seed 20260924) independently resample calibration tasks and evaluation tasks, retaining method pairing within each resample, and recompute the thresholds and rejection rates. The same resampled calibration thresholds are used across the five effect settings within a bootstrap replication. The reported two-stage MCSE is the standard deviation across these replications, and intervals are pointwise percentile intervals for Monte Carlo uncertainty. Binomial MCSEs conditional on the realized thresholds are also retained in the project repository. At cells with no observed failures to reject or no observed paired disagreement, empirical MCSEs and bootstrap intervals can be zero-width; these describe the retained simulation outcomes rather than asserting exact unit power or identical population rejection probabilities.

Tables~\ref{tab:power-critical}--\ref{tab:power-paired} give the calibration cutoffs, held-out rejection rates and common-data contrasts, respectively. They retain both conditional power and all-attempt detection for the weak- and larger-effect comparisons interpreted in the main article.

\begin{table}[htbp]
\centering\small
\caption{Independent null-calibration results from 5,000 attempted data sets per method. The interval for the critical value reflects 1,000 resamples of the calibration tasks.}
\label{tab:power-critical}
\begin{tabular}{lrrr}
\toprule
Method & Valid & Critical value & Monte Carlo 95\% interval \\ \midrule
OLS & 5000 & 1.669 & [1.636, 1.712] \\
Huber-FIX & 5000 & 1.938 & [1.885, 1.982] \\
Huber-SEL & 5000 & 1.988 & [1.929, 2.045] \\
Proposed & 4841 & 1.967 & [1.926, 2.026] \\
\bottomrule\end{tabular}
\end{table}

\clearpage
\begin{table}[htbp]
\centering\small
\caption{Held-out PNIE rejection probabilities. Each effect uses 1,000 new data sets. Nominal and calibrated rates condition on valid fits; Report/reject uses all attempts. MCSE includes both calibration and evaluation uncertainty. At $\beta_2=0$ the rates estimate type~I error, and elsewhere they estimate power.}
\label{tab:power-calibrated}
\begin{tabular}{rlrrrrr}
\toprule
$\beta_2$ & Method & Valid & Nominal & Calibrated & MCSE & Report/reject \\ \midrule
0.0 & OLS & 1000 & 0.018 & 0.054 & 0.008 & 0.054 \\
0.0 & Huber-FIX & 1000 & 0.048 & 0.057 & 0.010 & 0.057 \\
0.0 & Huber-SEL & 1000 & 0.055 & 0.052 & 0.008 & 0.052 \\
0.0 & Proposed & 965 & 0.055 & 0.052 & 0.008 & 0.050 \\
\midrule
0.1 & OLS & 1000 & 0.043 & 0.112 & 0.012 & 0.112 \\
0.1 & Huber-FIX & 1000 & 0.644 & 0.652 & 0.018 & 0.652 \\
0.1 & Huber-SEL & 1000 & 0.730 & 0.721 & 0.018 & 0.721 \\
0.1 & Proposed & 962 & 0.928 & 0.927 & 0.009 & 0.892 \\
\midrule
0.2 & OLS & 1000 & 0.174 & 0.323 & 0.018 & 0.323 \\
0.2 & Huber-FIX & 1000 & 0.990 & 0.992 & 0.003 & 0.992 \\
0.2 & Huber-SEL & 1000 & 0.999 & 0.999 & 0.001 & 0.999 \\
0.2 & Proposed & 960 & 1.000 & 1.000 & 0.000 & 0.960 \\
\midrule
0.3 & OLS & 1000 & 0.473 & 0.632 & 0.018 & 0.632 \\
0.3 & Huber-FIX & 1000 & 1.000 & 1.000 & 0.000 & 1.000 \\
0.3 & Huber-SEL & 1000 & 1.000 & 1.000 & 0.000 & 1.000 \\
0.3 & Proposed & 967 & 1.000 & 1.000 & 0.000 & 0.967 \\
\midrule
0.4 & OLS & 1000 & 0.631 & 0.789 & 0.015 & 0.789 \\
0.4 & Huber-FIX & 1000 & 1.000 & 1.000 & 0.000 & 1.000 \\
0.4 & Huber-SEL & 1000 & 1.000 & 1.000 & 0.000 & 1.000 \\
0.4 & Proposed & 967 & 1.000 & 1.000 & 0.000 & 0.967 \\
\bottomrule\end{tabular}
\end{table}

\clearpage
\begin{table}[htbp]
\centering\small
\caption{Calibrated rejection contrasts on data sets where all four methods succeed. Difference is proposed minus comparator. The Monte Carlo interval uses 1,000 paired, two-stage bootstrap resamples and describes simulation uncertainty, not uncertainty in an individual mediation analysis.}
\label{tab:power-paired}
\begin{tabular}{rlrrrr}
\toprule
$\beta_2$ & Comparator & Common valid & Difference & MCSE & 95\% interval \\ \midrule
0.0 & OLS & 965 & 0.000 & 0.012 & [-0.019, 0.026] \\
0.0 & Huber-FIX & 965 & -0.003 & 0.011 & [-0.021, 0.024] \\
0.0 & Huber-SEL & 965 & 0.000 & 0.010 & [-0.018, 0.020] \\
\midrule
0.1 & OLS & 962 & 0.814 & 0.015 & [0.782, 0.842] \\
0.1 & Huber-FIX & 962 & 0.269 & 0.018 & [0.236, 0.301] \\
0.1 & Huber-SEL & 962 & 0.204 & 0.017 & [0.172, 0.238] \\
\midrule
0.2 & OLS & 960 & 0.677 & 0.019 & [0.642, 0.714] \\
0.2 & Huber-FIX & 960 & 0.006 & 0.003 & [0.002, 0.013] \\
0.2 & Huber-SEL & 960 & 0.001 & 0.001 & [0.000, 0.003] \\
\midrule
0.3 & OLS & 967 & 0.364 & 0.018 & [0.327, 0.401] \\
0.3 & Huber-FIX & 967 & 0.000 & 0.000 & [0.000, 0.000] \\
0.3 & Huber-SEL & 967 & 0.000 & 0.000 & [0.000, 0.000] \\
\midrule
0.4 & OLS & 967 & 0.206 & 0.015 & [0.177, 0.239] \\
0.4 & Huber-FIX & 967 & 0.000 & 0.000 & [0.000, 0.000] \\
0.4 & Huber-SEL & 967 & 0.000 & 0.000 & [0.000, 0.000] \\
\bottomrule\end{tabular}
\end{table}

The complete records, selected Huber thresholds, warning and failure messages, calibration cutoffs, paired summaries and two-stage bootstrap output are retained in the project repository. Independent checks reconstruct data without the study's generator wrapper, reproduce all four methods on ten prespecified audit tasks, and verify all reported counts and summary denominators. Estimator-level tests compare Huber IRLS with an independent optimizer, check response-shift invariance, and verify that a sufficiently large threshold reproduces OLS coefficients and slope covariance.

\FloatBarrier
\subsection{Starting-value sensitivity}

Table~\ref{tab:algorithm_sensitivity} varies the coefficient perturbation scale using 300 common data sets per mixture at $n=200$. It records valid-fit estimates and success rates separately; unsuccessful density-score fits remain failures.

\begin{table}[htbp]
\centering
\caption{Sensitivity to the coefficient perturbation scale at $n=200$, based on 300 common data sets per error distribution. Success uses all attempts; other summaries use numerically successful fits}
\label{tab:algorithm_sensitivity}
\small
\begin{tabular}{lrrrrrrr}
\toprule
Error & Scale & Reps. & Success & Mean PNIE & SD & Roots M & Roots Y \\
\midrule
Asymmetric mixture & 0.05 & 300 & 0.957 & -0.320 & 0.037 & 1.01 & 1.00 \\
Asymmetric mixture & 0.10 & 300 & 0.900 & -0.319 & 0.037 & 1.00 & 1.01 \\
Asymmetric mixture & 0.20 & 300 & 0.860 & -0.321 & 0.029 & 1.00 & 1.01 \\
Symmetric bimodal & 0.05 & 300 & 0.950 & -0.319 & 0.029 & 1.00 & 1.01 \\
Symmetric bimodal & 0.10 & 300 & 0.940 & -0.319 & 0.029 & 1.00 & 1.04 \\
Symmetric bimodal & 0.20 & 300 & 0.910 & -0.318 & 0.075 & 1.00 & 1.09 \\
\bottomrule
\end{tabular}
\end{table}

\FloatBarrier

\clearpage

\section{Variance-misspecification sensitivity}
\label{sec:variance-sensitivity}

This supplementary experiment retains the main-study conditional means,
\[
M=0.2+0.4T+0.3X+s(T,X)U_M,\qquad
Y=0.5T-0.8M+TM+0.4X+s(T,X)U_Y.
\]
Here $X\sim N(0,1)$, $T\sim\operatorname{Bernoulli}(0.5)$, and the standardized innovations $U_M,U_Y$ are independent of one another and of $(T,X)$. We use the Gaussian and asymmetric-mixture densities in Section~\ref{sec:simulation-results}, with no change to their component parameters. Each error has conditional mean zero. For covariate-dependent variance,
\[
s_X^2(X)=\exp(\kappa X-\kappa^2/2),
\]
and for treatment-dependent variance,
\[
s_T^2(T)=\exp\{\kappa(2T-1)\}/\cosh(\kappa).
\]
Both have population expectation one. Normalization uses the known population law, not the realized sample. For $\kappa=0.15,0.30,0.60$, the covariate mechanism's variance ratio between the 75th and 25th percentiles of $X$ is approximately $1.22,1.50,2.25$; the treatment-to-control variance ratio is $1.35,1.82,3.32$. These ratios describe different contrasts and should not be treated as identical strengths of heterogeneity.

To see why causal identification is preserved in this design, define $M(t)$ by replacing $T$ with $t$ in the mediator equation and define $Y(t,m)$ by replacing $(T,M)$ with $(t,m)$ in the outcome equation. Conditional on $X$, the two potential variables depend on independent innovations; randomized $T$ is independent of both. This structural construction supplies sequential ignorability, which is stronger than conditional mean-zero restrictions alone. For positive $\kappa$, the observed errors can depend on the regressors through their scales. The population effects nevertheless remain PNIE $=-0.32$, TNIE $=0.08$, PNDE $=0.70$, TNDE $=1.10$ and TE $=0.78$, because the structural conditional means are unchanged.

We use $n=300$ and 500 attempts per innovation-by-variance configuration. Sharing the constant-variance reference gives seven configurations per density and 7,000 data sets. Seed 20260917 generates the stored task seeds. Configurations within a replication share $X,T,U_M,U_Y$; replications are independent. We compare OLS with its stacked empirical sandwich against the unchanged proposed estimator, retaining every fit status, including failures.

Table~\ref{tab:variance-sensitivity} reports PNIE and TE; the project repository retains all five effects. Estimation and interval metrics use successful fits. Bias MCSE is the empirical standard deviation divided by the square root of the successful-fit count; coverage MCSE is the binomial standard error with that denominator. Coverage 0.95 with 500 successful fits has MCSE 0.0097. Numerical success and report-and-cover use all 500 attempts; paired summaries use common successful replications.

\begin{figure}[htbp]
\centering
\includegraphics[width=0.96\textwidth]{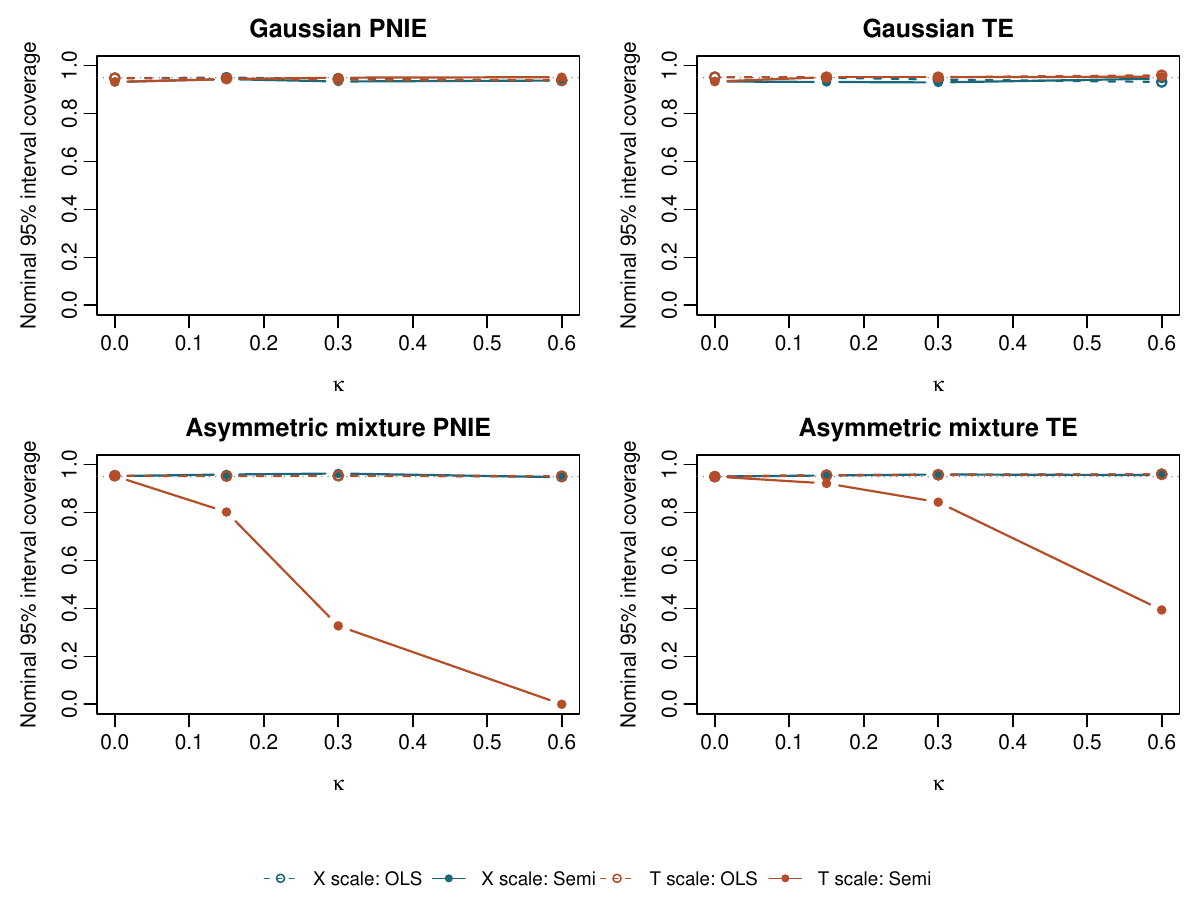}
\caption{Coverage conditional on numerical success in the variance-misspecification experiment. Each point summarizes 500 attempted replications; the shared constant-variance reference is at $\kappa=0$. Colors distinguish covariate- and treatment-dependent scales, and line types distinguish OLS from the original proposed estimator. The dotted horizontal line marks the nominal 0.95 reference; Table~\ref{tab:variance-sensitivity} reports coverage MCSEs and numerical success.}
\label{fig:variance-sensitivity}
\end{figure}

Figure~\ref{fig:variance-sensitivity} displays the mechanism-specific coverage patterns discussed in the main article. Table~\ref{tab:variance-sensitivity} adds bias, RMSE, interval length and success rates, allowing coverage changes to be read with estimation error and fit availability.

\clearpage
\begingroup
\let\small\footnotesize
\renewcommand{\arraystretch}{0.88}
\begingroup\small\setlength{\tabcolsep}{3pt}
\begin{longtable}{llrllrrrrrr}
\caption{Variance-misspecification sensitivity at $n=300$, with 500 attempted replications per configuration. Coverage and precision are conditional on numerical success. Bias and coverage MCSEs are in parentheses; Success uses all attempts.}\label{tab:variance-sensitivity}\\
\toprule
Error & Scale & $\kappa$ & Effect & Method & Bias (MCSE) & RMSE & Coverage (MCSE) & Length & Success & R+C\\
\midrule\endfirsthead
\toprule
Error & Scale & $\kappa$ & Effect & Method & Bias (MCSE) & RMSE & Coverage (MCSE) & Length & Success & R+C\\
\midrule\endhead
\bottomrule\endfoot
AM & None & 0.00 & PNIE & OLS & -0.004 (0.004) & 0.097 & 0.954 (0.009) & 0.388 & 1.000 & 0.954\\
AM & None & 0.00 & PNIE & Semi & -0.002 (0.001) & 0.028 & 0.952 (0.010) & 0.097 & 0.968 & 0.922\\
AM & None & 0.00 & TE & OLS & 0.003 (0.006) & 0.129 & 0.950 (0.010) & 0.527 & 1.000 & 0.950\\
AM & None & 0.00 & TE & Semi & 0.001 (0.003) & 0.070 & 0.950 (0.010) & 0.266 & 0.968 & 0.920\\
AM & $X$ & 0.15 & PNIE & OLS & -0.004 (0.004) & 0.097 & 0.954 (0.009) & 0.388 & 1.000 & 0.954\\
AM & $X$ & 0.15 & PNIE & Semi & -0.002 (0.001) & 0.026 & 0.959 (0.009) & 0.098 & 0.976 & 0.936\\
AM & $X$ & 0.15 & TE & OLS & 0.003 (0.006) & 0.129 & 0.956 (0.009) & 0.527 & 1.000 & 0.956\\
AM & $X$ & 0.15 & TE & Semi & -0.000 (0.003) & 0.066 & 0.955 (0.009) & 0.266 & 0.976 & 0.932\\
AM & $X$ & 0.30 & PNIE & OLS & -0.004 (0.004) & 0.097 & 0.954 (0.009) & 0.389 & 1.000 & 0.954\\
AM & $X$ & 0.30 & PNIE & Semi & -0.002 (0.001) & 0.026 & 0.963 (0.009) & 0.100 & 0.978 & 0.942\\
AM & $X$ & 0.30 & TE & OLS & 0.002 (0.006) & 0.129 & 0.958 (0.009) & 0.527 & 1.000 & 0.958\\
AM & $X$ & 0.30 & TE & Semi & 0.000 (0.003) & 0.065 & 0.959 (0.009) & 0.266 & 0.978 & 0.938\\
AM & $X$ & 0.60 & PNIE & OLS & -0.004 (0.004) & 0.100 & 0.950 (0.010) & 0.394 & 1.000 & 0.950\\
AM & $X$ & 0.60 & PNIE & Semi & -0.002 (0.001) & 0.027 & 0.948 (0.010) & 0.106 & 0.920 & 0.872\\
AM & $X$ & 0.60 & TE & OLS & 0.002 (0.006) & 0.129 & 0.960 (0.009) & 0.525 & 1.000 & 0.960\\
AM & $X$ & 0.60 & TE & Semi & 0.000 (0.003) & 0.067 & 0.957 (0.010) & 0.269 & 0.920 & 0.880\\
AM & $T$ & 0.15 & PNIE & OLS & -0.004 (0.004) & 0.097 & 0.952 (0.010) & 0.387 & 1.000 & 0.952\\
AM & $T$ & 0.15 & PNIE & Semi & 0.026 (0.001) & 0.036 & 0.802 (0.018) & 0.099 & 0.972 & 0.780\\
AM & $T$ & 0.15 & TE & OLS & 0.002 (0.006) & 0.127 & 0.956 (0.009) & 0.518 & 1.000 & 0.956\\
AM & $T$ & 0.15 & TE & Semi & -0.025 (0.003) & 0.069 & 0.922 (0.012) & 0.261 & 0.972 & 0.896\\
AM & $T$ & 0.30 & PNIE & OLS & -0.004 (0.004) & 0.098 & 0.954 (0.009) & 0.387 & 1.000 & 0.954\\
AM & $T$ & 0.30 & PNIE & Semi & 0.065 (0.001) & 0.070 & 0.327 (0.021) & 0.103 & 0.972 & 0.318\\
AM & $T$ & 0.30 & TE & OLS & 0.002 (0.006) & 0.124 & 0.958 (0.009) & 0.510 & 1.000 & 0.958\\
AM & $T$ & 0.30 & TE & Semi & -0.054 (0.004) & 0.112 & 0.844 (0.016) & 0.262 & 0.972 & 0.820\\
AM & $T$ & 0.60 & PNIE & OLS & -0.005 (0.004) & 0.099 & 0.952 (0.010) & 0.387 & 1.000 & 0.952\\
AM & $T$ & 0.60 & PNIE & Semi & 0.180 (0.001) & 0.182 & 0.000 (0.000) & 0.109 & 0.798 & 0.000\\
AM & $T$ & 0.60 & TE & OLS & 0.001 (0.005) & 0.120 & 0.960 (0.009) & 0.494 & 1.000 & 0.960\\
AM & $T$ & 0.60 & TE & Semi & -0.157 (0.013) & 0.306 & 0.393 (0.024) & 0.277 & 0.798 & 0.314\\
\midrule
G & None & 0.00 & PNIE & OLS & -0.001 (0.005) & 0.102 & 0.948 (0.010) & 0.385 & 1.000 & 0.948\\
G & None & 0.00 & PNIE & Semi & 0.002 (0.005) & 0.105 & 0.932 (0.011) & 0.408 & 1.000 & 0.932\\
G & None & 0.00 & TE & OLS & 0.003 (0.006) & 0.135 & 0.952 (0.010) & 0.527 & 1.000 & 0.952\\
G & None & 0.00 & TE & Semi & 0.003 (0.006) & 0.141 & 0.934 (0.011) & 0.549 & 1.000 & 0.934\\
G & $X$ & 0.15 & PNIE & OLS & -0.001 (0.005) & 0.102 & 0.950 (0.010) & 0.385 & 1.000 & 0.950\\
G & $X$ & 0.15 & PNIE & Semi & 0.001 (0.005) & 0.105 & 0.944 (0.010) & 0.406 & 1.000 & 0.944\\
G & $X$ & 0.15 & TE & OLS & 0.003 (0.006) & 0.135 & 0.950 (0.010) & 0.526 & 1.000 & 0.950\\
G & $X$ & 0.15 & TE & Semi & 0.003 (0.006) & 0.143 & 0.932 (0.011) & 0.548 & 1.000 & 0.932\\
G & $X$ & 0.30 & PNIE & OLS & -0.001 (0.005) & 0.102 & 0.946 (0.010) & 0.386 & 1.000 & 0.946\\
G & $X$ & 0.30 & PNIE & Semi & 0.000 (0.005) & 0.106 & 0.934 (0.011) & 0.411 & 1.000 & 0.934\\
G & $X$ & 0.30 & TE & OLS & 0.003 (0.006) & 0.136 & 0.942 (0.010) & 0.526 & 1.000 & 0.942\\
G & $X$ & 0.30 & TE & Semi & 0.004 (0.006) & 0.145 & 0.930 (0.011) & 0.555 & 1.000 & 0.930\\
G & $X$ & 0.60 & PNIE & OLS & -0.001 (0.005) & 0.104 & 0.938 (0.011) & 0.390 & 1.000 & 0.938\\
G & $X$ & 0.60 & PNIE & Semi & -0.000 (0.005) & 0.109 & 0.938 (0.011) & 0.416 & 1.000 & 0.938\\
G & $X$ & 0.60 & TE & OLS & 0.003 (0.006) & 0.137 & 0.932 (0.011) & 0.525 & 1.000 & 0.932\\
G & $X$ & 0.60 & TE & Semi & 0.006 (0.006) & 0.145 & 0.946 (0.010) & 0.558 & 1.000 & 0.946\\
G & $T$ & 0.15 & PNIE & OLS & -0.000 (0.005) & 0.102 & 0.946 (0.010) & 0.385 & 1.000 & 0.946\\
G & $T$ & 0.15 & PNIE & Semi & 0.001 (0.005) & 0.107 & 0.944 (0.010) & 0.420 & 1.000 & 0.944\\
G & $T$ & 0.15 & TE & OLS & 0.003 (0.006) & 0.132 & 0.952 (0.010) & 0.518 & 1.000 & 0.952\\
G & $T$ & 0.15 & TE & Semi & 0.003 (0.006) & 0.139 & 0.952 (0.010) & 0.558 & 1.000 & 0.952\\
G & $T$ & 0.30 & PNIE & OLS & 0.000 (0.005) & 0.102 & 0.942 (0.010) & 0.385 & 1.000 & 0.942\\
G & $T$ & 0.30 & PNIE & Semi & 0.001 (0.005) & 0.111 & 0.950 (0.010) & 0.455 & 1.000 & 0.950\\
G & $T$ & 0.30 & TE & OLS & 0.002 (0.006) & 0.130 & 0.952 (0.010) & 0.509 & 1.000 & 0.952\\
G & $T$ & 0.30 & TE & Semi & 0.001 (0.006) & 0.142 & 0.952 (0.010) & 0.599 & 1.000 & 0.952\\
G & $T$ & 0.60 & PNIE & OLS & 0.001 (0.005) & 0.102 & 0.940 (0.011) & 0.386 & 1.000 & 0.940\\
G & $T$ & 0.60 & PNIE & Semi & 0.003 (0.006) & 0.132 & 0.952 (0.010) & 0.553 & 1.000 & 0.952\\
G & $T$ & 0.60 & TE & OLS & 0.001 (0.006) & 0.125 & 0.960 (0.009) & 0.494 & 1.000 & 0.960\\
G & $T$ & 0.60 & TE & Semi & -0.000 (0.007) & 0.165 & 0.954 (0.009) & 0.716 & 1.000 & 0.954\\
\end{longtable}
\noindent G: Gaussian; AM: asymmetric mixture; Semi: the unchanged proposed estimator. R+C is the fraction of all attempted data sets yielding an interval covering the truth, not conventional coverage. Scale $X$ denotes covariate-dependent variance and scale $T$ treatment-dependent variance; None is their common constant-variance reference. All five effects and paired-fit summaries are retained in the project repository.
\endgroup

\endgroup

\clearpage
\clearpage
\section{Reduced-form confounding-sensitivity extension}
\label{sec:confounding-supp}

\subsection{Identification and attribution}

This section develops the density-score estimation and covariance propagation for the fixed-error-correlation sensitivity identity of Imai, Keele and Yamamoto (2010), as cited in the main article. Let $m_t(\bX)=\alpha_2+\beta_2t+\bm\xi_2\trans\bX$ and $q_t(\bX)=a_t+\bm d_t\trans\bX$. A sufficient structural formulation is
\[
M(t)=m_t(\bX)+U_t,\qquad
Y(t,m)=q_t(\bX)+b_t\{m-m_t(\bX)\}+E_t.
\]
Assume treatment exchangeability conditional on baseline $\bX$, consistency and positivity; $U_t,E_t$ have conditional mean zero; and $E_t$ is invariant to the intervention value $m$. The sensitivity model allows mediator--outcome dependence through the joint law of $U_t$ and $E_t$. The observed reduced-form error is $W_t=b_tU_t+E_t$. Its conditional variance and covariance with $U_t$ obey
\[
\omega_t=b_tv_M+\rho_t\sqrt{v_M\Var(E_t)},\quad
v_t=b_t^2v_M+\Var(E_t)+2b_t\rho_t\sqrt{v_M\Var(E_t)}.
\]
Solving with positive $\Var(E_t)$ yields the main article's sensitivity slope. The natural indirect effect is $b_t\E\{M(1)-M(0)\}=b_t\beta_2$, since the additive $E_t$ cancels in the difference of nested potential outcomes. The total effect is $\E(q_1-q_0)$, giving the direct effects by the two decomposition identities. Thus the linear structural form supplies the causal interpretation and defines the sensitivity class: additive mediator--outcome confounding with an intervention-invariant structural error and a stratum-specific constant mediator slope.

For estimation, $U=M-m_T(\bX)$ has a common law independent of $(T,\bX)$, and the joint law of $(U,W_t)$ is stable across $\bX$ within each stratum. These assumptions support pooled-mediator and stratum-specific reduced-outcome fits with unspecified marginal densities. The figures follow the common-$\rho$ path through $(\rho_0,\rho_1)$.

\subsection{Estimating equations and covariance}

Let $S_M$ be the marginal regression score for $M$ on $(1,T,\bX\trans)$ and $S_{W_t}$ the score for $Y$ on $(1,\bX\trans)$ within $T=t$. The stacked population function is
\[
\Psi_i(\theta)=\begin{pmatrix}
S_{M,i}\\
1(T_i=0)S_{W_0,i}\\
1(T_i=1)S_{W_1,i}\\
1(T_i=0)(U_iW_{0i}-\omega_0)\\
1(T_i=1)(U_iW_{1i}-\omega_1)\\
\bX_i-\bmu_X
\end{pmatrix}.
\]
The first three blocks use the density-score equations and numerical settings in the main article. The OLS benchmark replaces each regression block by $(D e, e^2/v-1)$, with parameters $(\beta,\log v)$, so both methods estimate the same coefficient and variance nuisance parameters. Cross-covariances are the sample averages of fitted residual products within treatment groups. The parameter domain requires $v_Mv_t-\omega_t^2>0$; fitted values outside this domain are recorded transparently as numerical failures rather than altered by truncation.

Let $\widehat A=n^{-1}\sum_i\partial\widehat\Psi_i/\partial\theta\trans$ and $\widehat B=n^{-1}\sum_i\widehat\Psi_i\widehat\Psi_i\trans$. Within-stratum regression bread blocks are multiplied by $n_t/n$. The cross-product row for stratum $t$ has derivatives
\[
-n^{-1}\sum_{i:T_i=t}D_{Mi}\widehat W_{ti},\qquad
-n^{-1}\sum_{i:T_i=t}D_{Wi}\widehat U_i,
\]
with respect to the mediator and corresponding reduced-outcome coefficients, respectively, and derivative $-n_t/n$ with respect to $\omega_t$. It has zero direct derivative with respect to log variances. The baseline-mean block has bread $-I$. Empirical off-diagonal bread entries are retained even where their population expectations vanish. The full meat includes cross-covariance between the marginal regression scores and residual-product moments. The implemented covariance is $\widehat A^{-1}\widehat B\widehat A^{-\mathsf T}/n$.

\paragraph{Conditional asymptotic justification.}
Assume positive treatment probabilities and nonsingular bread. Also require finite score second moments and fourth error moments, and $v_Mv_t>\omega_t^2$. Each fitted marginal regression is assumed to satisfy the marginal analogues of Conditions~(\ref{cond:score-model})--(\ref{cond:score-derivative}) in Section~\ref{sec:regularity}. Require covariance consistency as in Condition~(\ref{cond:covariance}) for the full augmented stack, retaining its cross-block dependence rather than imposing the independent-error factorization used for the original mediation model. On any fixed finite collection of sensitivity values strictly inside $(-1,1)$, the marginal expansions hold jointly. Taylor expansion of the appended residual-product equations and empirical baseline mean then gives the stacked expansion $\varphi_\theta=-A^{-1}\Psi$. Applying the chain rule gives $\varphi_{g,\rho}=G_\rho\varphi_\theta$ and the displayed sandwich covariance. Positive effect variance is required for a scalar Wald interval. Thus standard estimating-equation and delta-method theory supplies pointwise asymptotic inference under the stated marginal-estimator regularity. The efficiency interpretation is attached to the marginal score blocks; the displayed joint representation characterizes the assembled estimator in the reduced-form working model.

\subsection{Analytic gradients and zero boundaries}

Put $\ell_t=\omega_t/v_M$, $A_t=v_t/v_M$, $d_t=\sqrt{A_t-\ell_t^2}$ and $h_t=\rho_t/\sqrt{1-\rho_t^2}$. Then $b_t=\ell_t-h_td_t$ and
\begin{align*}
\frac{\partial b_t}{\partial\omega_t}&=\frac{1+h_t\ell_t/d_t}{v_M},\\
\frac{\partial b_t}{\partial\log v_M}&=-\ell_t+\frac{h_t(A_t-2\ell_t^2)}{2d_t},\\
\frac{\partial b_t}{\partial\log v_t}&=-\frac{h_tA_t}{2d_t}.
\end{align*}
Thus $d\delta_t=b_t\,d\beta_2+\beta_2\,db_t$. For the total effect,
\[
d\tau=d(a_1-a_0)+\bmu_X\trans d(\bm d_1-\bm d_0)
+(\bm d_1-\bm d_0)\trans d\bmu_X.
\]
Direct-effect gradients are obtained by subtraction. The code supports separate $(\rho_0,\rho_1)$ even though the figures use their common-value path.

For $\beta_2\ne0$, $\delta_t(\rho_t)=0$ exactly at $r_t=\omega_t/\sqrt{v_Mv_t}$. Its derivatives are $\partial r_t/\partial\omega_t=1/\sqrt{v_Mv_t}$ and $\partial r_t/\partial\log v_M=\partial r_t/\partial\log v_t=-r_t/2$. We form a delta interval for $\operatorname{atanh}(r_t)$ and transform its endpoints back with $\tanh$. At $\beta_2=0$, the complete indirect-effect curve equals zero, producing a set-valued rather than unique effect-zero threshold. The confidence interval for a unique zero boundary addresses uncertainty in the point-estimate threshold, while the statistical-significance boundary along the curve addresses pointwise interval exclusion of zero.

Under the working error independence from regressors, the first-order derivatives of residual second moments with respect to the mean coefficients vanish. Their influence terms reduce to $U^2-v_M$, $1(T=t)(W_t^2-v_t)/p_t$ and $1(T=t)(UW_t-\omega_t)/p_t$. Consequently the OLS-RF and DS-RF zero-boundary estimators share the same first-order residual-moment representation under this model, explaining their close finite-sample boundaries. Density-score gains for the indirect-effect curve arise instead through the pooled mediator treatment coefficient and its joint covariance with the remaining nuisance estimates.

For the contour calibration, Imai's additional shared-confounder variance decomposition expresses $U$ and $E_t$ as linear contributions of a hypothetical pretreatment confounder plus uncorrelated remainder errors, with the corresponding orthogonality to the confounder. It gives $\rho_t^2=R_M^{2*}R_{Y,t}^{2*}$, with the sign specified separately. The outcome fraction refers to the \emph{structural outcome error}, which is distinct from the observed reduced-form outcome residual. The figure uses this moment calibration to map hypothetical latent variance fractions into sensitivity values without requiring a complete latent distribution. Accordingly, its axes are labeled as residual-variance fractions rather than total-variance tilde quantities.

\subsection{Simulation design and verification}

The main article's confounding experiment uses seed 20260918, $n=300$ and 500 replications per innovation-by-correlation cell. For independent verification, the stratum-$t$ covariance is $-0.8+t+\rho_{\rm true}$ and the reduced-outcome variance is $(-0.8+t+\rho_{\rm true})^2+1-\rho_{\rm true}^2$; these determine the population zero boundaries. The original natural effects are retained. Evaluation at the generating correlation and at zero includes valid-fit, paired-fit and all-attempted summaries, with every fit status recorded.

Table~\ref{tab:confounding-simulation} gives performance at the generating correlation; the full-curve and zero-boundary assessments follow below. JOBS II curves are grouped with the application in Section~\ref{sec:jobs}.

\begin{table}[htbp]\centering\small\setlength{\tabcolsep}{3pt}
\caption{Reduced-form sensitivity estimation at the true structural correlation: $n=300$, 500 attempts per cell. Coverage and precision use valid fits; Success uses all attempts.}
\label{tab:confounding-simulation}
\begin{tabular}{lrllrrrrr}\toprule
Error & $\rho$ & Effect & Method & Bias & RMSE & Coverage (MCSE) & Length & Success\\\midrule
AM & -0.3 & PNIE & OLS-RF & 0.017 & 0.067 & 0.966 (0.008) & 0.257 & 1.000\\
AM & -0.3 & PNIE & DS-RF & 0.000 & 0.066 & 0.958 (0.009) & 0.260 & 0.996\\
AM & -0.3 & TE & OLS-RF & -0.011 & 0.145 & 0.950 (0.010) & 0.560 & 1.000\\
AM & -0.3 & TE & DS-RF & -0.011 & 0.145 & 0.946 (0.010) & 0.560 & 0.996\\
AM & 0.0 & PNIE & OLS-RF & 0.003 & 0.070 & 0.946 (0.010) & 0.283 & 1.000\\
AM & 0.0 & PNIE & DS-RF & 0.002 & 0.049 & 0.952 (0.010) & 0.198 & 1.000\\
AM & 0.0 & TE & OLS-RF & -0.004 & 0.128 & 0.964 (0.008) & 0.527 & 1.000\\
AM & 0.0 & TE & DS-RF & -0.004 & 0.127 & 0.968 (0.008) & 0.527 & 1.000\\
AM & 0.3 & PNIE & OLS-RF & 0.011 & 0.079 & 0.974 (0.007) & 0.324 & 1.000\\
AM & 0.3 & PNIE & DS-RF & 0.001 & 0.044 & 0.946 (0.010) & 0.170 & 0.998\\
AM & 0.3 & TE & OLS-RF & -0.002 & 0.124 & 0.960 (0.009) & 0.490 & 1.000\\
AM & 0.3 & TE & DS-RF & -0.002 & 0.124 & 0.958 (0.009) & 0.490 & 0.998\\
\midrule
G & -0.3 & PNIE & OLS-RF & 0.004 & 0.108 & 0.926 (0.012) & 0.418 & 1.000\\
G & -0.3 & PNIE & DS-RF & 0.005 & 0.110 & 0.924 (0.012) & 0.434 & 1.000\\
G & -0.3 & TE & OLS-RF & 0.000 & 0.146 & 0.932 (0.011) & 0.558 & 1.000\\
G & -0.3 & TE & DS-RF & 0.000 & 0.146 & 0.934 (0.011) & 0.558 & 1.000\\
G & 0.0 & PNIE & OLS-RF & 0.004 & 0.102 & 0.946 (0.010) & 0.397 & 1.000\\
G & 0.0 & PNIE & DS-RF & 0.003 & 0.105 & 0.946 (0.010) & 0.413 & 1.000\\
G & 0.0 & TE & OLS-RF & 0.001 & 0.140 & 0.940 (0.011) & 0.529 & 1.000\\
G & 0.0 & TE & DS-RF & 0.001 & 0.140 & 0.942 (0.010) & 0.529 & 1.000\\
G & 0.3 & PNIE & OLS-RF & 0.005 & 0.098 & 0.940 (0.011) & 0.385 & 1.000\\
G & 0.3 & PNIE & DS-RF & 0.005 & 0.102 & 0.938 (0.011) & 0.402 & 1.000\\
G & 0.3 & TE & OLS-RF & -0.000 & 0.119 & 0.962 (0.009) & 0.490 & 1.000\\
G & 0.3 & TE & DS-RF & -0.000 & 0.119 & 0.962 (0.009) & 0.491 & 1.000\\
\bottomrule\end{tabular}
\par\smallskip\begin{minipage}{0.98\textwidth}\footnotesize G: Gaussian; AM: asymmetric mixture. OLS-RF and DS-RF use the same reduced-form models with OLS and marginal density-score regression, respectively. They are not the conditional-outcome estimators in the original simulation. All five effects, zero-rho analyses, threshold inference and paired comparisons are retained in the project repository.\end{minipage}
\end{table}

The independent tests compare analytic and numerical gradients for all five effects, verify both exact decompositions and covariance symmetry, reconstruct the Imai covariance identity at known population parameters, and check zero-boundary derivatives and separate-correlation inputs. Output checks independently recompute performance summaries, compare paired replication identifiers, check all attempted/valid denominators, and reconstruct retained OLS fits from the recorded seeds. The existing core estimator is unchanged.

\clearpage
\subsection{Full-curve Monte Carlo assessment}

The full-curve evaluation reconstructs the same 3,000 data sets and estimator settings from the stored seeds, as recorded in \path{CURVE_SIMULATION_PLAN_20260916.md}. Each fit supplies five effects on $\{-0.8,-0.75,\ldots,0.8\}$. With generating correlation $r$, the moments $v_M=1$, $\omega_t=-0.8+t+r$ and $v_t=\omega_t^2+1-r^2$ give the main article's population curve. The total effect is 0.78 throughout; direct effects follow by subtraction.

For assumed $\rho$, bias, RMSE and coverage target the population sensitivity functional; \texttt{CausalBias} and \texttt{CausalInclusion} target the generating causal effect. Paired ratios use common valid replication identifiers. Monte Carlo calculations preserve within-curve dependence and use the valid-fit or all-attempt denominators defined in Section~\ref{sec:paired-audit}.

For a common assumed correlation, write $h=\rho/\sqrt{1-\rho^2}$. Each fitted effect has the exact form
\[
\widehat g(\rho)=\widehat A+h\widehat B,\qquad
\widehat\Var\{\widehat g(\rho)\}
=\widehat V_A+2h\widehat C_{AB}+h^2\widehat V_B.
\]
Here $A=g(\theta;0)$ and $B=g(\theta;1/\sqrt2)-g(\theta;0)$, with gradients obtained from the previously displayed effect map. The covariance terms follow jointly from the complete nuisance sandwich. The project repository stores these five coefficients for every effect and attempted fit, allowing all grid values and intervals to be reconstructed exactly for any plotting grid without refitting, smoothing or interpolating confidence limits. Unsuccessful fits retain missing coefficients and a diagnostic message.

\begin{figure}[p]
\centering
\includegraphics[width=0.98\textwidth]{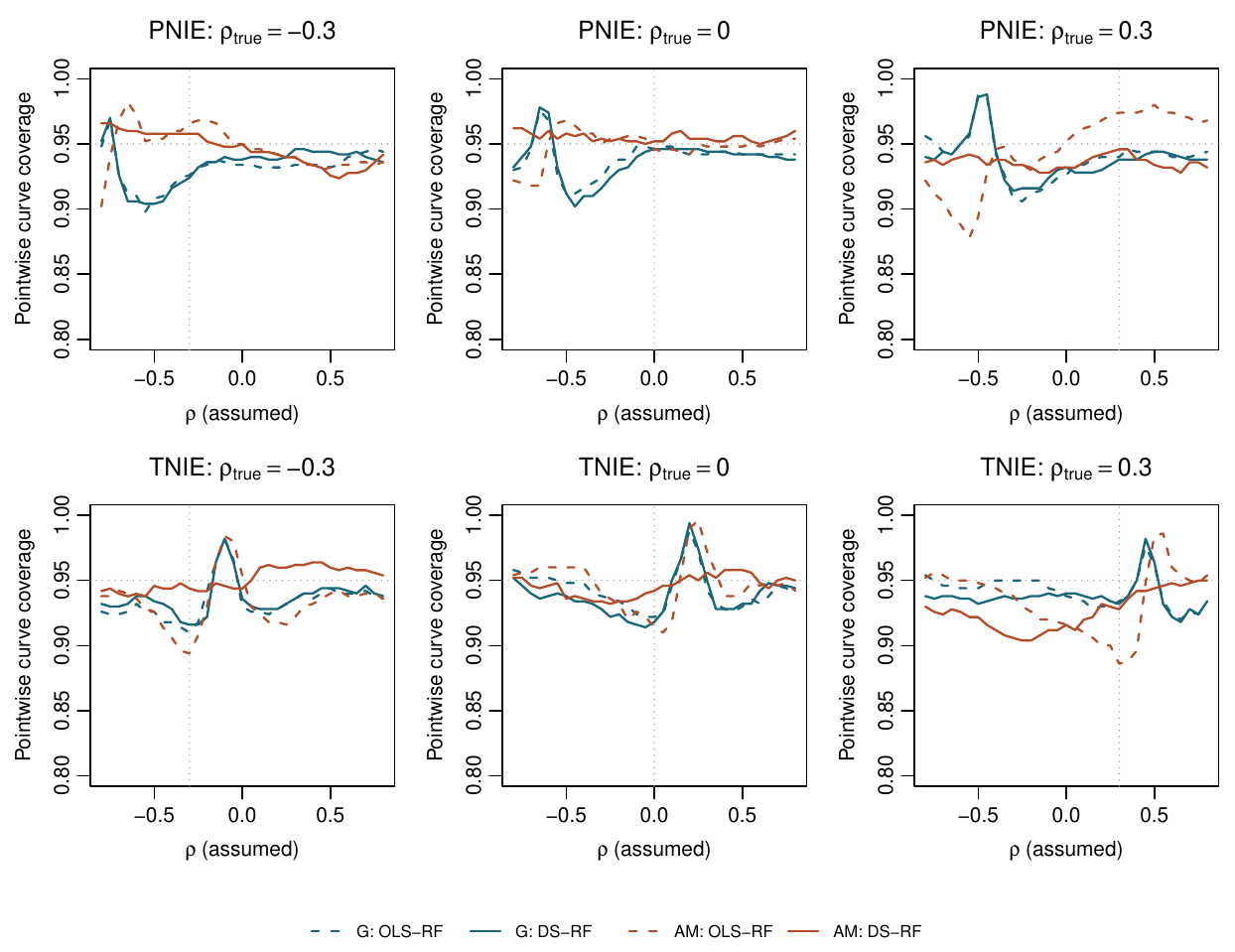}
\caption{Pointwise coverage of the population sensitivity functional across the full assumed-correlation grid, conditional on numerical success. Each configuration starts with 500 attempts. G denotes Gaussian innovations and AM asymmetric-mixture innovations. The horizontal line is 0.95 and the vertical line is the generating correlation. At other correlations, coverage targets the corresponding sensitivity functional; at the generating correlation, that functional equals the generating causal effect. Complete Monte Carlo standard errors and report-and-cover fractions are retained in the project repository.}
\label{fig:curve-coverage}
\end{figure}

\begin{figure}[p]
\centering
\includegraphics[width=0.98\textwidth]{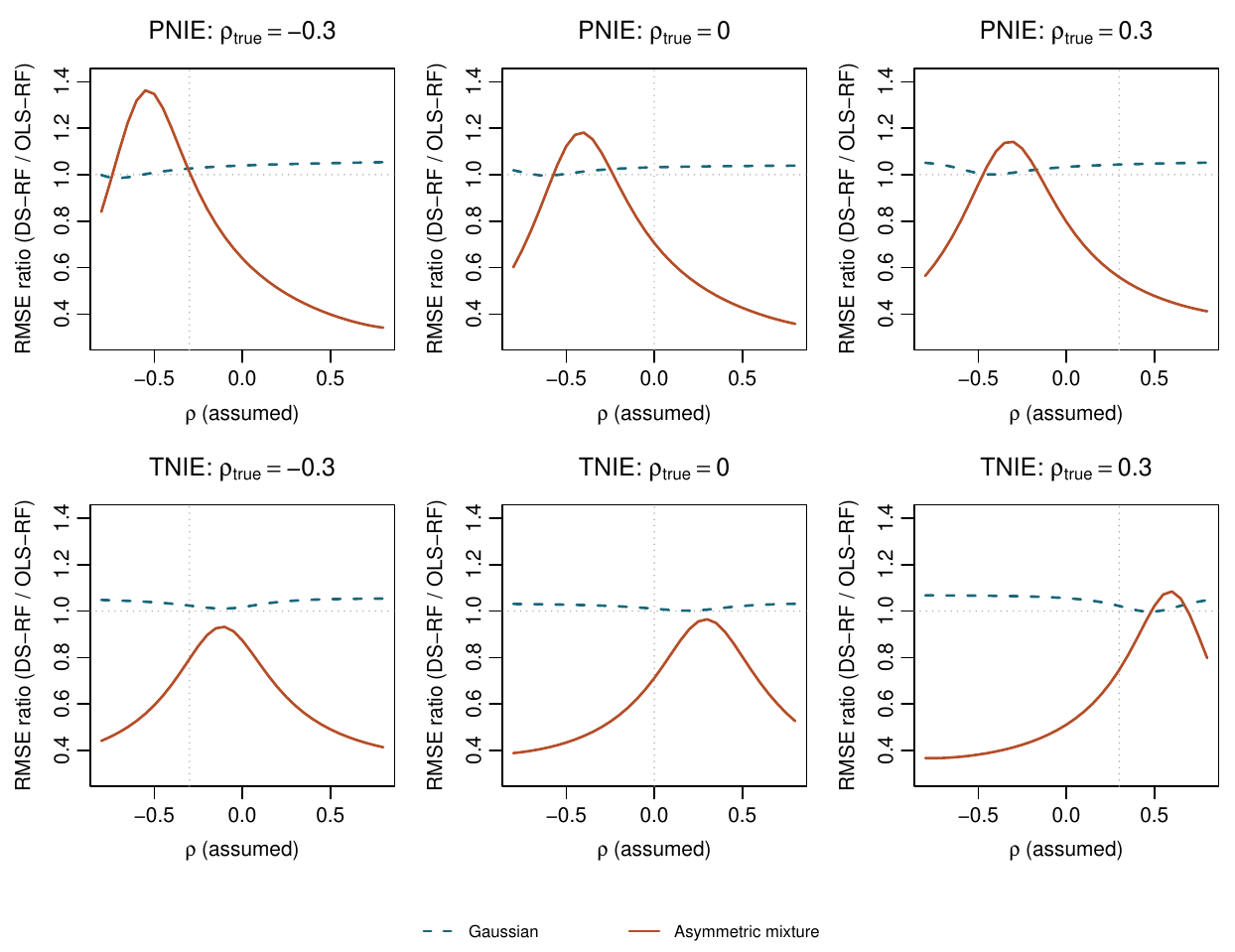}
\caption{Paired DS-RF/OLS-RF RMSE ratios for the population PNIE/TNIE sensitivity curves. Ratios below one favor DS-RF in RMSE. The same successful data sets are used in the numerator and denominator, and all assumed correlations and generating configurations are retained. Pointwise coverage and numerical success provide the complementary calibration and reliability measures.}
\label{fig:curve-ratios}
\end{figure}

The main figure's central 90\% empirical ranges describe the sampling distribution of the effect estimates across valid fits. Figure~\ref{fig:curve-coverage} separately evaluates each fit's nominal 95\% pointwise interval, and Figure~\ref{fig:curve-ratios} compares RMSE on paired valid data sets. Reading the RMSE ratios together with pointwise coverage is particularly informative where the estimated curve is near zero. Table~\ref{tab:curve-thresholds} adds a distinct evaluation of the estimated point-effect zero boundary.

\begin{table}[htbp]\centering\small\setlength{\tabcolsep}{4pt}
\caption{Zero-boundary inference in the confounding simulation. The same retained data sets are used; these are not additional independent replications. Coverage and length refer to the Fisher-transform interval for $\rho_t^\dagger$.}
\label{tab:curve-thresholds}
\begin{tabular}{lrllrrrr}\toprule
Error & $\rho_{\rm true}$ & Effect & Method & Bias & RMSE & Coverage (MCSE) & Length\\\midrule
AM & -0.3 & PNIE & OLS-RF & 0.003 & 0.068 & 0.958 (0.009) & 0.276\\
AM & -0.3 & PNIE & DS-RF & 0.004 & 0.068 & 0.960 (0.009) & 0.275\\
AM & -0.3 & TNIE & OLS-RF & 0.001 & 0.078 & 0.940 (0.011) & 0.311\\
AM & -0.3 & TNIE & DS-RF & 0.002 & 0.077 & 0.944 (0.010) & 0.311\\
AM & 0.0 & PNIE & OLS-RF & 0.006 & 0.072 & 0.958 (0.009) & 0.290\\
AM & 0.0 & PNIE & DS-RF & 0.006 & 0.072 & 0.960 (0.009) & 0.290\\
AM & 0.0 & TNIE & OLS-RF & 0.008 & 0.081 & 0.946 (0.010) & 0.311\\
AM & 0.0 & TNIE & DS-RF & 0.007 & 0.081 & 0.944 (0.010) & 0.312\\
AM & 0.3 & PNIE & OLS-RF & -0.002 & 0.079 & 0.950 (0.010) & 0.306\\
AM & 0.3 & PNIE & DS-RF & -0.002 & 0.079 & 0.950 (0.010) & 0.305\\
AM & 0.3 & TNIE & OLS-RF & -0.003 & 0.080 & 0.936 (0.011) & 0.301\\
AM & 0.3 & TNIE & DS-RF & -0.004 & 0.079 & 0.936 (0.011) & 0.301\\
\midrule
G & -0.3 & PNIE & OLS-RF & 0.002 & 0.058 & 0.950 (0.010) & 0.215\\
G & -0.3 & PNIE & DS-RF & 0.002 & 0.058 & 0.948 (0.010) & 0.215\\
G & -0.3 & TNIE & OLS-RF & 0.002 & 0.080 & 0.954 (0.009) & 0.311\\
G & -0.3 & TNIE & DS-RF & 0.002 & 0.080 & 0.952 (0.010) & 0.311\\
G & 0.0 & PNIE & OLS-RF & 0.006 & 0.063 & 0.946 (0.010) & 0.243\\
G & 0.0 & PNIE & DS-RF & 0.006 & 0.063 & 0.946 (0.010) & 0.243\\
G & 0.0 & TNIE & OLS-RF & 0.005 & 0.081 & 0.944 (0.010) & 0.307\\
G & 0.0 & TNIE & DS-RF & 0.005 & 0.081 & 0.948 (0.010) & 0.307\\
G & 0.3 & PNIE & OLS-RF & 0.008 & 0.069 & 0.954 (0.009) & 0.274\\
G & 0.3 & PNIE & DS-RF & 0.008 & 0.069 & 0.954 (0.009) & 0.274\\
G & 0.3 & TNIE & OLS-RF & -0.000 & 0.069 & 0.946 (0.010) & 0.274\\
G & 0.3 & TNIE & DS-RF & -0.000 & 0.069 & 0.944 (0.010) & 0.274\\
\bottomrule\end{tabular}
\par\smallskip\begin{minipage}{0.98\textwidth}\footnotesize G: Gaussian; AM: asymmetric mixture. Each cell starts with 500 attempts. Success denominators and failure records are the same as for Table~\ref{tab:confounding-simulation}. A zero boundary of the point-estimate curve is not an interval-zero boundary.\end{minipage}
\end{table}

\clearpage

\clearpage
\section{Additional JOBS II results}
\label{sec:jobs}

The \texttt{jobs} object in \textsf{R} package \texttt{mediation} version 4.5.1 supplies all 899 complete observations used in the main article. The mediator \texttt{job\_seek} has 25 distinct values on $[1,5]$; the outcome \texttt{depress2} has 50 on $[1,4.909]$. All analyses retain the main article's treatment--mediator interaction and four-covariate adjustment unless stated otherwise.

\subsection{Natural effects and regression estimates}

Table~\ref{tab:application_effects_full} gives the complete five-method comparison underlying the main article's application figure; Table~\ref{tab:application_regression} records the linked-regression fits. The diagnostic checks below accompany these working-model estimates.

\begin{table}[htbp]
\centering
\caption{Complete JOBS II natural-effect estimates and nominal 95\% limits for five methods ($n=899$). Negative values indicate lower post-intervention depressive symptoms under assignment to the intervention}
\label{tab:application_effects_full}
\small
\begin{tabular}{llrrr}
\toprule
Effect & Method & Estimate & Lower & Upper \\
\midrule
PNIE & OLS delta & -0.0136 & -0.0360 & 0.0087 \\
PNIE & Imai quasi-Bayes & -0.0138 & -0.0391 & 0.0080 \\
PNIE & Imai bootstrap & -0.0136 & -0.0373 & 0.0077 \\
PNIE & \texttt{medflex} NEM & -0.0136 & -0.0358 & 0.0087 \\
PNIE & Proposed & -0.0109 & -0.0205 & -0.0013 \\
\addlinespace[2pt]
TNIE & OLS delta & -0.0097 & -0.0261 & 0.0067 \\
TNIE & Imai quasi-Bayes & -0.0098 & -0.0272 & 0.0060 \\
TNIE & Imai bootstrap & -0.0097 & -0.0289 & 0.0052 \\
TNIE & \texttt{medflex} NEM & -0.0098 & -0.0263 & 0.0068 \\
TNIE & Proposed & -0.0072 & -0.0144 & -0.0001 \\
\addlinespace[2pt]
PNDE & OLS delta & -0.0365 & -0.1167 & 0.0437 \\
PNDE & Imai quasi-Bayes & -0.0357 & -0.1155 & 0.0451 \\
PNDE & Imai bootstrap & -0.0367 & -0.1150 & 0.0451 \\
PNDE & \texttt{medflex} NEM & -0.0369 & -0.1162 & 0.0425 \\
PNDE & Proposed & -0.0272 & -0.0892 & 0.0349 \\
\addlinespace[2pt]
TNDE & OLS delta & -0.0326 & -0.1116 & 0.0464 \\
TNDE & Imai quasi-Bayes & -0.0317 & -0.1128 & 0.0483 \\
TNDE & Imai bootstrap & -0.0327 & -0.1102 & 0.0473 \\
TNDE & \texttt{medflex} NEM & -0.0330 & -0.1113 & 0.0452 \\
TNDE & Proposed & -0.0235 & -0.0835 & 0.0364 \\
\addlinespace[2pt]
TE & OLS delta & -0.0462 & -0.1281 & 0.0357 \\
TE & Imai quasi-Bayes & -0.0454 & -0.1278 & 0.0370 \\
TE & Imai bootstrap & -0.0464 & -0.1278 & 0.0357 \\
TE & \texttt{medflex} NEM & -0.0466 & -0.1278 & 0.0346 \\
TE & Proposed & -0.0344 & -0.0962 & 0.0274 \\
\bottomrule
\end{tabular}
\end{table}

\FloatBarrier
\begin{table}[htbp]
\centering
\caption{Regression estimates for the JOBS II analysis. Variable names follow the reproducibility code}
\label{tab:application_regression}
\footnotesize
\begin{tabular}{lllrr}
\toprule
Equation & Method & Term & Estimate & SE \\
\midrule
Mediator & OLS delta & (Intercept) & 4.0172 & 0.1458 \\
Mediator & OLS delta & T & 0.0601 & 0.0490 \\
Mediator & OLS delta & depress1 & -0.2633 & 0.0462 \\
Mediator & OLS delta & econ\_hard & 0.1026 & 0.0262 \\
Mediator & OLS delta & sex & 0.0241 & 0.0476 \\
Mediator & OLS delta & age & 0.0041 & 0.0024 \\
Mediator & Proposed & (Intercept) & 3.9798 & 0.0734 \\
Mediator & Proposed & T & 0.0725 & 0.0288 \\
Mediator & Proposed & depress1 & -0.1065 & 0.0274 \\
Mediator & Proposed & econ\_hard & 0.0507 & 0.0146 \\
Mediator & Proposed & sex & 0.0248 & 0.0264 \\
Mediator & Proposed & age & 0.0013 & 0.0013 \\
Outcome & OLS delta & (Intercept) & 1.6210 & 0.2220 \\
Outcome & OLS delta & T & -0.2998 & 0.2450 \\
Outcome & OLS delta & M & -0.2269 & 0.0491 \\
Outcome & OLS delta & depress1 & 0.4165 & 0.0363 \\
Outcome & OLS delta & econ\_hard & 0.0684 & 0.0217 \\
Outcome & OLS delta & sex & 0.0566 & 0.0382 \\
Outcome & OLS delta & age & 0.0012 & 0.0017 \\
Outcome & OLS delta & T:M & 0.0658 & 0.0582 \\
Outcome & Proposed & (Intercept) & 1.7148 & 0.1553 \\
Outcome & Proposed & T & -0.2286 & 0.1868 \\
Outcome & Proposed & M & -0.1504 & 0.0331 \\
Outcome & Proposed & depress1 & 0.2621 & 0.0334 \\
Outcome & Proposed & econ\_hard & 0.0432 & 0.0152 \\
Outcome & Proposed & sex & 0.0369 & 0.0267 \\
Outcome & Proposed & age & 0.0003 & 0.0014 \\
Outcome & Proposed & T:M & 0.0504 & 0.0433 \\
\bottomrule
\end{tabular}
\end{table}

\FloatBarrier

\subsection{Numerical and baseline-adjustment checks}

Table~\ref{tab:application_sensitivity} records estimates, starting-value choices, root counts and score norms across the three perturbation scales.

\begin{table}[htbp]
\centering
\caption{JOBS II algorithm sensitivity across perturbation scales}
\label{tab:application_sensitivity}
\scriptsize
\begin{tabular}{rrrccccrr}
\toprule
Scale & PNIE & 95\% CI & Start M & Start Y & Roots M & Roots Y & Score M & Score Y \\
\midrule
0.05 & -0.011 & [-0.020, -0.001] & 1 & 1 & 1 & 1 & 2.05e-12 & 1.16e-12 \\
0.10 & -0.011 & [-0.020, -0.001] & 1 & 1 & 1 & 1 & 2.05e-12 & 1.16e-12 \\
0.20 & -0.011 & [-0.020, -0.001] & 1 & 1 & 1 & 1 & 2.05e-12 & 1.16e-12 \\
\bottomrule
\end{tabular}
\end{table}

Table~\ref{tab:application_covariates} removes pretreatment depression from both regressions while retaining all other terms, observations and numerical settings. The companion file \texttt{application\_covariate\_sensitivity.csv} contains all five effects for both estimators. Baseline-depression adjustment remains primary because pretreatment symptoms are relevant to post-intervention depression and mediator--outcome confounding.

\begin{table}[htbp]
\centering
\small
\caption{JOBS II sensitivity to adjustment for pretreatment depressive symptoms. The sample, remaining covariates and treatment--mediator interaction are unchanged. The specification including baseline depression is primary; all five effects are retained in the accompanying CSV}
\label{tab:application_covariates}
\begin{tabular}{lllrrr}
\toprule
Baseline depression & Method & Effect & Estimate & Lower & Upper \\
\midrule
Included & OLS delta & PNIE & -0.0136 & -0.0360 & 0.0087 \\
Included & OLS delta & TE & -0.0462 & -0.1281 & 0.0357 \\
Included & Proposed & PNIE & -0.0109 & -0.0205 & -0.0013 \\
Included & Proposed & TE & -0.0344 & -0.0962 & 0.0274 \\
Excluded & OLS delta & PNIE & -0.0198 & -0.0495 & 0.0100 \\
Excluded & OLS delta & TE & -0.0560 & -0.1460 & 0.0340 \\
Excluded & Proposed & PNIE & -0.0122 & -0.0230 & -0.0014 \\
Excluded & Proposed & TE & -0.0328 & -0.0964 & 0.0309 \\
\bottomrule
\end{tabular}
\end{table}

\FloatBarrier

\subsection{Residual diagnostics and working-model interpretation}

\begin{figure}[htbp]
\centering
\includegraphics[width=0.90\textwidth]{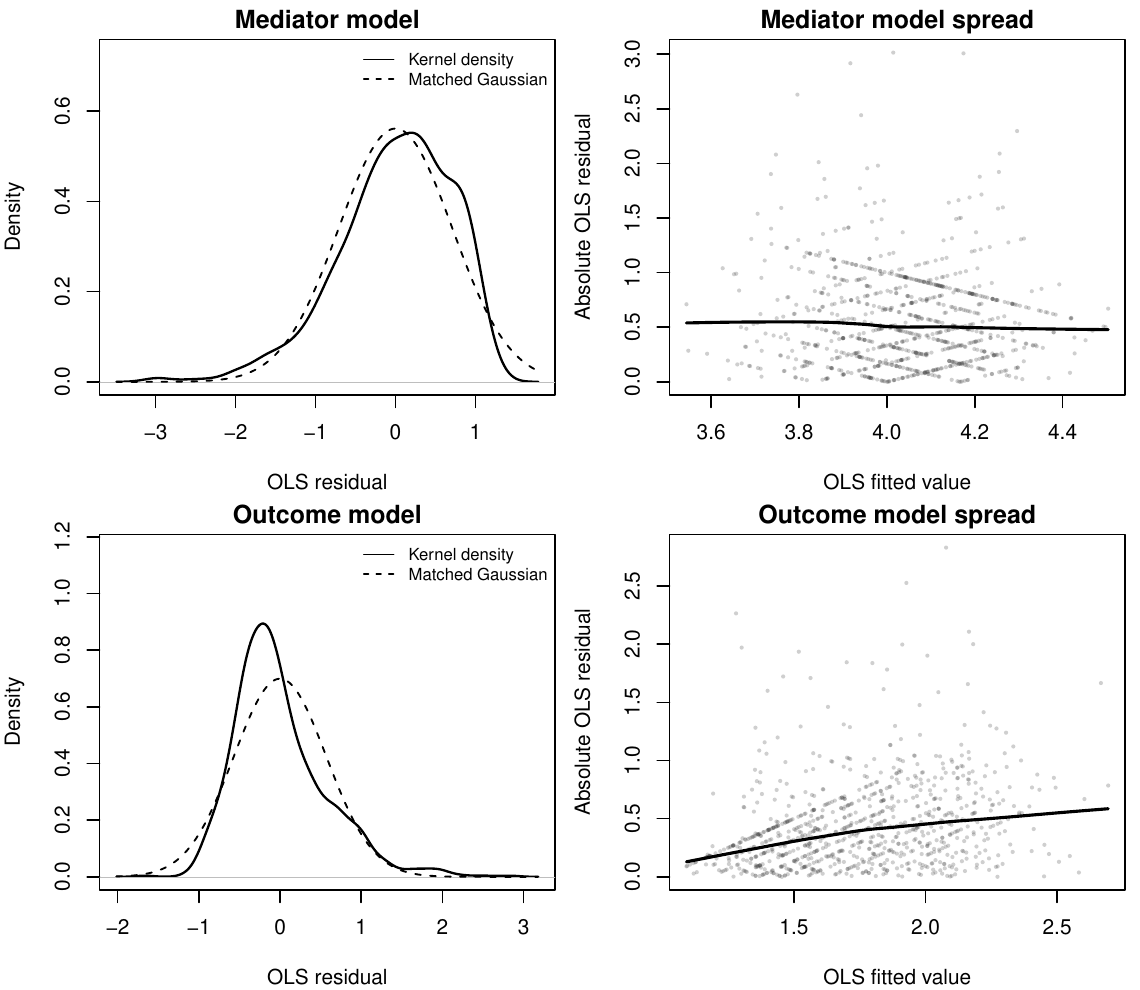}
\caption{Baseline-adjusted OLS residual diagnostics for JOBS II. Left panels compare kernel residual densities with Gaussians having the same residual mean and standard deviation. Right panels plot absolute residuals against fitted values with lowess smooths; the upper and lower rows correspond to the mediator and outcome equations}
\label{fig:jobs-residuals}
\end{figure}

For the residual-spread diagnostic, squared OLS residuals are regressed on all non-intercept design columns. The statistic $nR^2$ uses a chi-squared reference with 5 and 7 degrees of freedom for the mediator and outcome models, respectively. Table~\ref{tab:application_diagnostics} also reports skewness, excess kurtosis and one-dimensional spread tests using fitted values; Figure~\ref{fig:jobs-residuals} visualizes density shape and spread.

\begin{table}[htbp]
\centering
\caption{Baseline-adjusted OLS residual diagnostics. Normality $p$ is from Shapiro--Wilk. Fitted $p$ uses squared residuals regressed on fitted values; full LM is $nR^2$ using all original non-intercept design columns, with the displayed degrees of freedom (df). Tests are diagnostic checks, not tests of causal identification}
\label{tab:application_diagnostics}
\small
\begin{tabular}{lrrrrrrr}
\toprule
Equation & Skew. & Ex. kurt. & Normality $p$ & Fitted $p$ & Full LM & df & Full $p$ \\
\midrule
Mediator & -0.781 & 0.856 & 1.81e-14 & 3.00e-02 & 13.06 & 5 & 2.28e-02 \\
Outcome & 1.083 & 1.875 & 1.04e-18 & 8.43e-08 & 30.50 & 7 & 7.67e-05 \\
\bottomrule
\end{tabular}
\end{table}

\FloatBarrier

\subsection{Confounding-sensitivity illustration}

\begin{figure}[htbp]
\centering
\includegraphics[width=0.96\textwidth]{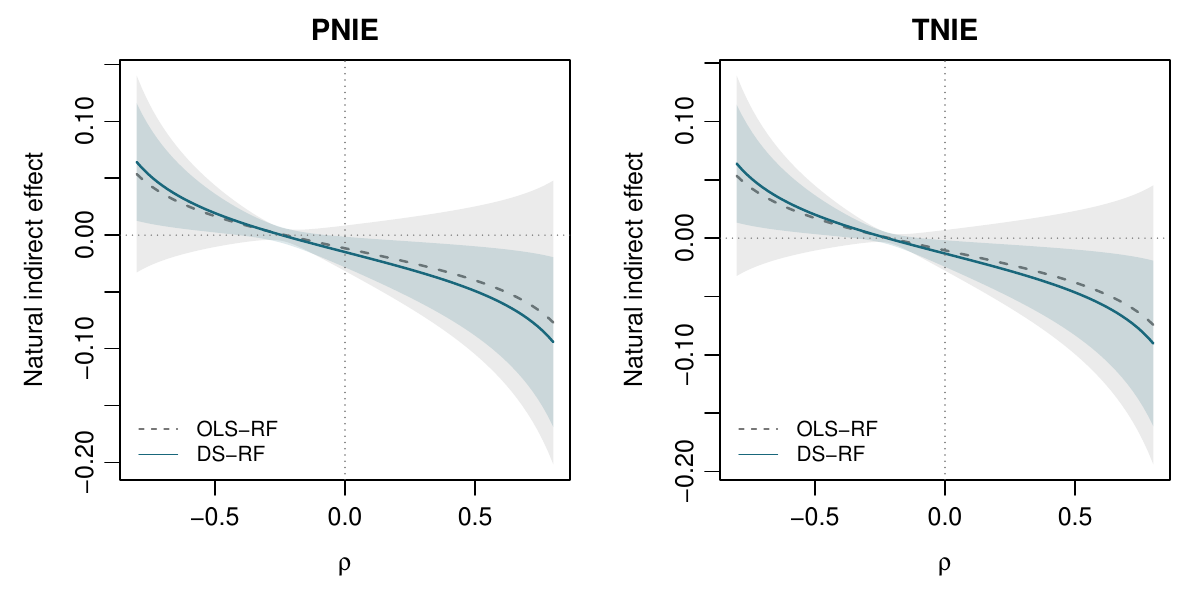}
\caption{JOBS II reduced-form PNIE/TNIE estimates over a common analyst-specified correlation $\rho$. Shading gives pointwise nominal 95\% intervals for each sensitivity value. The vertical reference is $\rho=0$ and the horizontal reference is effect zero. The curves display the range of results over the specified confounding correlations.}
\label{fig:confounding-curves}
\end{figure}

Figure~\ref{fig:confounding-curves} complements the main article's contours with pointwise PNIE/TNIE curves on the same sample and adjustment set. Separate reduced-form diagnostics assess the marginal-error restrictions in Section~\ref{sec:confounding-supp}. The project repository retains both signs of the residual-$R^2$ grid, zero-boundary intervals and regression diagnostics.

Interval widths describe precision at a specified correlation; point-estimate zero crossings describe the confounding value needed to reverse the estimated effect. Section~\ref{sec:confounding-supp} gives their distinct influence-function representations.
\FloatBarrier

\clearpage
\section{Reproducibility record}

The source code, JOBS II data object, retained numerical results and figure-generation scripts are available at \url{https://github.com/mijeong-kim/semi_causal_med}. The repository's \texttt{README.md} describes the required software, file organization and study-specific commands; the retained seeds, fit-status records, package versions and session information document the reported analyses.

From the repository root, \texttt{make smoke} and \texttt{make validate} check the implementation and retained results, and \texttt{make assets} rebuilds the tables and figures from those results without refitting the analyses. The validation scripts check the effect maps and gradients, both total-effect decompositions, numerical summaries and paired-fit denominators. Separate simulation, application and document-build session records preserve the provenance of the retained results.

\end{document}